%% file: paper.tex
\documentclass[sigconf]{acmart}

\copyrightyear{2020} 
\acmYear{2020} 
\setcopyright{acmlicensed}
\acmConference[ASIA CCS '20]{Proceedings of the 15th ACM Asia Conference on Computer and Communications Security}{October 5--9, 2020}{Taipei, Taiwan}
\acmBooktitle{Proceedings of the 15th ACM Asia Conference on Computer and Communications Security (ASIA CCS '20), October 5--9, 2020, Taipei, Taiwan}
\acmPrice{15.00}
\acmDOI{10.1145/3320269.3384764}
\acmISBN{978-1-4503-6750-9/20/06}


\usepackage{graphicx}
\usepackage{subcaption}
\usepackage{tabularx}
\usepackage{listings}
\usepackage{caption}
\usepackage{float}
\usepackage[ruled,vlined,linesnumbered]{algorithm2e}
\usepackage{tikz,pgfplots,pgfplotstable}
\usepackage{arydshln}
\usepackage{booktabs}
\usepackage[english]{babel}
\usepackage{makecell}
\pgfplotsset{compat=1.14} 
\usepackage[htt]{hyphenat}
\usepackage{amsmath}

\begin{document} 

\title[The Taint Rabbit]{The Taint Rabbit: Optimizing Generic Taint Analysis with Dynamic Fast Path Generation}

\author{John Galea}
\affiliation{
  \department{Department of Computer Science}              
  \institution{University of Oxford \\ \href{mailto:john.galea@cs.ox.ac.uk}{john.galea@cs.ox.ac.uk}}
}

\author{Daniel Kroening}
\affiliation{
  \department{Department of Computer Science}            
  \institution{University of Oxford \\ \href{mailto:daniel.kroening@gmail.com}{daniel.kroening@gmail.com}}
}

\begin{abstract}
\input{abstract.tex}

\end{abstract}
 
\maketitle

\input{introduction.tex}

\input{overview.tex}

\input{background.tex}

\input{generic_taint_analysis.tex}

\input{optimization.tex}

\input{evaluation.tex}

\input{related_work.tex}

\input{conclusion.tex}

\input{acknowledgement.tex}

\bibliography{../my_bib} 

\pagebreak
\input{appendix.tex}

\end{document}

%% file: abstract.tex
Generic taint analysis is a pivotal technique in software security. However, it
suffers from staggeringly high overhead.  In this paper, we explore the
hypothesis whether just-in-time (JIT) generation of fast paths for tracking
taint can enhance the performance.  To this end, we present the \emph{Taint
Rabbit}, which supports highly customizable user-defined taint policies and
combines a JIT with fast context switching.  Our experimental results
suggest that this combination outperforms notable existing implementations
of generic taint analysis and bridges the performance gap to specialized
trackers.  For instance, Dytan incurs an average overhead of 237x, while the
Taint Rabbit achieves 1.7x on the same set of benchmarks.  This compares
favorably to the 1.5x overhead delivered by the bitwise, non-generic, taint
engine LibDFT.

%% file: introduction.tex
\section{Introduction}

Dynamic taint analysis~\cite{schwartz2010all} is an enabling technique in
software security for tracking information flows.  Typical applications
include malware analysis~\cite{yin2007panorama, korczynski2017capturing,
bayer2009scalable}, vulnerability discovery~\cite{rawat2017vuzzer,
caballero2012undangle, chen2018angora} and runtime attack
detection~\cite{newsome2005dynamic, lerner2010empirical, vogt2007cross}. 
The key feature is the tracking of memory locations and CPU registers that
store ``interesting'' or ``suspicious'' data.  Data of this kind is called
\emph{tainted}.  Taint is checked at particular points during program
execution to determine whether certain runtime properties hold, e.g.,
to detect if the instruction pointer could be controlled by an
attacker~\cite{newsome2005dynamic}, or to identify which parts of the user
input influence path conditions to optimize fuzzing~\cite{rawat2017vuzzer,
chen2018angora}.

Most taint analyzers, e.g., LibDFT~\cite{kemerlis2012libdft}, implement
single, byte-sized tags and often just track whether data is tainted or not. 
This setup supports efficient propagation of taint (using a \textit{bitwise
or}) and efficient querying of a location's taint status.  However, many
interesting applications that build upon taint analysis require richer
propagation logic and more complex taint labels.  For instance,
VUzzer~\cite{rawat2017vuzzer} propagates sets that contain offsets of input
bytes, and Undangle~\cite{caballero2012undangle} tracks heap pointers,
storing taint information in a composite data structure.  To support these
use cases, previous work proposed to extend the single tags and
deliver what is called \emph{generic taint analysis}.  Generic taint engines
track richer labels (say via a 32-bit pointer) and support
user-defined taint propagation policies.  The first notable generic taint
engine is Dytan~\cite{clause2007dytan}.

The key problem is that the versatility of Dytan comes at a price: the
authors of Dytan report a staggering runtime overhead of $\sim$30x on
\texttt{gzip}, which has led to the perception that generic taint analysis
is, in essence, impractical.  We challenge this perception, and explore the
hypothesis whether generic taint analysis can be delivered with a
runtime overhead that is low enough for practical security applications.

We argue that a combination of two optimizations is able to deliver taint
tracking that is both versatile and sufficiently fast.
We present an implementation of our ideas in a tool called the \emph{Taint Rabbit}. 
The Taint Rabbit achieves an overhead that is significantly lower than that of
the generic taint analyzer Dytan.  On the CPU-bounded benchmarks that Dytan 
manages to run, we observe an overhead of 237x compared to native execution.
By~contrast, the Taint Rabbit incurs only 1.7x.  This is close to
what can be expected: LibDFT, the leading bitwise taint engine,
achieves an overhead of 1.5x on the same benchmarks.
Therefore, our approach reduces the conflict between
performance and versatility significantly.

\textbf{The Taint Rabbit is Generic.} Our taint propagation is not specific
to a fixed taint policy. We map a 32-bit word (or pointer) to every tainted
byte, enabling the storage of a reference to a custom taint label data structure. 
The Taint Rabbit propagates the pointers efficiently, and supports custom
handlers, provided by the user, to update the taint labels according to a
desired taint policy.  Section~\ref{high-level_tb} details our algorithms
for generic taint analysis.

\textbf{The Taint Rabbit is Optimized.} The key idea behind the Taint
Rabbit's high performance is to optimize taint analysis for dynamic binary
instrumentation (DBI)~\cite{bruening2012transparent}.  This approach is
standard in leading bitwise taint analyzers, such as
LibDFT~\cite{kemerlis2012libdft}, but has not yet been thoroughly
investigated for generic taint analysis, which is much harder to optimize. 
In~particular, it is not possible to build instrumented instruction
handlers for taint propagation using simple \textit{bitwise or} operations;
the taint propagation has to be optimized for a given, custom taint
policy.  We~investigate two techniques to speed up generic taint analysis. 
First, we reduce analysis overhead just-in-time by dynamically generating
fast paths according to \textit{in} and \textit{out} taint states of basic
blocks.  Second, our generic analysis avoids expensive context-switching
by limiting function calls.  Section~\ref{optimisated_imp} details these
optimizations.

The Taint Rabbit's generic capabilities are assessed using three security
applications, which employ different taint policies.  The applications have
been proposed previously but have not yet leveraged our efficient taint
engine.  Specifically, we evaluate an exploit detector~\cite{newsome2005dynamic}, 
a Use-After-Free debugger~\cite{caballero2012undangle} and a fuzzer~\cite{rawat2017vuzzer}.

To measure performance, we use SPEC CPU 2017~\cite{bucek2018spec},
command-line utilities, PHP and Apache as benchmarks.  As baselines for
comparison, we conduct the same experiments on a wide range of alternative
trackers, including LibDFT~\cite{kemerlis2012libdft},
DataTracker~\cite{stamatogiannakis2014looking},
DataTracker-EWAH~\cite{rawat2017vuzzer}, Traintgrind~\cite{weitaintgrind},
 BAP-PinTraces~\cite{brumley2011bap}, Triton~\cite{SSTIC2015-Saudel-Salwan},
Dr.~Memory~\cite{bruening2011practical}, DECAF~\cite{henderson2017decaf} and Dytan~\cite{clause2007dytan}. 
Our results show that the Taint Rabbit is the fastest generic taint
tracker among those evaluated.

In summary, we make the following contributions:
\begin{enumerate} 

\item \textbf{Optimized and generic taint analysis.} While optimized
taint analyses have been proposed, none support extensible propagation logic, 
and thus lack versatility.  Meanwhile, existing generic taint engines
incur prohibitively high overheads.  Our
contribution bridges this gap via dynamic fast path generation and
instrumentation that avoids calls.

\item \textbf{The Taint Rabbit.} We introduce a framework
for building security applications based on dynamic taint analysis. The
Taint Rabbit and the tools built upon it are all available at \href{https://github.com/Dynamic-Rabbits/Dynamic-Rabbits}{https://github.com/Dynamic-Rabbits/Dynamic-Rabbits}.
  
\item \textbf{An extensive evaluation.} The Taint Rabbit is evaluated on
several relevant benchmarks, including SPEC CPU 2017, and is compared with
\textit{nine} other taint-based systems.  Three security
applications are also assessed to demonstrate the versatility of the Taint Rabbit.

\end{enumerate}

%% file: overview.tex
\section{Overview}

\begin{figure}[t]
\centering
 \includegraphics[width=0.45\textwidth]{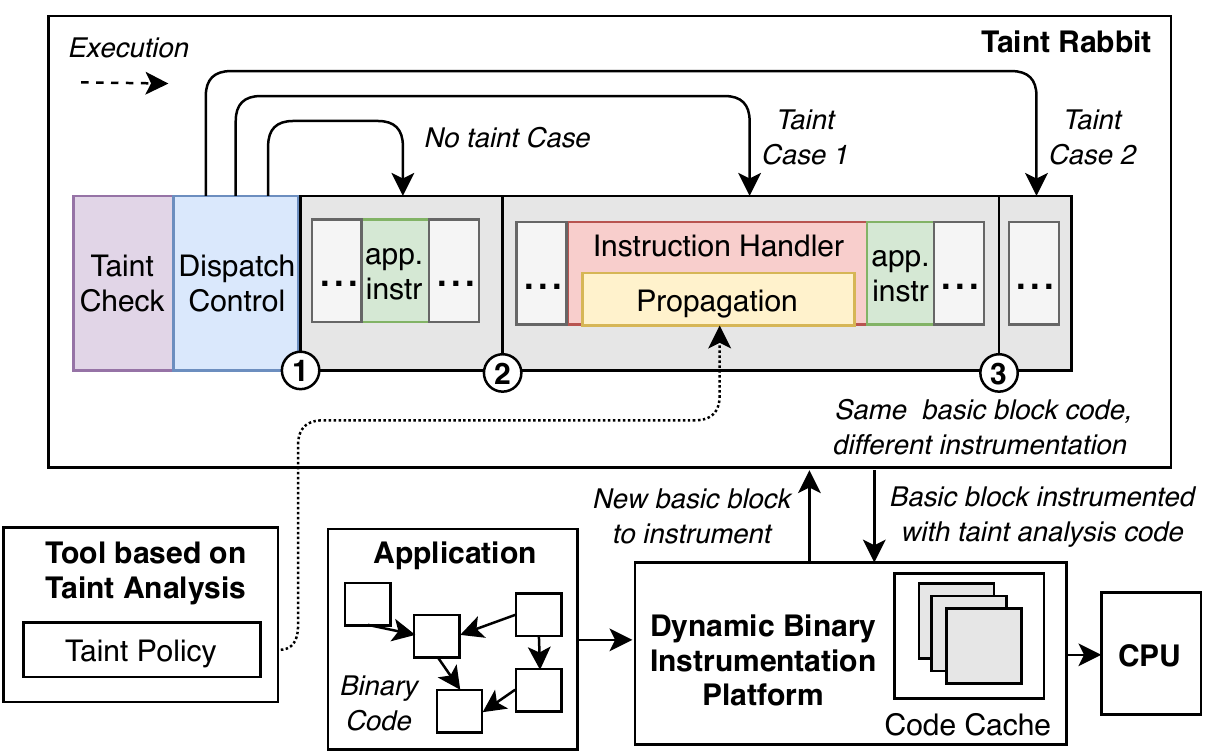}
\caption{High-level design of the Taint Rabbit}
\label{fig:high_level_design}
\end{figure}

Figure~\ref{fig:high_level_design} illustrates the high-level design of our
approach.  The DBI platform passes every new basic block observed during
runtime to the Taint Rabbit for instrumentation.  The Taint Rabbit
weaves efficient instruction handlers, responsible for propagating taint
generically, into the application's code.  Instruction handlers are
implemented in assembly to limit context-switching done by transparent
function calls.

The Taint Rabbit employs a JIT approach to adaptively enhance the
performance of generic taint analysis.  It~generates copies of original
basic blocks but leaves them uninstrumented to establish fast
paths.  In particular, the uninstrumented basic block is executed when all of
its input and output registers/memory are not tainted.  Otherwise, the slow
path is taken, implementing full-blown taint analysis.  The basic variant of
this scheme has been proposed previously and implemented in
Lift~\cite{qin2006lift}, but the Taint Rabbit can do more: it~also
\textbf{dynamically} generates fast paths for the case when taint is present.  If~the Taint Rabbit
encounters a set of \textit{in} and \textit{out} taint states that are
frequently executed at runtime, the basic block is duplicated again and
instrumented specifically to handle the particular case.  Irrelevant instructions that
do not deal with taint in the given case are safely elided from instrumentation. 
Therefore, fully-instrumented code is executed less often than in
conventional approaches as, owing to the additional fast paths,
control is not always blindly directed to it when taint is encountered.
Our technique is based on the hypothesis that
basic blocks are usually executed with the same taint states. Therefore,
the cost of generating fast paths for these states pays off.

Our approach allows the user to focus on defining the desired taint
propagation policy, while the Taint Rabbit facilitates fast and generic
taint analysis.  Inspired by previous work~\cite{yin2010temu,
chang2008efficient}, the user provides code describing how labels are
merged and derived, without delving into intricacies of the internals of
the engine.

%% file: background.tex
\section{Background}
\label{sec:background}

\input{taint_analysis_applications.tex}

\input{taint_analysis_DBI.tex}

%% file: taint_analysis_applications.tex
\subsection{Applications of Taint Analysis}
\label{sect:uses}

There are numerous use-cases for taint analysis.  We~give three example
applications and emphasize that their taint policies and taint propagation
logic differ.

\label{fuzzing_prev_work}

\begin{example}
\textbf{Control-Flow Hijacking.} Previous work~\cite{newsome2005dynamic} has
shown that taint analysis can detect control-flow hijacking attacks.  Since
the analysis only has to taint check control data, \textit{bitwise or}
operations suffice for propagating taint status flags.

\end{example}

\begin{example}\label{sect:uaf_policy}
\textbf{UAF Detection.}  Use-after-free (UAF) bugs are exploitable \cite{serna2012info}. 
Undangle~\cite{caballero2012undangle} debugs such vulnerabilities by
tracking heap pointers via taint analysis.  Undangle monitors allocations
and deallocations and assigns the pointer status stored in the taint labels to
\textit{LIVE} and \textit{DANGLING}, respectively.  Taint is propagated when
pointers are copied either directly or arithmetically, and a location is
untainted if it is no longer a pointer.  For instance, the subtraction of
two pointers yields a taint-free distance even though both the sources are
tainted.

A \textit{bitwise or} operation is not suitable for pointer tracking;
a~location may be associated with one of three states, namely
\textit{NO-TAINT}, \textit{LIVE}, and \textit{DANGLING}, and their merging
cannot be appropriately done with the operation.  Moreover, apart from
the status of the pointer, Undangle's labels also contain debugging data, e.g., PCs of
pointer creations, and thus are of composite type.  Instead of using a
\textit{bitwise or}, Algorithm~\ref{algo:2} (in the appendix) gives an
implementation for propagating such labels via conditional statements.
\end{example}

\begin{example} \textbf{Fuzzing.} VUzzer~\cite{rawat2017vuzzer} uses taint
analysis to discover interesting input bytes to mutate.  The label is a bit
set, where each bit corresponds to a byte of the input file.  Since
registers or memory may be influenced by multiple bytes, propagation
performs a union operation.  A~\textit{bitwise or} is sufficient if bit sets
fit within the operand size; however, this is unlikely as the input files of
interest may be several kilobytes large.  VUzzer therefore uses a bit array,
which implies that the union operations require branching.
\end{example}

While bitwise tainting is appropriate for some applications, others require
richer capabilities. Yet, many taint engines are tuned solely for the
former~\cite{cheng2006tainttrace, qin2006lift, bosman2011minemu}.  Our approach is more versatile and suitable for all use cases.

%% file: taint_analysis_DBI.tex
\subsection{Taint Analysis via DBI}

Similar to previous research~\cite{cheng2006tainttrace, kemerlis2012libdft,
clause2007dytan, qin2006lift}, we focus on an online analysis that is
implemented using dynamic binary instrumentation
(DBI)~\cite{bruening2012transparent}.  In DBI, basic blocks of the
application under analysis are instrumented and stored in a
code cache at runtime.  The inserted code needs to be transparent so that it
does not affect the execution of the application.  To simplify tool
development, DBI frameworks~\cite{nethercote2007valgrind}, such as
Pin~\cite{luk2005pin} and DynamoRIO~\cite{bruening2012transparent}, allow
the insertion of transparent calls, known as \textit{clean
calls}~\cite{drdoc}, which invoke a given function at runtime. Essentially, these functions
implement the taint analysis. However, before the call, a context switch is performed, which
creates a dedicated stack and comprehensively spills/restores the CPU
registers~\cite{uh2006analyzing}.  Since taint analysis requires
instrumenting many instructions to track data movements, these context
switches incur high overheads of at least $\sim$15x\footnote{To~quantify this overhead, we
ran the DynamoRIO tool \texttt{inscount} that uses clean calls to count the
number of instructions executed by an application.  We see a slowdown of
$\sim$15x on SPEC CPU 2017 (Figure~\ref{fig:ccvsin} in the appendix). \href{https://github.com/DynamoRIO/dynamorio/blob/master/api/samples/inscount.cpp}{https://github.com/DynamoRIO/dynamorio/blob/master/api/samples/inscount.cpp}}.

Consequently, DBI frameworks attempt to avoid clean calls and automatically
inline analysis code with the application's instructions.  Ideally, the
context switches only spill/restore live registers used by the routines and
therefore are cheaper than full clean calls.  Figure~\ref{fig:ccvsin} (in
the appendix) shows that this optimization reduces the overhead to
$\sim$3.3x.  Routines are inlined by DBI frameworks only if they are \textit{simple}, i.e., 
they are small, avoid control-flow and perform no function calls themselves~\cite{pindoc, drdoc}.

LibDFT exploits the inline optimization.  Listing~\ref{code:libdfthandle} in
the appendix shows one of its taint propagation routines.  
Essentially, propagation is done by bitwise tainting, which avoids
long complicated code with conditional branches. 
Notably, the use of bit flags as taint labels, combined with
bitwise operations for propagation, yields simple
routines, thus activating the inline optimization.

However, the propagation supported by LibDFT is limited.  Previous
work~\cite{stamatogiannakis2014looking} has extended LibDFT to track input
file offsets.  The work increased LibDFT's versatility,
resulting in a new taint engine called DataTracker.  With some modifications,
DataTracker is used by VUzzer.  However, the changes made in
DataTracker break the original inline optimization. 
Listing~\ref{code:datatrackerhandle} (in the appendix) gives the instruction
handler that corresponds to the one in listing~\ref{code:libdfthandle}. 
Because of the function calls and the branching in the instruction handler,
Pin fails to inline and the performance drops.  We~ran DataTracker and
confirmed the failure to inline by inspecting the logs produced by Pin. 
Our results on \texttt{bzip2} also show that LibDFT is faster than
DataTracker: LibDFT has an overhead of 2.6x, while DataTracker incurs 36x
over native execution.

Although existing optimizations for propagating taint are effective, many
are dependent on specific policies and taint label structures.  LibDFT's
inlining approach is mainly suitable for bitwise tainting.  We believe that
optimizations not tied to particular policies are desirable as they are
more useful to the community who use taint analysis for a broad 
range of applications.

%% file: generic_taint_analysis.tex
\section{The Taint Rabbit}

\label{high-level_tb}

Generic taint analysis enables user-defined taint policies.  The support for
custom merging of labels during propagation removes the
need to change the internals of the taint engine for a particular application. 
We now describe the Taint Rabbit's high-level algorithms.  Our optimizations
are then detailed in the next section.

\textbf{Binary Analysis.} We scope our analysis to x86 binaries.  The code
that performs propagation considers the semantics of the instructions. 
This avoids tainting output locations unnecessarily, e.g., tainting stack
pointers.

\textbf{Generic Label Structure.} The unit of meta-data that the Taint
Rabbit uses as a label is a 32-bit word.  The word may itself store tags or
act as a pointer to a larger taint label data structure\footnote{In contrast
to the Taint Rabbit, Dytan~\cite{clause2007dytan} uses a bit vector as
its label structure instead of a generic pointer. The number of bits is
configurable at compile time.}.  A \texttt{NULL} value
represents ``no taint''.

\textbf{Byte Granularity.} Meta-data is mapped to every byte in memory and
registers; e.g, a \texttt{mov} \texttt{eax,} \texttt{ebx} propagates four
labels, one for each byte in \texttt{ebx}.  Labels are stored in shadow
memory~\cite{zhao2010umbra}. We do not label the x86 flag register to avoid
\textit{taint explosion}~\cite{zhu2011tainteraser}.

\label{sect:taint_algo}

\textbf{Generic Taint Propagation.} As illustrated in
Figure~\ref{fig:primitives}, taint labels are propagated via user-defined
code called \textit{taint primitives}.  A~taint primitive is a building
block for taint propagation, and is responsible for deriving a taint label
from a set of source taint labels.  During propagation, the Taint Rabbit fetches
the labels of the source operands, applies the appropriate primitives with
respect to the semantics of the x86 instructions, and assigns the resulting
labels to the destination operands.

\newlength{\twosubht}
\newsavebox{\twosubbox}

\begin{figure*}[htp]

\sbox\twosubbox{%
  \resizebox{\dimexpr.70\textwidth-1em}{!}{%
    \includegraphics[height=1cm]{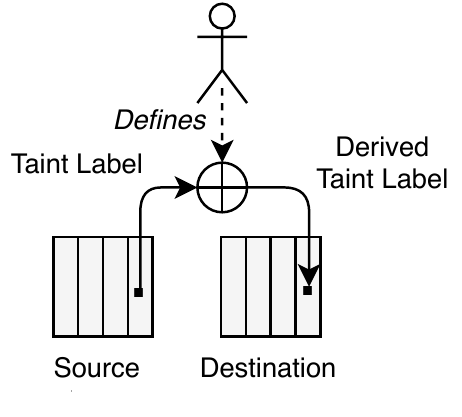}%
    \includegraphics[height=1cm]{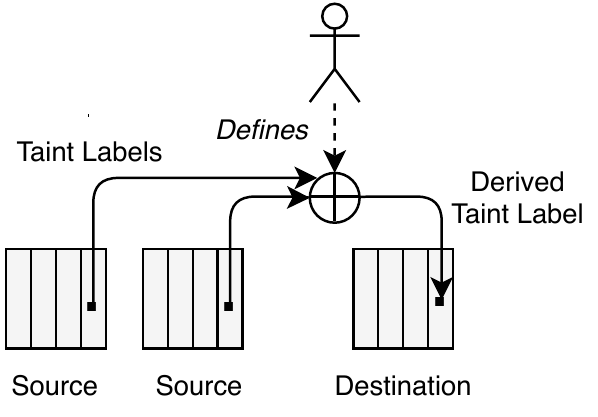}%
    \includegraphics[height=1cm]{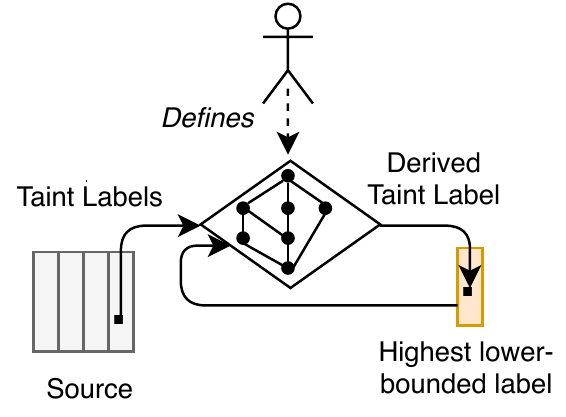}%
  }%
}
\setlength{\twosubht}{\ht\twosubbox}

\centering

\subcaptionbox{Primitive $\mathit{src} \rightarrow \mathit{dst}$}{%
  \includegraphics[height=\twosubht]{images/srcdst.pdf}%
}\quad
\subcaptionbox{Primitive $\mathit{src}, \mathit{src} \rightarrow \mathit{dst}$}{%
  \includegraphics[height=\twosubht]{images/srcsrcdst.pdf}%
}\quad
\subcaptionbox{Primitive $\mathit{src}, \mathit{src} \rightarrow_{M} \mathit{meet}$}{%
  \includegraphics[height=\twosubht]{images/meet.pdf}%
}
\caption{User-defined primitives invoked by the Taint Rabbit to propagate taint}
\label{fig:primitives}

\end{figure*}

Three user-defined taint primitives are currently required for our supported
instructions, and are informally defined as
(1)~$\mathit{src} \rightarrow \mathit{dst}$,
(2)~$\mathit{src}, \mathit{src} \rightarrow \mathit{dst}$, and
(3)~$\mathit{src}, \mathit{src} \rightarrow_{M} \mathit{meet}$.
The first two primitives produce a label to associate with a destination
byte from one and two sources respectively.  For example, Algorithm~\ref{algo:2}
(in the appendix) is a $\mathit{src}, \mathit{src} \rightarrow
\mathit{dst}$ primitive.  Meanwhile, inspired by previous
work~\cite{chang2008efficient}, the third primitive computes the highest
lower-bounded label for two given labels in a lattice.

The taint primitives are the interface between the user-defined taint policy
and the Taint Rabbit.  We found that these three taint primitives are
sufficient for our instruction handlers to track taint effectively and with
reasonable precision, even for complex x86 instructions such as
\texttt{punpckldq} and \texttt{pmaddwd}.

The primitives~(2) and (3) serve very different purposes despite the fact
that they have the same signature.
\begin{itemize}
\item The primitive $\mathit{src}, \mathit{src} \rightarrow \mathit{dst}$ is
used to combine the labels of two bytes stemming from different sources.

\item The primitive $\mathit{src}, \mathit{src} \rightarrow_{M} \mathit{meet}$ is
used to combine the labels of two bytes found within one source.
\end{itemize}
Essentially, the $\mathit{meet}$ primitive provides means to compute a
single label that summarizes the taint of a multi-byte operand.  The two are
incomparable.  We recall the pointer tracking use-case
(Example~\ref{sect:uaf_policy}) to illustrate this point.  Given two labels
that represent a \textit{LIVE} and \textit{DANGLING} status, respectively,
their \textit{meet} is \textit{DANGLING}, while the combination of two
pointers is \textit{NO-TAINT}.

Algorithm~\ref{algo:over_approx_prim2} specifies how the Taint Rabbit uses
primitives~(2) and~(3) to compute the taint label for an instruction with
two operands. The algorithm first iterates over the bytes of each operand
separately and applies the $\mathit{meet}$ primitive on these bytes.
This yields one taint label for each of the two operands, denoted by
$\mathit{meet\_label}_1$ and $\mathit{meet\_label}_2$, respectively.

In Line~11, the algorithm then iterates over the bytes of the destination
operand, and uses the $\mathit{src}, \mathit{src} \rightarrow \mathit{dst}$
primitive to combine the labels of the two source operands.  The combined
label is then assigned to the destination\footnote{We remark that the
primitive may be stateful, and hence, its invocation is not hoisted out of
the loop.}.

\begin{algorithm}[t]
 \SetAlgoLined
 \KwData{ID $\mathit{dst}$, ID $\mathit{src}_{1}$, ID $\mathit{src}_{2}$, Integer $\mathit{opnd\_size}$ }
 
 $\mathit{meet\_label}_{1}$ $\gets$ $\mathit{NULL}$\;
 
 \For{$i \gets 0$ to $\mathit{opnd\_size} - 1$}{
    $\mathit{label}$ $\gets$ lookup\_label($\mathit{src}_{1} + i$)\;
    $\mathit{meet\_label}_{1}$ $\gets$ meet$_\mathit{primitive}$($meet\_label_{1}$, $\mathit{label}$)
 } 
   
 $\mathit{meet\_label}_{2}$ $\gets$ $\mathit{NULL}$\;

 \For{$i \gets 0$ to $\mathit{opnd\_size} - 1$}{
    $\mathit{label}$ $\gets$ lookup\_label($\mathit{src}_{2} + i$)\;
    $\mathit{meet\_label}_{2}$ $\gets$ meet$_\mathit{primitive}$($meet\_label_{2}$, $\mathit{label}$)
  }

 \For{$i \gets 0$ to $\mathit{opnd\_size} - 1$}{

    $\mathit{dst\_label}$ $\gets$ src\_src\_dst$_\mathit{primitive}$($\mathit{meet\_label}_{1}$, $\mathit{meet\_label}_{2}$)\;

    set\_label($\mathit{dst} + i$, $\mathit{dst\_label}$)\;
  }

\caption{Computing the taint label for a two-operand instruction}
\label{algo:over_approx_prim2}
\end{algorithm}

\textbf{Optimization.} For many x86 instructions, resulting bytes
are independent.  Instances of this are most transfer instructions
(e.g.~\texttt{mov}) and many bit-manipulating instructions (e.g.~\texttt{or},
\texttt{xor}).  The semantics of these instructions guarantee that byte $i$
of the result only depends on byte $i$ of the first and byte $i$ of the
second operand.  For this case, the two loops that merge the taint labels
can be dropped.  Algorithm~\ref{algo:under_approx_prim2} gives the resulting
specialized instruction handler.  Algorithm~\ref{algo:under_approx_prim2} is
both faster than Algorithm~\ref{algo:over_approx_prim2} and produces a
result that is more precise.

\begin{algorithm}[t]
 \SetAlgoLined
 \KwData{ID $\mathit{dst}$, ID $\mathit{src}_{1}$, ID $\mathit{src}_{2}$, Integer $\mathit{opnd\_size}$ }
  
 \For{$i \gets 0$ to $\mathit{opnd\_size} - 1$}{
    $\mathit{src\_label}_{1}$ $\gets$ lookup\_label($\mathit{src}_{1} + i$)\;
    $\mathit{src\_label}_{2}$ $\gets$ lookup\_label($\mathit{src}_{2} + i$)\;

    $\mathit{dst\_label}$ $\gets$ src\_src\_dst$_\mathit{primitive}$($\mathit{src\_label}_{1}$, $\mathit{src\_label}_{2}$)\;
    
    set\_label($\mathit{dst} + i$, $\mathit{dst\_label}$)\;
}
\caption{Optimized tainting for instructions with independent bytes}
\label{algo:under_approx_prim2}
\end{algorithm}

LibDFT uses the approach taken in Algorithm~\ref{algo:under_approx_prim2}
even in cases when bytes may affect each other (e.g.~\texttt{add}); it
therefore under-approximates and may lose taint in return for a
performance gain.

Algorithms~\ref{algo:over_approx_prim1} and~\ref{algo:under_approx_prim1},
which illustrate the usage of the $\mathit{src} \rightarrow \mathit{dst}$
primitive, are in the appendix.  They are similar to
Algorithms~\ref{algo:over_approx_prim2} and~\ref{algo:under_approx_prim2},
but only accept one source operand.  We implement them to propagate taint
for instructions such as \texttt{mov}, \texttt{inc} and \texttt{bswap}.

%% file: optimization.tex
\section{Optimized Design}

\label{optimisated_imp}

The previous section describes the Taint Rabbit's high-level algorithms for
generic taint analysis.  We now focus on the Taint Rabbit's design optimized
for DBI.

\subsection{Challenges}

We address the following non-trivial challenges: 

\textbf{High Tracking Rate.} Dynamic tainting incurs overhead due to
the high execution rate of instruction handlers.  On a test run, we measured
that at least 73\% (over 8 billion) of the instructions executed by
\texttt{bzip2} conventionally require instrumentation (excluding instructions such as \texttt{jmp} and \texttt{cmp}).  We address this challenge in Section~\ref{sect:contrib1}.
  
\textbf{Expensive Context Switching.} Unlike bitwise tainting, generic taint
propagation is more complex, e.g., because of complex control flow.  This
leads to expensive context-switching incurred by clean calls.  We address
this challenge in Section~\ref{sect:contrib2}.

\subsection{Dynamic Fast Path Generation} 
\label{sect:contrib1}

The Taint Rabbit generates fast paths to reduce the execution of instruction
handlers. We now detail the actual process of the Taint Rabbit.  A code
example is given in Figure~\ref{fig:dfpg_stages} in the appendix.

\textbf{Truncation.} When a new basic block is provided by the DBI platform,
the Taint Rabbit begins by identifying any memory addresses that cannot be
determined at the start of the basic block due to non-static dependencies. 
Such addresses are problematic as their taint status cannot be checked
prior to entering a fast path at runtime.  The issue is mitigated by truncating
basic blocks at points where memory dereferences are calculated based on
register values that are inconsistent with their starting
values\footnote{Our implementation reduces the impact of dynamic
dependencies using constant propagation.  For example, instead of
truncating upon \texttt{push} and \texttt{pop} instructions,
the Taint Rabbit patches operands with the
offsets calculated by decrementing/incrementing the stack pointer.}.  The
cut-off code is no longer considered at this point, but is treated as a
new separate basic block that undergoes its own analysis.  The input and
output operands are then retrieved and stored in a set by simply
inspecting the remaining instructions found before the cut-off point.

\textbf{Code Duplication.} Next, the basic block is copied to produce
multiple adjacent instances of it.  A global map $\mathcal{M}$ associates a
basic block ID with meta-data specifying the different cases of
instrumentation; ergo, the number of cases determines the total number of
basic block instances.  By default, this meta-data is initialized with two
defined cases where all or none of the basic block's inputs and outputs are
tainted.  An entry label is inserted prior to each instance, and direct
jumps are inserted at the ends to span over the code of other instances and
exit.  To maintain the one-exit-point property of basic blocks, control-flow
instructions in the analyzed code of the application are not duplicated.

\textbf{Taint Checks and Control Dispatch.} The Taint Rabbit proceeds
by inserting initial code to determine the \textit{in} and \textit{out}
runtime taint states of a basic block at point of entry.  The result is
encoded as a mask where each bit indicates whether or not an input/output is
tainted.  Compare and branch code sequences check the encoded mask with the
masks of the defined cases (retrieved via $\mathcal{M}$) and direct control
to appropriate basic block instances.  Fall-through implies that a new
case is encountered, which is an opportunity for fast path generation. 
An unhandled case defaults to the execution of the fully instrumented basic
block.

Placed in the common path, taint checking is performance critical.  The
dispatcher \textit{must} direct control fast.  Determining the taint status
of an input/output by inspecting all of its pointer-sized tags one-by-one is
costly because of a large number of comparison instructions and cache pollution.  The Taint
Rabbit alleviates this issue by quickly checking registers via an
over-approximation where a taint status bit is tracked for each register (as
opposed to each byte in each register).  Apart from conducting generic taint
analysis, our instrumented paths also maintain these status bits. 
Therefore, a lot of the dispatcher's checking process is shifted down to
paths that are less critical, away from the uninstrumented fast path.  The
idea of using over-approximate tags is similar to~\cite{saxena2008efficient},
but the Taint Rabbit cleverly uses the \texttt{pext}
instruction~\cite{intel2014intel} to construct the mask quickly.  Although
checks are imprecise owing to the higher granularity (sub-registers may
considered as tainted when they are not), taint propagation is still
performed by our byte-precise instruction handlers.  Profiling done during
development showed that the use of shared tags alone led to
a speed-up of $\sim$0.6x over native execution. Although shared tags are
only associated to registers, the Taint Rabbit leverages SIMD instructions 
to efficiently test multiple labels simultaneously when taint checking
memory.

\textbf{Data-flow Analysis and Instrumentation.} The basic blocks are then
instrumented with taint propagation code.  The paths for the two default
cases are established by creating one fully instrumented basic block and
maintaining another without any instrumentation at all.  To handle other
cases identified at runtime, forward data-flow analysis is performed on the
basic block to determine which instructions deal with tainted operands. 
Such instructions propagate taint at runtime and are therefore instrumented, while
others are elided.  Naturally, the initial in-set for data-flow analysis
includes the \textit{in} and \textit{out} taint states of the particular
case.

Inlining instrumentation code which is based on user-defined taint
primitives may result in large code fragments that stresses the instruction
cache and the encoding to the DBI cache. This issue is exacerbated by the
instrumentation of duplicated basic blocks.  As a mitigation, the Taint
Rabbit outlines instruction handlers to shared code caches at the user's
discretion. Note that outlining does not use clean calls but trampolines.

\begin{figure}[t]
\centering
 \includegraphics[width=0.41\textwidth]{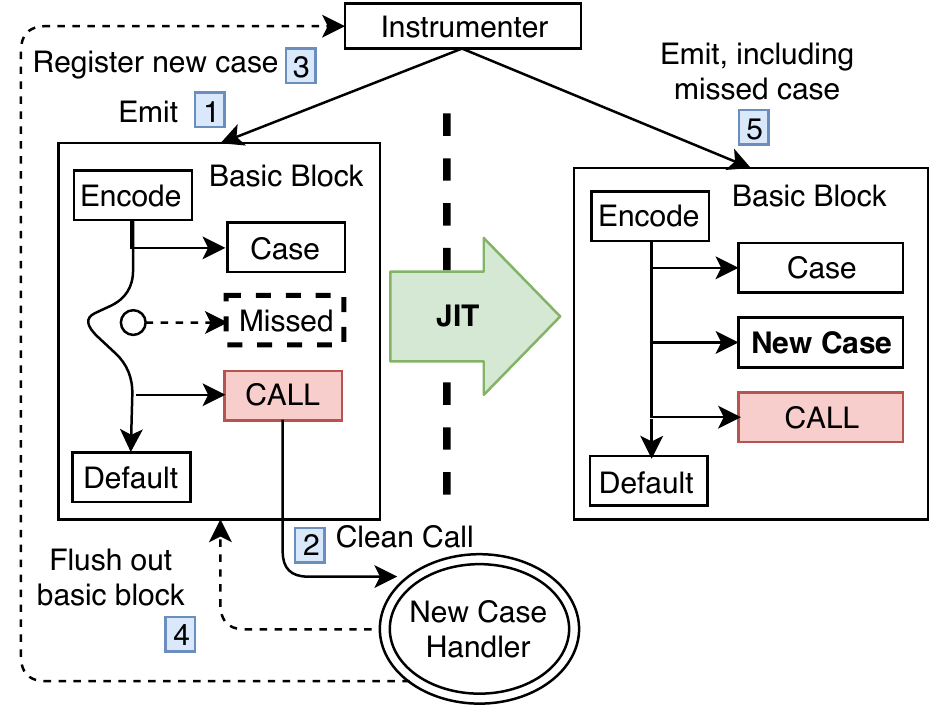}
\caption{Dynamic Fast Path Generation}
\label{fig:fast_path_generation_fault}
\end{figure}

\textbf{Fast Path Generation.} Figure~\ref{fig:fast_path_generation_fault}
describes the process of dynamic fast path generation.  A clean call is
performed (infrequently) when no fast path exists for a new set of
\textit{in} and \textit{out} taint states.  The mask of the unhandled case
is retrieved and registered by updating~$\mathcal{M}$.  The existing code
fragment is then flushed out from the DBI cache, and instrumentation is
re-triggered; \textbf{now} with the inclusion of the missed path.  The Taint
Rabbit also has a stopping mechanism that prevents basic blocks from
attempting generation if intended fast paths do not actually elide any
instructions.  Finally, rather than immediately triggering dynamic fast path
generation, we employ conventional JIT
heuristics~\cite{schilling2003simplest}, based on execution count, to reduce
the latency induced by flushing.

\subsection{Efficient Instruction Handlers} 
\label{sect:contrib2}

Building a generic taint engine with a high-level programming language
renders instruction handlers, responsible for propagating taint, too complex
to be automatically inlined by a DBI tool.  Therefore, instruction handlers
are built using hand-crafted x86 assembly code.  Although previous work
\cite{cheng2006tainttrace,bruening2011practical} take a similar approach,
their instruction handlers are coded for bitwise tainting, using simple
\textit{or} operations, rather than for generic taint analysis.  Many
instructions are supported, including SIMD.  We designed the code to follow
known practices for optimization to the best of our
abilities~\cite{intel2014intel}.  Iteration and branching are reduced with
instruction handlers, amounting to over 970 in count, specialized not only
to different opcodes but also to operand sizes and types.  The loops in
Algorithms~\ref{algo:over_approx_prim2} and~\ref{algo:under_approx_prim2}
are unrolled.

Instruction handlers do not use a stack but rely on thread-local storage and
registers for memory.  Although Algorithm~\ref{algo:over_approx_prim2} is
slower than Algorithm~\ref{algo:under_approx_prim2}, owing to the
\textit{meet} primitives, it is only used for certain instructions, as
described in Section~\ref{sect:taint_algo}.  The taint primitives are given
memory operands that refer to taint labels and two general purpose
(GP) scratch registers for their implementation.  Spillage is performed
if the primitive requires additional registers.

\subsection{Other Optimizations}
\label{sec:previous_opti}

We also adopt previously proposed optimizations~\cite{bruening2011practical,
qin2006lift, kemerlis2012libdft}.  First, live register analysis is done
to only spill/restore register values that are relied upon by subsequent
application instructions.  Second, we optimize taint checks by minimizing
redundant shadow address translations when memory operands share the same
base address.  Third, space overhead is reduced by creating shadow memory
on demand, with the first write, detected via special faults.  Lastly,
memory dereferences are minimized by using addressable thread local storage
to access frequent fields, e.g., registers' shadow memory.

\subsection{Implementation}
\label{dynamic_bunnies}

The Taint Rabbit is the core of the Dynamic Rabbits, a suite of binary
analysis libraries for building taint-based tools.  The Dynamic Rabbits are
built upon DynamoRIO and Dr.~Memory.  They consist of over 70,000 lines of
C code (including tests) and their source is available at
\href{https://github.com/Dynamic-Rabbits/Dynamic-Rabbits}{https://github.com/Dynamic-Rabbits/Dynamic-Rabbits}. 
Furthermore, Dr.~Memory's shadow memory library Umbra~\cite{zhao2010umbra} 
was enhanced to handle 32-bit tags.  We~also implemented a new 
DynamoRIO library called \texttt{drbbdup}, which duplicates the code of 
basic blocks.  In turn, \texttt{drbbdup} is used to implement fast path 
generation. The majority of drbbdup's code has been merged into
DynamoRIO's repository\footnote{\href{https://github.com/DynamoRIO/dynamorio/commit/6195c00}{https://github.com/DynamoRIO/dynamorio/commit/6195c00}}.
Lastly, several tools, including Perf~\cite{de2010new}, were leveraged to profile
the Taint Rabbit.  Analysis results, visualized via flame
graphs~\cite{Gregg}, are given in the appendix (Figure~\ref{perf_example}).

\subsection{Limitations}
\label{sect:limitations}

Currently, the Taint Rabbit does not analyze 64-bit binaries.  The main
reason is that many existing engines, particularly LibDFT, only support
32-bit and a like-for-like experimental comparison reduces the threat to
validity.  Moreover, the Taint Rabbit does not support some of the FPU
instructions.  Table~\ref{table:supported_instr} (in the appendix) provides
a comprehensive list of the supported instructions.  When an unsupported
instruction is encountered, all destinations are untainted to avoid false
positives.  To penalize the Taint Rabbit, this process is done via a clean
call.

Our approach is more versatile than bitwise tainting.  However, while the
instruction handlers are call-free, user-defined taint primitives could
prevent optimization.  These include primitives that perform a call to
allocate dynamic memory.  We mitigate this issue with an inline custom
allocator that performs clean calls in a slow-path only when requesting
additional memory for management.
 
Truncation of basic blocks removes the need for a static whole-program
pointer analysis.  However, this is not a perfect solution as the number of
basic blocks increases as a consequence.  This, in turn, increases the number
of taint checks done by the dispatcher.  Moreover, \texttt{rep}
instructions, which deal with many bytes, are not checked, as determining
their \textit{in} and \textit{out} taint states could be expensive. 
Therefore, these instructions are treated as potential taint sources for
 data-flow analysis, and are always instrumented.

The Taint Rabbit uses additional memory, and the memory overhead may cause
issues when analyzing large applications.  The memory overhead is primarily
caused by Taint Rabbit's shadow memory where a 32-bit pointer is mapped to
each application byte.  To address this challenge, we implemented a simple
garbage collector that is triggered when memory is low.  The collector
iterates over shadow memory blocks and checks whether they store any tainted
data.  If an entire block is found to store only untainted data, i.e., NULL
values, it is deallocated.

%% file: evaluation.tex
\section{Evaluation}
\label{evaluation}

We performed an experimental evaluation to answer the following
research questions.
\begin{itemize}

\item \texttt{RQ1}: How much does call-avoiding instrumentation and dynamic fast path
generation improve the performance of generic taint analysis?
  
\item \texttt{RQ2}: With these techniques, is the performance of generic taint
analysis comparable to the state of the art of bitwise taint analysis?
  
\item \texttt{RQ3}: Can the improved generic taint analysis scale to real-world
target applications?

\item \texttt{RQ4}: Do taint primitives enable generic taint
analysis?
\end{itemize}

We ran the experiments on 32-bit Ubuntu 14.04 machines, each equipped with
an 8~core 2.60\,GHz Intel Core i7-6700HQ CPU and 32\,GB RAM.  Full results
with numerical
figures\footnote{\href{https://docs.google.com/spreadsheets/d/1gAm7GJBB3Rl4bfTwWq-ITcNyVuRtTH2vQYYaS3n-OUk}{https://docs.google.com/spreadsheets/d/1gAm7GJBB3Rl4bfTwWq-ITcNyVuRtTH2vQYYaS3n-OUk}},
along with scripts for running many of our
experiments\footnote{\href{https://github.com/Dynamic-Rabbits/Taint-Evaluator/commit/e594963}{https://github.com/Dynamic-Rabbits/Taint-Evaluator/commit/e594963}},
are available online.  The specific version of the Dynamic Rabbits that we
used for our experiments is available as
well\footnote{\href{https://github.com/Dynamic-Rabbits/Dynamic-Rabbits/commit/56f9e2b9}{https://github.com/Dynamic-Rabbits/Dynamic-Rabbits/commit/56f9e2b9}}.
  
\subsection{The Taint Rabbit Engines}

The Taint Rabbit (TR) offers two generic taint engines.  As a baseline,
\texttt{TR-CC} has instruction handlers implemented in C and uses clean
calls.  The second engine, \texttt{TR-RAW}, has its instruction handlers
implemented in assembly without clean calls.  When combined with fast paths,
 these variants are referred to as \texttt{TR-CC-FP} and
\texttt{TR-RAW-FP}. The engineering effort required to implement another
taint engine, namely \texttt{TR-CC}, as our baseline was worthwhile
to answer \texttt{RQ1}.

We perform our experiments using two taint policies.  The first policy
(\texttt{TR-ID}) assigns a new numerical ID to each destination byte
whenever taint propagation occurs.  This policy could serve as a basis for a
static single assignment trace generator.  ID~assignment is achieved by
using $\mathit{src} \rightarrow \mathit{dst}$ and $\mathit{src},\mathit{src}
\rightarrow \mathit{dst}$ primitives that increment a counter if any source
is tainted. The 32-bit tags contain the IDs and are not used as pointers.

The second policy (\texttt{TR-BV}) propagates bit vectors similar to the
multi-tag policy adopted by Dytan.  Instead of mapping a separate bit vector
to each tag, which results in high memory usage, our policy represents bit
vectors concisely.  We use a global reduced binary decision tree, similar to
previous work \cite{chen2018angora}.  However, the algorithms presented
previously are recursive and would break our call-free optimization upon
union operations.  Therefore, we devised iterative variants where clean
calls are done only when inserting a new allocated node to the tree. 
Through this memoization, inserted nodes only represent bit vectors that
have not been encountered previously.  The $\mathit{src} \rightarrow
\mathit{dst}$ primitive simply transfers a source's pointer referring to a
node in the tree, while the other primitives efficiently perform unions via
inlined hash-lookups.  Note, these two taint policies cannot be
implemented with a bitwise taint engine.

\subsection{Other Taint-Based Systems}

To answer \texttt{RQ1} and \texttt{RQ2}, we ran nine other taint
analyzers on our benchmarks as baselines for comparison. 
Table~\ref{table:taint_engine} (in the appendix) gives a summary of their
main features.  LibDFT 3.1415 alpha~\cite{kemerlis2012libdft}   inlines bitwise taint
analysis, while Dytan\footnote{\href{https://github.com/dytan-taint-tracking/dytan-taint-tracking/commit/5211823d575}{https://github.com/dytan-taint-tracking/dytan-taint-tracking/commit/5211823d575}}\footnote{We modified Dytan by removing failing assertions for unsupported x86 instructions.}~\cite{clause2007dytan} performs user-defined
operations on bit vectors that contain multiple tags. 
Triton 0.6~\cite{SSTIC2015-Saudel-Salwan} is a dynamic binary analysis framework
which is set up to use Pin 2.14 for tracing.  DataTracker\footnote{\href{https://github.com/m000/dtracker/commit/dc729dca8}{https://github.com/m000/dtracker/commit/dc729dca8}}~\cite{stamatogiannakis2014looking} focuses on data
provenance; a variant, named DataTracker-EWAH\footnote{\href{https://github.com/vusec/vuzzer/commit/f6f7d593a}{https://github.com/vusec/vuzzer/commit/f6f7d593a}
}~\cite{rawat2017vuzzer},
records input offsets to optimize fuzzing.  We also ran
BAP-PinTraces\footnote{\href{https://github.com/BinaryAnalysisPlatform/bap-pintraces/commit/bed2b108}{https://github.com/BinaryAnalysisPlatform/bap-pintraces/commit/bed2b108}}~\cite{brumley2011bap}, which generates execution logs of
instructions that deal with taint.  Its taint propagation routines are not
implemented specifically to the semantics of the instructions, but instead leverage
the IR of the DBI to determine the source and destination operands. 
DECAF\footnote{\href{https://github.com/decaf-project/DECAF/commit/1de4ed7c95}{https://github.com/decaf-project/DECAF/commit/1de4ed7c95}}~\cite{henderson2017decaf} is a QEMU-based taint tracker that inlines
precise bitwise propagation into Tiny Code Generator (TCG) instructions, and
Taintgrind 3.15.0~\cite{weitaintgrind} is a taint engine built upon
Valgrind~\cite{nethercote2007valgrind}.  Moreover, the
Dr.~Memory 2.1.17972\footnote{\href{https://github.com/DynamoRIO/drmemory/commit/5b988e31}{https://github.com/DynamoRIO/drmemory/commit/5b988e31}}~\cite{bruening2011practical} debugger builds on DynamoRIO to
check the addressability of memory using bitwise tainting.  Unfortunately,
we are unable to assess its taint analysis separately as it is tightly
coupled with other components.  Therefore, its reported overhead also
includes memory checks.  However, we did remove code in DataTracker-EWAH and
BAP-PinTraces that concerns logging to file to reduce the overhead.  DBI
overhead was also measured separately without taint analysis.  We give
results for Pin 2.12, DynamoRIO 7.1 and Valgrind 3.13.0, labeled as
\texttt{Pin-Null}\footnote{We ran the same Pin-Null tool provided by
LibDFT 3.1415 alpha~\cite{kemerlis2012libdft}.}, \texttt{DR-Null} and
\texttt{Nullgrind}, respectively. DECAF, just using its virtual machine
introspection and with no taint analysis, is labeled as
\texttt{DECAF-VMI}.

\subsection{Performance}
\label{sect:perf}

\input{charts/binutil.tex}

To answer \texttt{RQ1}, \texttt{RQ2} and \texttt{RQ3},
we measure the Taint Rabbit's performance on benchmarks
relating to data compression, PHP, image parsing, Apache and SPEC CPU.
For these experiments, we configure the Taint Rabbit to taint
all data read from files, sockets, command-line arguments and
environment variables. While we envisage better performance with less taint
introduction, our methodology aims for a grounded evaluation, measuring the
worst performance cases where taking optimal fast paths is difficult due to
taint prevalence\footnote{Figure \ref{fig:sampling_test} in the appendix gives performance results
when taint is sampled.}.  No taint is introduced when running other tools, which
nevertheless instrument instructions even though no taint is propagated. However,
our setup benefits instruction handlers that perform efficiently when 
dealing with untainted data only. There are exceptions, e.g., Dr.~Memory,
which automatically tags memory. We set BAP-PinTraces to
taint command-line arguments to circumvent its fast-forward mechanism
and measure the overhead of its taint analysis.

\textbf{Data Compression.} First, we evaluate the Taint Rabbit on
well-known compression utilities, including \texttt{gzip} 1.6, \texttt{bzip2} 1.0.6,
\texttt{pigz} 2.3 and \texttt{pxz} 4.999.9 beta.  Results are given in
Figures~\ref{fig:tools_gzip}--\ref{fig:tools-pxz}.  As input, all
applications were given a file that is 9.6M large and contains random
data.  We note that \texttt{TR-CC-ID} and \texttt{TR-CC-BV} are
substantially slower than their RAW counterparts.  For instance,
\texttt{TR-CC-BV} and \texttt{TR-RAW-BV} incur overheads of 258x and 3.7x
respectively compared to native runs.  The use of fast paths further
enhances performance: \texttt{TR-RAW-BV-FP} reduces the overhead from 3.7x
down to 2.3x.  Results also show the positive impact of fast paths with
respect to expensive clean-call implementations; \texttt{TR-CC-BV-FP}
achieves 42.6x overhead.  Consequently, fast paths also benefit users who
 implement taint primitives using a high-level programming language.

Comparing to existing tools, Dytan incurs a 308x overhead and is
significantly surpassed by \texttt{TR-CC-BV-FP} and \texttt{TR-RAW-BV-FP}. 
All other generic engines, including DataTracker, are also slower than
\texttt{TR-RAW-BV-FP}.  For instance, DataTracker and DataTracker-EWAH incur
overheads of 6.7x and 40.5x on \texttt{gzip}.  By contrast,
\texttt{TR-RAW-BV-FP} only results in an overhead of 1.3x.  Triton suffers
from the heaviest slowdown; we aborted Triton's run on \texttt{bzip2} after
60 hours and therefore no result is reported.  Owing to efficient bitwise
tainting, LibDFT is faster than \texttt{TR-RAW-BV-FP} on average. 
It~achieves 1.9x overhead, as opposed to 2.3x achieved by
\texttt{TR-RAW-BV-FP}.

\textbf{PHP.} We have applied the Taint Rabbit to PHP 7.2.4, driven by PHPBench\footnote{\href{https://github.com/phpbench/phpbench/commit/04a1a1b}{https://github.com/phpbench/phpbench/commit/04a1a1b}} 0.15.0~\cite{phpbench},
a framework that provides a collection of micro/macro benchmarks for measuring
performance.  We ran the following benchmarks with 10,000 revolutions each:
\texttt{container}, \texttt{hashing}, \texttt{kde} and \texttt{statistics}. 
The results are presented in Figures~\ref{fig:tools_php_container}--\ref{fig:tools_php_statistics}. 
An improvement is achieved by \texttt{TR-RAW-BV-FP} when compared to \texttt{TR-CC-BV};
the former achieves 94.9x while the latter incurs a staggering 806.4x overhead relative to
native execution time. Fast paths also enhance
performance for the clean call implementation, with \texttt{TR-CC-BV-FP} resulting in 187.5x.

Several taint engines, including Dytan and \texttt{TR-CC-BV},
crashed on the \texttt{kde} benchmark. The crash is caused by an integer overflow
error, and suggests that the slowdown imposed by taint analysis is to
blame.  We do not encounter this error when benchmarking performant engines
such as LibDFT, \texttt{TR-RAW-BV} and \texttt{TR-RAW-ID}.

Another observation is that \texttt{TR-RAW-BV} and \texttt{TR-RAW-ID} perform
faster than \texttt{TR-RAW-BV-FP} and \texttt{TR-RAW-ID-FP}, despite the use of fast paths. 
The issue is that \texttt{TR-RAW-ID-FP} fails to amortize many of its initial
overheads, such as those posed by the dispatcher and the generation of fast
paths pertaining to untaint cases.  Since the majority of the PHP benchmarks
take less than a second to execute natively, the optimizations of
\texttt{TR-RAW-ID-FP} do not have enough time to be effective.  Overall,
this increases the overhead from to 62.2x to 89.6x.
Nevertheless, \texttt{TR-RAW-ID-FP} still outperforms all other existing generic taint analyzers. 
For example, it is faster than DataTracker, which incurs 250.1x overhead.

\textbf{Image Parsing.} We consider \texttt{djpeg}, version 9c, and \texttt{gif2png} 2.5.8 as two exemplars of image parsing. 
Results are given in Figures~\ref{fig:tools_djpeg}--\ref{fig:tools_gif}. 
Similar to the results presented so far, we again observe  the high
overheads incurred by existing generic taint engines.  For instance, when running \texttt{djpeg},
DataTracker, DataTracker-EWAH and Dytan achieve slowdowns of 2.7x, 13.8x
and 24.3x respectively over native execution time.  Meanwhile, \texttt{TR-RAW-BV-FP} yields better
results with just an overhead of 1.3x.  Interestingly, fast paths do not
contribute to performance on this benchmark for RAW implementations as they are not significantly
taken due to the large amount of tainted data.  The short one-second runtime of
\texttt{gif2png} also renders generation difficult to amortize.  Nevertheless,
\texttt{TR-CC-BV-FP} achieves a lower overhead of 222.6x on \texttt{gif2png} in comparison to \texttt{TR-CC-BV}, 
which incurs 292x. Since clean call based instruction handlers
are expensive, their elision is more effective in
improving performance than those implemented in efficient assembly code.

\textbf{Apache.} Figures~\ref{fig:tools-apache-1}--\ref{fig:tools-apache-2} depict our
results on Apache~2.4.33.  The benchmark tool \texttt{ab} \cite{apacheab} was used to send
10,000 and 100,000 requests to Apache. Results 
again show that fast paths speed up taint engines implemented using clean calls.  
\texttt{TR-CC-BV-FP} results in 29.5x overhead over native runtime execution, which is 
less than the overheads of 51.4x and 122.8x incurred by \texttt{TR-CC-BV} and Dytan 
respectively. Moreover, \texttt{TR-RAW-ID-FP} is only slightly slower than LibDFT;
the former incurs 7x overhead, while the latter achieves 6.1x. 
In our experiments, Apache primarily performs I/O and process forks to handle requests,
 and hence Triton is able to complete the experiment, albeit with 334x overhead. 
DataTracker and DataTracker-EWAH both time-out after 3~hours.

\input{charts/spec.tex}

\textbf{SPEC CPU 2017.}  The average overheads observed on the SPEC-rate 2017 Integer
benchmark 1.0.2\footnote{The new SPECint 2017 does not support 32-bit systems.}
are given in Figure~\ref{fig:spec_cpu}.  \texttt{TB-RAW-ID} achieves
an overhead of 36.7x over native execution, which is reduced to 17.9x when fast paths are
enabled. It fails to out-perform specialized bitwise taint engines such as LibDFT,
which achieves 10.5x. However, this is expected given its trade-off for versatility.
Moreover, the Taint Rabbit is significantly faster than Taintgrind and DECAF, which
incur overheads of 278x and 74.1x respectively.

We excluded the \texttt{gcc} and \texttt{x264} benchmarks when calculating
results because of known limitations of the Taint Rabbit.  First, the Taint
Rabbit ran out of memory on \texttt{gcc} for both \text{TR-ID} and \text{TR-BV}.  This problem can be mitigated by
using a 64-bit architecture.  Second, \texttt{TB-RAW-BV} times-out on
\texttt{x264} because of high overhead.  Moreover, we terminated
experimentation with \texttt{TB-CC-ID} as the duration of the first benchmark
(i.e., perlbench) exceeded 24 hours.  Because of the unmanageable overhead
of the clean-call implementations of the taint analyzers, we focused on the
optimized versions when running the SPECrate benchmark.  DataTracker also
faced issues as it crashed on all benchmarks except for one.  The remaining
benchmark, namely \texttt{x264}, exceeded 24 hours.

\subsection{Dynamic Fast Path Generation}
\label{sect:gen_stat}

Table~\ref{table:stats_fp} (in the appendix) gives insight about the benefit of fast
path generation.  We gathered these measurements mainly by executing the training
sets of several SPEC CPU 2017 benchmarks.  The second column gives the
percentage of basic blocks where dynamic path generation is applicable. 
For instance, we exclude basic blocks consisting of just one instruction or
those that do not contain data-flow.  The third column details the average
basic block size after truncation.  The fourth column gives an approximation
of the average number of instructions per basic block that elided
instrumentation due to fast path generation.  The fifth and sixth column
indicate the total number of fast paths dynamically generated and reverts. 
The next three columns denote the execution counts of paths with no
instrumentation, adaptive instrumentation, and full instrumentation
respectively.  Finally, the last two columns depict the timelines when fast
paths were generated and executed during runtime.

The results show that execution dominantly takes fast paths.  Although
the most commonly executed path is the \textit{no taint} case, generated
fast paths are executed frequently, particularly for
\texttt{mcf}.  As one would expect, the number of fast paths generated is
negligible when compared to the number of times they are executed.

\textbf{Static vs.~Dynamic Fast Paths.} Lift~\cite{qin2006lift} 
does not generate fast paths just-in-time.  It is fixed to only
consider fast paths that never engage in taint propagation represented by the
\textit{no taint} case.  If any input or output of a basic block is tainted,
execution leads to the slow fully-instrumented path.  We call this approach
\textit{static path generation}, because no other fast paths are
constructed at runtime.

To answer \texttt{RQ1}, we quantify the performance benefits of dynamic fast path
generation compared to the static variant by running the same set of experiments
described in Section~\ref{sect:perf}.  Unfortunately, Lift is not publicly
available.  Therefore, we modeled similar functionality by modifying the
Taint Rabbit and switching off dynamic fast path generation.  Average
results are given in Figure~\ref{fig:staticfp_test}.  \texttt{TR-CC-BV-FP-DYNAMIC}
outperforms \texttt{TR-CC-BV-FP-STATIC} on all considered benchmarks.  For instance,
results obtained using the compression benchmarks show that \texttt{TR-CC-BV-FP-DYNAMIC} 
achieves an overhead of 42.6x, while \texttt{TR-CC-BV-FP-STATIC} incurs 81.8x overhead relative to 
native execution time. Moreover, \texttt{TR-RAW-BV-FP-DYNAMIC} is faster
than \texttt{TR-RAW-BV-FP-STATIC} on the compression benchmarks and SPEC CPU.
\texttt{TR-RAW-BV-FP-STATIC} incurs 2.7x and 25x overheads on these
benchmarks, respectively.  Meanwhile, \texttt{TR-RAW-BV-FP-DYNAMIC} improves
performance with overheads of 2.3x and 22.4x.

\textbf{Performance impact of the number of Fast Paths.} In order to observe the
relationship between performance and the number of possible fast paths that
the Taint Rabbit can generate per basic block, we ran our experiments with
varying limits.  Once the limit is reached, the Taint Rabbit no longer
monitors and attempts fast path generation for the block.  Our results are
in Figures~\ref{fig:num_paths_start}--\ref{fig:num_paths_end} (in the
appendix).  They indicate that the performance impact of fast paths highly
depends on whether the costs of monitor checks and fast path generation are
amortized.  In particular, the generation of multiple fast paths gradually
improves the performance of the Taint Rabbit on the compute-intensive SPEC
CPU 2017 benchmark.  However, applications, such as PHP, that incur heavy
costs during the instrumentation process do not benefit.

\input{charts/staticfp.tex}

\subsection{Application-Specific Experiments}

\renewcommand\arraystretch{1}
\renewcommand{\tabcolsep}{3pt}

\makeatletter
\def\thickhline{%
  \noalign{\ifnum0=`}\fi\hrule \@height \thickarrayrulewidth \futurelet
   \reserved@a\@xthickhline}
\def\@xthickhline{\ifx\reserved@a\thickhline
               \vskip\doublerulesep
               \vskip-\thickarrayrulewidth
             \fi
      \ifnum0=`{\fi}}
\makeatother

\newlength{\thickarrayrulewidth}
\setlength{\thickarrayrulewidth}{1.1pt}

\renewcommand\arraystretch{1}
\renewcommand{\tabcolsep}{3pt}

Apart from performance, we aim to validate the versatility of the Taint Rabbit.
In particular, we show that our taint-primitive based approach
supports three different taint policies to answer \texttt{RQ4}.

\textbf{Control-Flow Hijacking Prevention.} The first use case is the
detection of control-flow hijacking attacks.  We configure the Taint Rabbit
to perform bitwise tainting similar to previous work~\cite{weitaintgrind}. 
The $\mathit{src} \rightarrow \mathit{dst}$ primitive does a move operation,
while the $\mathit{src},\mathit{src} \rightarrow \mathit{dst}$ and
$\mathit{src},\mathit{src} \rightarrow_{M} \mathit{meet}$ primitives perform
a \textit{bitwise or}.  Table~\ref{table:control_flow} presents the attacks
detected by our tool called \texttt{TR-CHECK}.  Although the short execution
times of our benchmarks make amortization difficult, the Taint Rabbit has a
faster mean detection time than our baseline.  \texttt{TR-CHECK-CC} and
\texttt{TR-CHECK-RAW-FP} result in average duration times of 1.46\,s and
0.97\,s, respectively.

\begin{table} [t]
\centering
\caption{Results for detecting control-flow attacks}
\label{table:control_flow}
\begin{small}
\begin{tabular}{@{}cccc@{}}
\textbf{Application}      & \textbf{CVE ID}  &  \thead{\textbf{TR-CHECK} \\ \textbf{(CC)}} & \thead{\textbf{TR-CHECK} \\ \textbf{(RAW-FP)}}
\vspace{-0.3cm} \\
 \\ \thickhline
\vspace{-0.2cm} \\

RTF2Latex & 2004-1293 & 3.75\,s            & 0.8\,s\hphantom{0}
\vspace{-0.2cm} \\ \\

rsync     & 2004-2093 & 0.26\,s            & 0.63\,s
\vspace{-0.2cm} \\ \\

Aeon      & 2005-1019 & 0.24\,s            & 0.5\,s\hphantom{0}
\vspace{-0.2cm} \\ \\

Nginx     & 2013-2028 & 1.6\,s\hphantom{0} & 1.94\,s
\vspace{-0.2cm} 
\\ \\ \thickhline

\end{tabular}
\end{small}
\end{table}

\textbf{Use-After-Free Debugging.} 
The second application, \texttt{TR-UAF}, uses taint analysis to
track pointers and debug use-after-free
vulnerabilities~\cite{caballero2012undangle}.  The 32-bit
tags represent pointers to composite labels containing debugging
information, and are propagated with primitives based on the policy
given as Example~\ref{sect:uaf_policy}.  These labels are shared with
tainted pointers derived from the same root address and reference counting is
employed to manage their memory. Taint propagation is also turned off when
entering allocation routines to prevent false triggers.  Results are given
in Table~\ref{table_uaf}.  We validated the results by checking public bug
reports and comparing alarms with those raised by Dr.~Memory.  In summary,
UAF bug detection can be performed using our generic taint analyzer.

Although our results indicate that Dr.~Memory is faster than \texttt{TR-UAF}, 
the detection mechanisms which they employ are different and therefore performance
cannot be directly compared. Unlike \texttt{TR-UAF}, 
Dr.~Memory does not tag pointers to identify UAF vulnerabilities, but instead
poisons freed memory regions. Dr.~Memory also
tracks addressable data via bitwise tainting.

\begin{table}\centering
\caption{Time taken to detect UAF Vulnerabilities}
\label{table_uaf}
\begin{small}
\begin{tabular}{@{}cccc@{}}
\textbf{Application}   & \textbf{CVE ID} & \textbf{Dr.~Memory} &  \textbf{TR-UAF}
\vspace{-0.3cm} \\
 \\ \thickhline
\vspace{-0.2cm} \\

V8     & 2017-5098  &  4.59\,s & 6.31\,s
\vspace{-0.2cm} \\ \\

Yara   & 2017-5924  &  0.67\,s & 1.57\,s
\vspace{-0.2cm} \\ \\

libzip & 2017-12858 &  1.51\,s & 1.24\,s
\vspace{-0.2cm} \\ \\

lrzip  & 2017-5924  &  1.84\,s & 2.47\,s
\vspace{-0.2cm} \\ \\

libzip  & 2017-12858  &  0.51\,s & 0.44\,s
\vspace{-0.2cm} \\ \\

libwebm  & 2018-6548  &  1.19\,s & 2.47\,s
\vspace{-0.2cm} \\ \\

libsass  & 2018-19827 & 2.88\,s & 3.68\,s
\vspace{-0.2cm} \\ \\

imagemagic  & 2019-19952  &  57.29\,s & 6.12\,s
\vspace{-0.2cm} 
\\ \\ \thickhline

\end{tabular}
\end{small}
\end{table}

\textbf{Fuzzing.}
Although a thorough evaluation of the application of Taint Rabbit to fuzzing
is beyond the scope of this paper, we demonstrate that our approach could be
used for this purpose.  We replaced VUzzer's taint engine with our own
custom tool that provides the same output, i.e., a list of file offsets that
affect \texttt{lea} and comparison instructions.  The rest of VUzzer's code,
e.g.~its mutator, is left untouched.  Table~\ref{table:fuzzing_results}
gives the bug counts obtained on LAVA~\cite{dolan2016lava} by VUzzer and our
version, \texttt{TR-Fuzz}, in 6 hour runs.

\begin{table}[t] \centering
\caption{Bug counts achieved by VUzzer and TR-Fuzz}
\label{table:fuzzing_results}
\begin{small}
\begin{tabular}{@{}cccc@{}}

\textbf{Application}   & \textbf{Total Bugs} & \textbf{VUzzer} & \textbf{TR-Fuzz}
\vspace{-0.3cm} \\
 \\ \thickhline
\vspace{-0.2cm} \\

base64 & \hphantom{00}44 & \hphantom{0}1 & \hphantom{0}1
\vspace{-0.2cm} \\ \\

uniq   & \hphantom{00}28 &            26 & 27
\vspace{-0.2cm} \\ \\

who    & 2136            &            33 & 55
\vspace{-0.2cm} 
\\ \\ \thickhline
\end{tabular}
\end{small}
\end{table}

\subsection{Research Questions}

\textbf{RQ1:} \textit{How much does call-avoiding instrumentation and
dynamic fast path generation improve the performance of generic taint
analysis?}
  
Our results show that the overhead of the Taint Rabbit is reduced
from 224x down to 3.5x when call-avoiding propagation is enabled on
benchmarks related to compression and image parsing.  The optimization is
effective because it essentially addresses the main bottleneck of expensive
context switching, which is experienced by existing generic taint engines. 
Furthermore, fast paths alone reduce the overhead from 224x to 68.8x.  The
benefit of this optimization is its broad applicability as it can be used with
clean call based instruction handlers, thus removing the need to write
low-level assembly code.  We observe positive synergy on these benchmarks
when the two optimizations are used together: the overhead is further
reduced to 3x.  However, fast paths are mainly effective for long-running,
CPU-bound applications where tainting does not comprehensively limit the
execution of fast paths, unlike witnessed when parsing images.  On SPEC CPU,
the overall overhead is reduced to 22.4x from 39.2x. 

\textbf{RQ2:} \textit{With these techniques, is the performance of generic
taint analysis comparable to the state of the art of bitwise taint
analysis?}

Inherently, specialized and generic taint engines have opposing performance
and versatility trade-offs.  The Taint Rabbit has to perform heavier
analyses to support custom taint propagation logic.  We therefore cannot
expect that the Taint Rabbit is faster than optimized bitwise tainting.  On
SPEC CPU, \texttt{TR-RAW-ID-FP} is slower than LibDFT with overheads of
17.8x and 10.5x respectively when compared to native runs.  Nevertheless,
the Taint Rabbit significantly reduces the performance gap that existed
between the two types of analyses.  On the CPU-bound benchmarks
(compression and image parsing) that Dytan manages to complete, Dytan incurs
237x overhead.  By contrast, LibDFT incurs 1.5x and the Taint Rabbit is
only slightly slower with an overhead of 1.7x.

\textbf{RQ3:} \textit{Can the improved generic taint analysis scale to real-world
target applications?}
 
We argue that our proposed optimizations increase the performance of generic
taint analysis to the point that real-world target applications can be
analyzed.  For instance, DataTracker fails to run SPEC CPU, while
\texttt{TR-RAW-BV-FP} achieves an overhead of 22.4x.  On smaller real-world
benchmarks relating to compression and image parsing, the Taint Rabbit has
an overhead of 2.8x, and therefore outperforms DataTracker significantly,
which incurs an overhead of 14.5x.  Unfortunately, like the existing generic
tools, we did encounter one case, namely SPEC CPU's \texttt{gcc} benchmark,
where the Taint Rabbit crashed because of memory limitations.  However, this
limitation is exacerbated by our current implementation, which is intended
to analyze 32-bit software.  Nevertheless, the Taint Rabbit provides a new
opportunity to better scale expensive dynamic analyses when applied to large
and CPU-bound applications.
 
\textbf{RQ4:} \textit{Do taint primitives enable generic taint
analysis?}

To answer \texttt{RQ4}, we demonstrate versatility by considering a variety of
exemplars.  Our results indicate that it is feasible for user-defined
primitives to support security applications concerning exploit detection,
UAF debugging and fuzzing, all of which rely on different taint policies.

\subsection{Threats to Validity}
\label{ttv}

Our experimental setup may have impacted our results.  Particularly, we
use old versions of Pin (i.e., 2.12 and 2.14) since LibDFT and Triton do not support
newer, potentially faster, versions.  To reduce this threat, we ran Pin-Null
with a recent version of Pin (3.7, released in 2018) on SPEC CPU.  The results
(given in the appendix) indicate no major changes in performance (with an
average overhead difference of \texttildelow 0.01x).
Moreover, unlike many other tools that use Pin as a DBI framework,
we use DynamoRIO. Therefore results cannot be compared directly.
However, the Taint Rabbit is faster than other generic taint engines with high
margins, which should exceed any performance benefits provided by the DBI
framework. We also built our own baseline, 
\texttt{TR-CC}, to mitigate this threat. 

Moreover, there might exist a taint policy that cannot be implemented
effectively using the interfaces between Taint Rabbit and the taint
primitives.  This would impact our claim that the Taint Rabbit delivers
generic taint analysis.  We are also exposed to the problem of benchmark
bias, i.e., our findings might not generalize to further benchmarks.  We
have addressed these two threats by considering three different taint-based
applications and a wide range of complex benchmarks.

%% file: charts/binutil.tex
\captionsetup[subfigure]{labelformat=parens, labelsep=space}
\addtocounter{figure}{+1}

\definecolor{tabutter}{rgb}{0.98824, 0.91373, 0.30980}		%
\definecolor{ta2butter}{rgb}{0.92941, 0.83137, 0}		%
\definecolor{ta3butter}{rgb}{0.76863, 0.62745, 0}		%

\definecolor{taorange}{rgb}{0.98824, 0.68627, 0.24314}		%
\definecolor{ta2orange}{rgb}{0.96078, 0.47451, 0}		%
\definecolor{ta3orange}{rgb}{0.80784, 0.36078, 0}		%

\definecolor{tachocolate}{rgb}{0.91373, 0.72549, 0.43137}	%
\definecolor{ta2chocolate}{rgb}{0.75686, 0.49020, 0.066667}	%
\definecolor{ta3chocolate}{rgb}{0.56078, 0.34902, 0.0078431}	%

\definecolor{tachameleon}{rgb}{0.54118, 0.88627, 0.20392}	%
\definecolor{ta2chameleon}{rgb}{0.45098, 0.82353, 0.086275}	%
\definecolor{ta3chameleon}{rgb}{0.30588, 0.60392, 0.023529}	%

\definecolor{taskyblue}{rgb}{0.44706, 0.56078, 0.81176}		%
\definecolor{ta2skyblue}{rgb}{0.20392, 0.39608, 0.64314}	%
\definecolor{ta3skyblue}{rgb}{0.12549, 0.29020, 0.52941}	%

\definecolor{taplum}{rgb}{0.67843, 0.49804, 0.65882}		%
\definecolor{ta2plum}{rgb}{0.45882, 0.31373, 0.48235}		%
\definecolor{ta3plum}{rgb}{0.36078, 0.20784, 0.4}		%

\definecolor{tascarletred}{rgb}{0.93725, 0.16078, 0.16078}	%
\definecolor{ta2scarletred}{rgb}{0.8, 0, 0}			%
\definecolor{ta3scarletred}{rgb}{0.64314, 0, 0}			%

\definecolor{taaluminium}{rgb}{0.93333, 0.93333, 0.92549}	%
\definecolor{ta2aluminium}{rgb}{0.82745, 0.84314, 0.81176}	%
\definecolor{ta3aluminium}{rgb}{0.72941, 0.74118, 0.71373}	%

\begin{figure*}[t]

\centering

\begin{tabularx}{1\textwidth}{X X X X}
\setlength{\tabcolsep}{80pt}

\begin{tikzpicture}[xscale=.51, yscale=.45]

\fill[fill=gray!0!white ] (0,0) rectangle (1.55,3.9);
\node[above] at (0.775,3.9){\footnotesize DBI};

\fill[fill=gray!15!white ] (1.55,0) rectangle (4,3.9);
\node[above] at (2.775,3.9){\footnotesize Others};

\fill[fill=gray!33!white ] (4,0) rectangle (5.1,3.9);
\node[above] at (4.55,3.9){\footnotesize TR};

\fill[fill=gray!50!white ] (5.1,0) rectangle (6.6,3.9);
\node[above] at (5.9,3.9){\footnotesize TR - FP};

    \begin{axis}[
        height = 5.5 cm,
        width= 8.2 cm,
        clip=false,
        major x tick style = transparent,
        ybar=2*\pgflinewidth,
        bar width=6.2pt,
        ymajorgrids = true,
        ymode=log,
        label style={font=\huge},
        tick label style={font=\huge}, 
        log basis y={10},
        ylabel = {slowdown},
        axis on top,
        xticklabels={,,,,,},
        log origin y=infty,
        scaled y ticks = true,
        xtick={1,2,3,4,5,6,7,8,9,10,11,12,13. 14, 15, 16, 17, 18, 19, 20, 21},
        x tick style={draw=none},
        enlarge x limits=0.1,
         every axis plot/.append style={
          bar width=.7,
          bar shift=0pt,
          fill
        },       
        legend style={draw={none},font=\huge},
        legend style={/tikz/every even column/.append style={column sep=0.2cm}},
        legend cell align=left,        
        legend style={at={(-0.2,1.61)},anchor=north west,
                     legend columns=10,
				     cells={align=left}},]
    \legend{PIN-Null, NullGrind, DR-Null, DECAF-VMI, LibDFT, DataTracker, DataTracker-EWAH, Dytan, Triton, BAP-Pintraces, TaintGrind,  Dr.~Memory, DECAF, TB-CC-BV, TR-RAW-BV, TR-CC-ID, TR-RAW-ID, TR-CC-FP-BV, TR-RAW-FP-BV, TR-CC-FP-ID, TR-RAW-FP-ID}

        \addplot[style={fill=tabutter,mark=none}]
            coordinates {(1,1.096918172)};

        \addplot[style={fill=taorange,mark=none}]
            coordinates {(2,1.31455898) };

        \addplot[style={fill=tachocolate,mark=none}]
            coordinates {(3,1.052497343) };

        \addplot[style={fill=tachameleon,mark=none}]
            coordinates {(4,1.279489904) };

        \addplot[style={fill=taskyblue,mark=none}]
            coordinates {(5, 1.616578108) };
            
        \addplot[style={fill=taplum,mark=none}]
            coordinates {(6,6.677364506) };
          
        \addplot[style={fill=tascarletred,mark=none}]
            coordinates {(7,40.53623804) };

        \addplot[style={fill=taaluminium,mark=none}]
            coordinates {(8, 88.87672689) };

        \addplot[style={fill=ta2butter,mark=none}]
            coordinates {(9,23946.30011)};

        \addplot[style={fill=ta2orange,mark=none}]
            coordinates {(10, 45.34899044) };

        \addplot[style={fill=ta2chocolate,mark=none}]
            coordinates {(11, 9.72901169) };

        \addplot[style={fill=ta2chameleon,mark=none}]
        coordinates {(12,4.322210414)};

        \addplot[style={fill=ta2skyblue,mark=none}]
            coordinates {(13,2.195111583) };

        \addplot[style={fill=ta2plum,mark=none}]
            coordinates {(14,81.74792774)};

        \addplot[style={fill=ta2scarletred,mark=none}]
            coordinates {(15,1.802763018)};

        \addplot[style={fill=ta2aluminium,mark=none}]
            coordinates {(16,80.77640808)};

        \addplot[style={fill=ta3butter,mark=none}]
            coordinates {(17,2.264824655)};

        \addplot[style={fill=ta3orange,mark=none}]
            coordinates {(18,24.5088204)};

        \addplot[style={fill=ta3chocolate,mark=none}]
            coordinates {(19,1.273113709)};

        \addplot[style={fill=ta3chameleon,mark=none}]
            coordinates {(20,24.69521785)};

        \addplot[style={fill=ta3skyblue,mark=none}]
            coordinates {(21,1.141126461)};

        \end{axis}
\end{tikzpicture}
\vspace*{-0.9cm}
\captionof{subfigure}{gzip}
\label{fig:tools_gzip}

&

\begin{tikzpicture}[xscale=.51, yscale=.45]

\fill[fill=gray!0!white ] (0,0) rectangle (1.55,3.9);
\node[above] at (0.775,3.9){\footnotesize DBI};

\fill[fill=gray!15!white ] (1.55,0) rectangle (4,3.9);
\node[above] at (2.775,3.9){\footnotesize Others};

\fill[fill=gray!33!white ] (4,0) rectangle (5.1,3.9);
\node[above] at (4.55,3.9){\footnotesize TR};

\fill[fill=gray!50!white ] (5.1,0) rectangle (6.6,3.9);
\node[above] at (5.9,3.9){\footnotesize TR - FP};

    \begin{axis}[
        height = 5.5 cm,
        width= 8.2 cm,
        clip=false,
        major x tick style = transparent,
        ybar=2*\pgflinewidth,
        bar width=6pt,
        ymajorgrids = true,
        log origin y=infty,
        ymode=log,
        label style={font=\huge},
        tick label style={font=\huge}, 
        log basis y={10},
        ylabel = {slowdown},
        axis on top,
        scaled y ticks = true,
        xtick={1,2,3,4,5,6,7,8,9,10,11,12,13},
        xticklabels={,,,,,},
        x tick style={draw=none},
        enlarge x limits=0.1,
         every axis plot/.append style={
          bar width=.7,
          bar shift=0pt,
          fill
        },]

        \addplot[style={fill=tabutter,mark=none}]
            coordinates {(1,1.00528289) };

        \addplot[style={fill=taorange,mark=none}]
            coordinates {(2,1.566121336)};

        \addplot[style={fill=tachocolate,mark=none}]
            coordinates {(3,0.9749488753)};

        \addplot[style={fill=tachameleon,mark=none}]
            coordinates {(4,4.724437628) };

        \addplot[style={fill=taskyblue,mark=none}]
            coordinates {(5,2.57941377)};
            
       \addplot[style={fill=taplum,mark=none}]
            coordinates {(6,35.95858896) };
          
        \addplot[style={fill=tascarletred,mark=none}]
            coordinates {(7,299.4333674) };
            
        \addplot[style={fill=taaluminium,mark=none}]
            coordinates {(8,607.9234833)};

        \addplot[style={fill=ta2orange,mark=none}]
            coordinates {(10, 307.8093047) };

        \addplot[style={fill=ta2chocolate,mark=none}]
            coordinates {(11,51.2240968)};

        \addplot[style={fill=ta2chameleon,mark=none}]
        coordinates {(12,2.348329925) };

        \addplot[style={fill=ta2skyblue,mark=none}]
            coordinates {(13,11.59935242) };

        \addplot[style={fill=ta2plum,mark=none}]
            coordinates {(14,627.7621677)};

       \addplot[style={fill=ta2scarletred,mark=none}]
            coordinates {(15,4.560327198)};

        \addplot[style={fill=ta2aluminium,mark=none}]
            coordinates {(16,633.4267212)};

       \addplot[style={fill=ta3butter,mark=none}]
            coordinates {(17,4.754430811)};

        \addplot[style={fill=ta3orange,mark=none}]
            coordinates {(18,43.15593047)};

        \addplot[style={fill=ta3chocolate,mark=none}]
            coordinates {(19,2.215235174)};

        \addplot[style={fill=ta3chameleon,mark=none}]
            coordinates {(20,42.28834356)};

        \addplot[style={fill=ta3skyblue,mark=none}]
            coordinates {(21,2.000170416)};
            
    \end{axis}

\end{tikzpicture}
\vspace*{-0.9cm}
\captionof{subfigure}{bzip2}

&

\begin{tikzpicture}[xscale=.51, yscale=.45]

\fill[fill=gray!0!white ] (0,0) rectangle (1.55,3.9);
\node[above] at (0.775,3.9){\footnotesize DBI};

\fill[fill=gray!15!white ] (1.55,0) rectangle (4,3.9);
\node[above] at (2.775,3.9){\footnotesize Others};

\fill[fill=gray!33!white ] (4,0) rectangle (5.1,3.9);
\node[above] at (4.55,3.9){\footnotesize TR};

\fill[fill=gray!50!white ] (5.1,0) rectangle (6.6,3.9);
\node[above] at (5.9,3.9){\footnotesize TR - FP};

    \begin{axis}[
        height = 5.5 cm,
        width= 8.2 cm,
        clip=false,
        major x tick style = transparent,
        ybar=2*\pgflinewidth,
        bar width=6pt,
        ymajorgrids = true,
        log origin y=infty,
        ymode=log,
        label style={font=\huge},
        tick label style={font=\huge}, 
        log basis y={10},
        ylabel = {slowdown},
        axis on top,
        scaled y ticks = true,
        xtick={1,2,3,4,5,6,7,8,9,10,11,12,13},
        xticklabels={,,,,,},
        x tick style={draw=none},
        enlarge x limits=0.1,
         every axis plot/.append style={
          bar width=.7,
          bar shift=0pt,
          fill
        },]

        \addplot[style={fill=tabutter,mark=none}]
            coordinates {(1,0.9339782756)};

        \addplot[style={fill=taorange,mark=none}]
            coordinates {(2,0.8879837067) };

        \addplot[style={fill=tachocolate,mark=none}]
            coordinates {(3,0.9249734325) };

        \addplot[style={fill=tachameleon,mark=none}]
            coordinates {(4,0.917109458) };

        \addplot[style={fill=taskyblue,mark=none}]
            coordinates {(5,0.9168363883)};
            
       \addplot[style={fill=taplum,mark=none}]
            coordinates {(6,2.192294637) };
          
        \addplot[style={fill=tascarletred,mark=none}]
            coordinates {(7,137.2070604) };

        \addplot[style={fill=taaluminium,mark=none}]
            coordinates {(8,227.0845909) };

        \addplot[style={fill=ta2butter,mark=none}]
            coordinates {(9,393.0135777) };

        \addplot[style={fill=ta2orange,mark=none}]
            coordinates {(10, 112.122539) };

        \addplot[style={fill=ta2chocolate,mark=none}]
            coordinates {(11,7.059063136) };

        \addplot[style={fill=ta2chameleon,mark=none}]
        coordinates {(12,1.205951116) };

        \addplot[style={fill=ta2skyblue,mark=none}]
            coordinates {(13,2.261211477) };

        \addplot[style={fill=ta2plum,mark=none}]
            coordinates {(14, 21.22640869)};

        \addplot[style={fill=ta2scarletred,mark=none}]
            coordinates {(15,2.30125594)};

        \addplot[style={fill=ta2aluminium,mark=none}]
            coordinates {(16,21.6739647)};

        \addplot[style={fill=ta3butter,mark=none}]
            coordinates {(17, 0.989137814)};

        \addplot[style={fill=ta3orange,mark=none}]
            coordinates {(18, 6.620672098)};

        \addplot[style={fill=ta3chocolate,mark=none}]
            coordinates {(19, 2.106415479)};

       \addplot[style={fill=ta3chameleon,mark=none}]
            coordinates {(20, 6.672946368)};

        \addplot[style={fill=ta3skyblue,mark=none}]
            coordinates {(21, 1.023421589)};

        \end{axis}
\end{tikzpicture}
\vspace*{-0.9cm}
\captionof{subfigure}{pigz}

&

\begin{tikzpicture}[xscale=.51, yscale=.45]

\fill[fill=gray!0!white ] (0,0) rectangle (1.55,3.9);
\node[above] at (0.775,3.9){\footnotesize DBI};

\fill[fill=gray!15!white ] (1.55,0) rectangle (4,3.9);
\node[above] at (2.775,3.9){\footnotesize Others};

\fill[fill=gray!33!white ] (4,0) rectangle (5.1,3.9);
\node[above] at (4.55,3.9){\footnotesize TR};

\fill[fill=gray!50!white ] (5.1,0) rectangle (6.6,3.9);
\node[above] at (5.9,3.9){\footnotesize TR - FP};

    \begin{axis}[
        height = 5.5 cm,
        width= 8.2 cm,
        clip=false,
        major x tick style = transparent,
        ybar=2*\pgflinewidth,
        bar width=6pt,
        ymajorgrids = true,
        log origin y=infty,
        ymode=log,
        label style={font=\huge},
        tick label style={font=\huge}, 
        log basis y={10},
        ylabel = {slowdown},
        axis on top,
        scaled y ticks = true,
        xtick={1,2,3,4,5,6,7,8,9,10,11,12,13},
        xticklabels={,,,,,},
        x tick style={draw=none},
        enlarge x limits=0.1,
         every axis plot/.append style={
          bar width=.7,
          bar shift=0pt,
          fill
        },]

        \addplot[style={fill=tabutter,mark=none}]
            coordinates {(1, 1.076355875)};

        \addplot[style={fill=taorange,mark=none}]
            coordinates {(2, 1.422140293) };

        \addplot[style={fill=tachocolate,mark=none}]
            coordinates {(3, 2.479914984) };

        \addplot[style={fill=tachameleon,mark=none}]
            coordinates {(4,6.521997875) };

        \addplot[style={fill=taskyblue,mark=none}]
            coordinates {(5, 2.495876589)};

        \addplot[style={fill=ta2orange,mark=none}]
            coordinates {(10, 133.2855341) };

        \addplot[style={fill=ta2chocolate,mark=none}]
            coordinates {(11,27.84748229) };

        \addplot[style={fill=ta2chameleon,mark=none}]
        coordinates {(12, 11.44038257) };

        \addplot[style={fill=ta2skyblue,mark=none}]
            coordinates {(13,15.70754516) };

        \addplot[style={fill=ta2plum,mark=none}]
            coordinates {(14, 301.2880567)};

        \addplot[style={fill=ta2scarletred,mark=none}]
            coordinates {(15, 4.859221888)};

        \addplot[style={fill=ta2aluminium,mark=none}]
            coordinates {(16, 298.9384884)};

        \addplot[style={fill=ta3butter,mark=none}]
            coordinates {(17, 4.147278549)};

        \addplot[style={fill=ta3orange,mark=none}]
            coordinates {(18, 96.16047346)};

        \addplot[style={fill=ta3chocolate,mark=none}]
            coordinates {(19, 3.785582614)};

        \addplot[style={fill=ta3chameleon,mark=none}]
            coordinates {(20,92.65877559)};

        \addplot[style={fill=ta3skyblue,mark=none}]
            coordinates {(21,2.806345202)};

        \end{axis}
\end{tikzpicture}
\vspace*{-0.9cm}
\captionof{subfigure}{pxz}
\label{fig:tools-pxz}

\\

\begin{tikzpicture}[xscale=.51, yscale=.45]

\fill[fill=gray!0!white ] (0,0) rectangle (1.55,3.9);
\node[above] at (0.775,3.9){\footnotesize DBI};

\fill[fill=gray!15!white ] (1.55,0) rectangle (4,3.9);
\node[above] at (2.775,3.9){\footnotesize Others};

\fill[fill=gray!33!white ] (4,0) rectangle (5.1,3.9);
\node[above] at (4.55,3.9){\footnotesize TR};

\fill[fill=gray!50!white ] (5.1,0) rectangle (6.6,3.9);
\node[above] at (5.9,3.9){\footnotesize TR - FP};

    \begin{axis}[
        height = 5.5 cm,
        width= 8.2 cm,
        clip=false,
        major x tick style = transparent,
        ybar=2*\pgflinewidth,
        bar width=6pt,
        ymajorgrids = true,
        ymode=log,
        log origin y=infty,
        label style={font=\huge},
        tick label style={font=\huge}, 
        log basis y={10},
        ylabel = {slowdown},
        axis on top,
        scaled y ticks = true,
        xtick={1,2,3,4,5,6,7,8,9,10,11,12,13},
        xticklabels={,,,,,},
        x tick style={draw=none},
        enlarge x limits=0.1,
         every axis plot/.append style={
          bar width=.7,
          bar shift=0pt,
          fill
        },]

        \addplot[style={fill=tabutter,mark=none}]
            coordinates {(1, 33.98167006) };

        \addplot[style={fill=taorange,mark=none}]
            coordinates {(2, 11.0101833) };
            
\addplot[style={fill=tachocolate,mark=none}]
            coordinates {(3, 5.185336049) };
            
        \addplot[style={fill=tachameleon,mark=none}]
            coordinates {(4, 29.90835031) };

        \addplot[style={fill=taskyblue,mark=none}]
            coordinates {(5, 90.16496945) };

        \addplot[style={fill=ta2butter,mark=none}]
            coordinates {(9, 32629.81263) };

        \addplot[style={fill=ta2orange,mark=none}]
            coordinates {(10, 1703.708758) };

        \addplot[style={fill=ta2chocolate,mark=none}]
            coordinates {(11, 226.3543788) };

        \addplot[style={fill=ta2skyblue,mark=none}]
            coordinates {(13,60.48268839) };

        \addplot[style={fill=ta2plum,mark=none}]
            coordinates {(14,1150.808554)};

        \addplot[style={fill=ta2scarletred,mark=none}]
            coordinates {(15,41.44806517)};

        \addplot[style={fill=ta2aluminium,mark=none}]
            coordinates {(16,1163.311609)};

        \addplot[style={fill=ta3butter,mark=none}]
            coordinates {(17,38.72301426)};

        \addplot[style={fill=ta3orange,mark=none}]
            coordinates {(18,171.4276986)};

         \addplot[style={fill=ta3chocolate,mark=none}]
            coordinates {(19,68.25661914)};

        \addplot[style={fill=ta3chameleon,mark=none}]
            coordinates {(20,166.8696538)};

        \addplot[style={fill=ta3skyblue,mark=none}]
            coordinates {(21,62.6395112)};
            
        \end{axis}
\end{tikzpicture}
\vspace*{-0.9cm}
\captionof{subfigure}{PHPBench -- container}
\label{fig:tools_php_container}

&

\begin{tikzpicture}[xscale=.51, yscale=.45]

\fill[fill=gray!0!white ] (0,0) rectangle (1.55,3.9);
\node[above] at (0.775,3.9){\footnotesize DBI};

\fill[fill=gray!15!white ] (1.55,0) rectangle (4,3.9);
\node[above] at (2.775,3.9){\footnotesize Others};

\fill[fill=gray!33!white ] (4,0) rectangle (5.1,3.9);
\node[above] at (4.55,3.9){\footnotesize TR};

\fill[fill=gray!50!white ] (5.1,0) rectangle (6.6,3.9);
\node[above] at (5.9,3.9){\footnotesize TR - FP};

    \begin{axis}[
        height = 5.5 cm,
        width= 8.2 cm,
        clip=false,
        major x tick style = transparent,
        ybar=2*\pgflinewidth,
        bar width=6pt,
        ymajorgrids = true,
        ymode=log,
        log origin y=infty,
        label style={font=\huge},
        tick label style={font=\huge}, 
        log basis y={10},
        ylabel = {slowdown},
        axis on top,
        scaled y ticks = true,
        xtick={1,2,3,4,5,6,7,8,9,10,11,12,13},
        xticklabels={,,,,,},
        x tick style={draw=none},
        enlarge x limits=0.1,
         every axis plot/.append style={
          bar width=.7,
          bar shift=0pt,
          fill
        },
        ]

        \addplot[style={fill=tabutter,mark=none}]
            coordinates {(1,22.23479189)};

        \addplot[style={fill=taorange,mark=none}]
            coordinates { (2,5.991462113)};

        \addplot[style={fill=tachocolate,mark=none}]
            coordinates { (3,3.045891142)};

        \addplot[style={fill=tachameleon,mark=none}]
            coordinates {(4,8.756670224) };

        \addplot[style={fill=taskyblue,mark=none}]
            coordinates {(5, 57.25933831) };

        \addplot[style={fill=taplum,mark=none}]
            coordinates {(6,64.14941302)};

        \addplot[style={fill=tascarletred,mark=none}]
            coordinates { (7,156.3543223)};

        \addplot[style={fill=taaluminium,mark=none}]
            coordinates { (8,364.0853789)};

        \addplot[style={fill=ta2orange,mark=none}]
            coordinates {(10, 637.2166489) };

        \addplot[style={fill=ta2chocolate,mark=none}]
            coordinates { (11, 51.94770544)};

\addplot[style={fill=ta2chameleon,mark=none}]
            coordinates {(12, 30.89434365)};
            
        \addplot[style={fill=ta2skyblue,mark=none}]
            coordinates {(13,16.74919957) };

        \addplot[style={fill=ta2plum,mark=none}]
            coordinates {(14, 170.8633938)};

        \addplot[style={fill=ta2scarletred,mark=none}]
            coordinates {(15, 18.83671291)};

        \addplot[style={fill=ta2aluminium,mark=none}]
            coordinates {(16,172.0800427)};

        \addplot[style={fill=ta3butter,mark=none}]
            coordinates {(17,15.69477054)};

        \addplot[style={fill=ta3orange,mark=none}]
            coordinates {(18,68.54002134)};

        \addplot[style={fill=ta3chocolate,mark=none}]
            coordinates {(19,40.45891142)};

        \addplot[style={fill=ta3chameleon,mark=none}]
            coordinates {(20,67.37139808)};

        \addplot[style={fill=ta3skyblue,mark=none}]
            coordinates {(21, 36.84098186)};
                        
        \end{axis}
\end{tikzpicture}
\vspace*{-0.9cm}
\captionof{subfigure}{PHPBench -- hashing}

&

\begin{tikzpicture}[xscale=.51, yscale=.45]

\fill[fill=gray!0!white ] (0,0) rectangle (1.55,3.9);
\node[above] at (0.775,3.9){\footnotesize DBI};

\fill[fill=gray!15!white ] (1.55,0) rectangle (4,3.9);
\node[above] at (2.775,3.9){\footnotesize Others};

\fill[fill=gray!33!white ] (4,0) rectangle (5.1,3.9);
\node[above] at (4.55,3.9){\footnotesize TR};

\fill[fill=gray!50!white ] (5.1,0) rectangle (6.6,3.9);
\node[above] at (5.9,3.9){\footnotesize TR - FP};

    \begin{axis}[
        height = 5.5 cm,
        width= 8.2 cm,
        clip=false,
        major x tick style = transparent,
        ybar=2*\pgflinewidth,
        bar width=6pt,
        ymajorgrids = true,
        ymode=log,
        log origin y=infty,
        label style={font=\huge},
        tick label style={font=\huge}, 
        log basis y={10},
        ylabel = {slowdown},
        axis on top,
        scaled y ticks = true,
        xtick={1,2,3,4,5,6,7,8,9,10,11,12,13},
        xticklabels={,,,,,},
        x tick style={draw=none},
        enlarge x limits=0.1,
         every axis plot/.append style={
          bar width=.7,
          bar shift=0pt,
          fill
        },]

        \addplot[style={fill=tabutter,mark=none}]
            coordinates {(1,10.88852279) };

        \addplot[style={fill=taorange,mark=none}]
            coordinates { (2,8.217462932)};

\addplot[style={fill=tachocolate,mark=none}]
            coordinates { (3,3.1394838)};

        \addplot[style={fill=tachameleon,mark=none}]
            coordinates {(4,54.29708951) };

        \addplot[style={fill=taskyblue,mark=none}]
            coordinates { (5,39.77210324)};

        \addplot[style={fill=taplum,mark=none}]
            coordinates { (6,294.630972)};

        \addplot[style={fill=ta2chocolate,mark=none}]
            coordinates { (11,576.7605711)};
            
\addplot[style={fill=ta2chameleon,mark=none}]
            coordinates {(12,32.91158704)};
            
        \addplot[style={fill=ta2skyblue,mark=none}]
            coordinates {(13,116.542559) };

       \addplot[style={fill=ta2scarletred,mark=none}]
            coordinates {(15,79.32674355)};

        \addplot[style={fill=ta3butter,mark=none}]
            coordinates {(17,84.62712795)};

       \addplot[style={fill=ta3orange,mark=none}]
            coordinates {(18,92.94343767)};

        \addplot[style={fill=ta3chocolate,mark=none}]
            coordinates {(19,29.13014827)};

        \addplot[style={fill=ta3chameleon,mark=none}]
            coordinates {(20, 90.276771)};

        \addplot[style={fill=ta3skyblue,mark=none}]
            coordinates {(21,27.97419001)};
                        
        \end{axis}
\end{tikzpicture}
\vspace*{-0.9cm}
\captionof{subfigure}{PHPBench -- kde}

& 

\begin{tikzpicture}[xscale=.51, yscale=.45]

\fill[fill=gray!0!white ] (0,0) rectangle (1.55,3.9);
\node[above] at (0.775,3.9){\footnotesize DBI};

\fill[fill=gray!15!white ] (1.55,0) rectangle (4,3.9);
\node[above] at (2.775,3.9){\footnotesize Others};

\fill[fill=gray!33!white ] (4,0) rectangle (5.1,3.9);
\node[above] at (4.55,3.9){\footnotesize TR};

\fill[fill=gray!50!white ] (5.1,0) rectangle (6.6,3.9);
\node[above] at (5.9,3.9){\footnotesize TR - FP};

    \begin{axis}[
        height = 5.5 cm,
        width= 8.2 cm,
        clip=false,
        major x tick style = transparent,
        ybar=2*\pgflinewidth,
        bar width=6pt,
        ymajorgrids = true,
        ymode=log,
        log origin y=infty,
        label style={font=\huge},
        tick label style={font=\huge}, 
        log basis y={10},
        ylabel = {slowdown},
        axis on top,
        scaled y ticks = true,
        xtick={1,2,3,4,5,6,7,8,9,10,11,12,13,14,15,16,17,18,19},
        xticklabels={,,,,,},
        x tick style={draw=none},
        enlarge x limits=0.1,
         every axis plot/.append style={
          bar width=.7,
          bar shift=0pt,
          fill
        },]

        \addplot[style={fill=tabutter,mark=none}]
            coordinates {(1,135.1382114)};

        \addplot[style={fill=taorange,mark=none}]
            coordinates {(2,40.19512195)};
         
\addplot[style={fill=tachocolate,mark=none}]
            coordinates {(3,18.86178862)};

        \addplot[style={fill=tachameleon,mark=none}]
            coordinates {(4,58.77235772) };

        \addplot[style={fill=taskyblue,mark=none}]
            coordinates {(5,348.1869919)};

        \addplot[style={fill=taplum,mark=none}]
            coordinates { (6,391.6178862)};

        \addplot[style={fill=tascarletred,mark=none}]
            coordinates {(7,967.3821138)};

        \addplot[style={fill=taaluminium,mark=none}]
            coordinates {(8,2278.00813)};

        \addplot[style={fill=ta2orange,mark=none}]
            coordinates {(10,3779.252033) };

        \addplot[style={fill=ta2chocolate,mark=none}]
            coordinates { (11,353.195122)};
                        
\addplot[style={fill=ta2chameleon,mark=none}]
            coordinates {(12,186.0650407)};
            
        \addplot[style={fill=ta2skyblue,mark=none}]
            coordinates {(13,115.2845528) };

       \addplot[style={fill=ta2plum,mark=none}]
            coordinates {(14,1097.601626)};

        \addplot[style={fill=ta2scarletred,mark=none}]
            coordinates {(15,131.5121951)};

         \addplot[style={fill=ta2aluminium,mark=none}]
            coordinates {(16,1147.780488)};

        \addplot[style={fill=ta3butter,mark=none}]
            coordinates {(17,109.8536585)};

       \addplot[style={fill=ta3orange,mark=none}]
            coordinates {(18,417.0650407)};

         \addplot[style={fill=ta3chocolate,mark=none}]
            coordinates {(19,241.6910569)};

        \addplot[style={fill=ta3chameleon,mark=none}]
            coordinates {(20,411.1056911)};

        \addplot[style={fill=ta3skyblue,mark=none}]
            coordinates {(21,231.0650407)};

        \end{axis}
\end{tikzpicture}
\vspace*{-0.9cm}
\captionof{subfigure}{PHPBench -- statistics}
\label{fig:tools_php_statistics}

\\

\begin{tikzpicture}[xscale=.51, yscale=.45]

\fill[fill=gray!0!white ] (0,0) rectangle (1.55,3.9);
\node[above] at (0.775,3.9){\footnotesize DBI};

\fill[fill=gray!15!white ] (1.55,0) rectangle (4,3.9);
\node[above] at (2.775,3.9){\footnotesize Others};

\fill[fill=gray!33!white ] (4,0) rectangle (5.1,3.9);
\node[above] at (4.55,3.9){\footnotesize TR};

\fill[fill=gray!50!white ] (5.1,0) rectangle (6.6,3.9);
\node[above] at (5.9,3.9){\footnotesize TR - FP};

    \begin{axis}[
        height = 5.5 cm,
        width= 8.2 cm,
        clip=false,
        major x tick style = transparent,
        ybar=2*\pgflinewidth,
        bar width=6pt,
        ymajorgrids = true,
        ymode=log,
        log origin y=infty,
        label style={font=\huge},
        tick label style={font=\huge}, 
        log basis y={10},
        ylabel = {slowdown},
        axis on top,
        scaled y ticks = true,
        xtick={1,2,3,4,5,6,7,8,9,10,11,12,13},
        xticklabels={,,,,,},
        x tick style={draw=none},
        enlarge x limits=0.1,
         every axis plot/.append style={
          bar width=.7,
          bar shift=0pt,
          fill
        },]

        \addplot[style={fill=tabutter,mark=none}]
            coordinates {(1,0.9818361092)};

        \addplot[style={fill=taorange,mark=none}]
            coordinates {(2,1.001128561)};
         
\addplot[style={fill=tachocolate,mark=none}]
            coordinates {(3,1.039713464)};

        \addplot[style={fill=tachameleon,mark=none}]
            coordinates {(4,0.2182874385) };

        \addplot[style={fill=taskyblue,mark=none}]
            coordinates { (5,1.068664021)};

        \addplot[style={fill=taplum,mark=none}]
            coordinates { (6,2.698662358)};

        \addplot[style={fill=tascarletred,mark=none}]
            coordinates {(7,13.80804961)};

        \addplot[style={fill=taaluminium,mark=none}]
            coordinates {(8,24.34144313)};

        \addplot[style={fill=ta2orange,mark=none}]
            coordinates {(10,11.58100691) };

        \addplot[style={fill=ta2chocolate,mark=none}]
            coordinates { (11,3.035543729)};
                        
\addplot[style={fill=ta2chameleon,mark=none}]
            coordinates {(12,1.09440709)};
           
\addplot[style={fill=ta2skyblue,mark=none}]
            coordinates {(13,0.6079854594) };

       \addplot[style={fill=ta2plum,mark=none}]
            coordinates {(14,22.72813562)};

       \addplot[style={fill=ta2scarletred,mark=none}]
            coordinates {(15,1.198294091)};

         \addplot[style={fill=ta2aluminium,mark=none}]
            coordinates {(16,20.38761909)};

        \addplot[style={fill=ta3butter,mark=none}]
            coordinates {(17,1.175461522)};

      \addplot[style={fill=ta3orange,mark=none}]
            coordinates {(18,19.45325382)};

         \addplot[style={fill=ta3chocolate,mark=none}]
            coordinates {(19,1.328862648)};

        \addplot[style={fill=ta3chameleon,mark=none}]
            coordinates {(20,19.57471073)};

        \addplot[style={fill=ta3skyblue,mark=none}]
            coordinates {(21,1.170804723)};
            
        \end{axis}
\end{tikzpicture}
\vspace*{-0.9cm}
\captionof{subfigure}{djpeg}
\label{fig:tools_djpeg}

& 

\begin{tikzpicture}[xscale=.51, yscale=.45]

\fill[fill=gray!0!white ] (0,0) rectangle (1.55,3.9);
\node[above] at (0.775,3.9){\footnotesize DBI};

\fill[fill=gray!15!white ] (1.55,0) rectangle (4,3.9);
\node[above] at (2.775,3.9){\footnotesize Others};

\fill[fill=gray!33!white ] (4,0) rectangle (5.1,3.9);
\node[above] at (4.55,3.9){\footnotesize TR};

\fill[fill=gray!50!white ] (5.1,0) rectangle (6.6,3.9);
\node[above] at (5.9,3.9){\footnotesize TR - FP};

    \begin{axis}[
        height = 5.5 cm,
        width= 8.2 cm,
        clip=false,
        major x tick style = transparent,
        ybar=2*\pgflinewidth,
        bar width=6pt,
        ymajorgrids = true,
        ymode=log,
        log origin y=infty,
        label style={font=\huge},
        tick label style={font=\huge}, 
        log basis y={10},
        ylabel = {slowdown},
        axis on top,
        scaled y ticks = true,
        xtick={1,2,3,4,5,6,7,8,9,10,11,12,13,14,15,16,17,18,19,20,21},
        xticklabels={,,,,,},
        x tick style={draw=none},
        enlarge x limits=0.1,
         every axis plot/.append style={
          bar width=.7,
          bar shift=0pt,
          fill
        },]

        \addplot[style={fill=tabutter,mark=none}]
            coordinates {(1,1.606815203)};

        \addplot[style={fill=taorange,mark=none}]
            coordinates {(2,1.322411533)};
         
\addplot[style={fill=tachocolate,mark=none}]
            coordinates {(3,0.9252948886)};

        \addplot[style={fill=tachameleon,mark=none}]
            coordinates {(4,7.903014417) };

        \addplot[style={fill=taskyblue,mark=none}]
            coordinates { (5,3.022280472)};

        \addplot[style={fill=taplum,mark=none}]
            coordinates { (6,24.7785059)};

        \addplot[style={fill=tascarletred,mark=none}]
            coordinates {(7,190.4338139)};

        \addplot[style={fill=ta2butter,mark=none}]
            coordinates {(9,63923.11927)};

        \addplot[style={fill=ta2orange,mark=none}]
            coordinates {(10,149.9292267) };

        \addplot[style={fill=ta2chocolate,mark=none}]
            coordinates { (11,31.40366972)};
                        
\addplot[style={fill=ta2chameleon,mark=none}]
            coordinates {(12,33.06946265)};
            
        \addplot[style={fill=ta2skyblue,mark=none}]
            coordinates {(13,12.45216252) };

       \addplot[style={fill=ta2plum,mark=none}]
            coordinates {(14,292.0550459)};

        \addplot[style={fill=ta2scarletred,mark=none}]
            coordinates {(15,6.462647444)};

        \addplot[style={fill=ta2aluminium,mark=none}]
            coordinates {(16,281.0013106)};

        \addplot[style={fill=ta3butter,mark=none}]
            coordinates {(17,4.719528178)};

        \addplot[style={fill=ta3orange,mark=none}]
            coordinates {(18,222.63827)};

        \addplot[style={fill=ta3chocolate,mark=none}]
            coordinates {(19,7.311926606)};

        \addplot[style={fill=ta3chameleon,mark=none}]
            coordinates {(20,216.3132372)};

        \addplot[style={fill=ta3skyblue,mark=none}]
            coordinates {(21, 5.72870249)};

        \end{axis}
\end{tikzpicture}
\vspace*{-0.9cm}
\captionof{subfigure}{gif2png}
\label{fig:tools_gif}

&

\begin{tikzpicture}[xscale=.51, yscale=.45]

\fill[fill=gray!0!white ] (0,0) rectangle (1.55,3.9);
\node[above] at (0.775,3.9){\footnotesize DBI};

\fill[fill=gray!15!white ] (1.55,0) rectangle (4,3.9);
\node[above] at (2.775,3.9){\footnotesize Others};

\fill[fill=gray!33!white ] (4,0) rectangle (5.1,3.9);
\node[above] at (4.55,3.9){\footnotesize TR};

\fill[fill=gray!50!white ] (5.1,0) rectangle (6.6,3.9);
\node[above] at (5.9,3.9){\footnotesize TR - FP};

    \begin{axis}[
        height = 5.5 cm,
        width= 8.2 cm,
        clip=false,
        major x tick style = transparent,
        ybar=2*\pgflinewidth,
        bar width=3pt,
        ymajorgrids = true,
        log origin y=infty,
        ymode=log,
        label style={font=\huge},
        tick label style={font=\huge}, 
        log basis y={10},
        ylabel = {slowdown},
        axis on top,
        scaled y ticks = true,
        xtick={1,2,3,4,5,6,7,8,9,10,11,12},
        xticklabels={,,,,,},
        x tick style={draw=none},
        enlarge x limits=0.1,
         every axis plot/.append style={
          bar width=.7,
          bar shift=0pt,
          fill
        },]

        \addplot[style={fill=tabutter,mark=none}]
            coordinates {(1, 2.557004418)};
        
        \addplot[style={fill=taorange,mark=none}]
            coordinates {(2, 2.292147641)};
            
        \addplot[style={fill=tachocolate,mark=none}]
            coordinates {(3, 2.392831694) };

        \addplot[style={fill=tachameleon,mark=none}]
            coordinates {(4,20.03071113) };

        \addplot[style={fill=taskyblue,mark=none}]
            coordinates {(5, 5.371882571)};

        \addplot[style={fill=taaluminium,mark=none}]
            coordinates {(8, 109.9208351)};

        \addplot[style={fill=ta2butter,mark=none}]
            coordinates {(9, 381.2625766)};

        \addplot[style={fill=ta2orange,mark=none}]
            coordinates {(10,69.98589141) };

        \addplot[style={fill=ta2chocolate,mark=none}]
            coordinates {(11, 12.67628616) };

    \addplot[style={fill=ta2chameleon,mark=none}]
            coordinates {(12, 370.4971498) };

        \addplot[style={fill=ta2skyblue,mark=none}]
            coordinates {(13,35.42689183) };

        \addplot[style={fill=ta2plum,mark=none}]
            coordinates {(14,44.1464301)};

       \addplot[style={fill=ta2scarletred,mark=none}]
            coordinates {(15,5.167878011)};

        \addplot[style={fill=ta2aluminium,mark=none}]
            coordinates {(16,45.72495368)};

        \addplot[style={fill=ta3butter,mark=none}]
            coordinates {(17,3.457959242)};

        \addplot[style={fill=ta3orange,mark=none}]
            coordinates {(18,27.33518598)};

        \addplot[style={fill=ta3chocolate,mark=none}]
            coordinates {(19,6.192746188)};

        \addplot[style={fill=ta3chameleon,mark=none}]
            coordinates {(20,26.08550663)};

       \addplot[style={fill=ta3skyblue,mark=none}]
            coordinates {(21,5.035057717)};

        \end{axis}
\end{tikzpicture}
\vspace*{-0.9cm}
\captionof{subfigure}{Apache -- 10,000 reqs}
\label{fig:tools-apache-1}

&

\begin{tikzpicture}[xscale=.51, yscale=.45]

\fill[fill=gray!0!white ] (0,0) rectangle (1.55,3.9);
\node[above] at (0.775,3.9){\footnotesize DBI};

\fill[fill=gray!15!white ] (1.55,0) rectangle (4,3.9);
\node[above] at (2.775,3.9){\footnotesize Others};

\fill[fill=gray!33!white ] (4,0) rectangle (5.1,3.9);
\node[above] at (4.55,3.9){\footnotesize TR};

\fill[fill=gray!50!white ] (5.1,0) rectangle (6.6,3.9);
\node[above] at (5.9,3.9){\footnotesize TR - FP};

    \begin{axis}[
        height = 5.5 cm,
        width= 8.2 cm,
        clip=false,
        major x tick style = transparent,
        ybar=2*\pgflinewidth,
        bar width=3pt,
        ymajorgrids = true,
        log origin y=infty,
        ymode=log,
        label style={font=\huge},
        tick label style={font=\huge}, 
        log basis y={10},
        ylabel = {slowdown},
        axis on top,        
        scaled y ticks = true,
        xtick={1,2,3,4,5,6,7,8,9,10,11,12},
        xticklabels={,,,,,},
        x tick style={draw=none},
        enlarge x limits=0.1,
         every axis plot/.append style={
          bar width=.7,
          bar shift=0pt,
          fill
        },]

        \addplot[style={fill=tabutter,mark=none}]
            coordinates {(1,2.593013166)};
          
        \addplot[style={fill=taorange,mark=none}]
            coordinates { (2,4.550502459)};

        \addplot[style={fill=tachocolate,mark=none}]
            coordinates {(3,1.41194144)};

        \addplot[style={fill=tachameleon,mark=none}]
            coordinates {(4,24.73651013) };

        \addplot[style={fill=taskyblue,mark=none}]
            coordinates {(5,7.002603268)};

        \addplot[style={fill=taaluminium,mark=none}]
            coordinates {(8,135.8171181)};

        \addplot[style={fill=ta2butter,mark=none}]
            coordinates {(9,286.7494332)};

        \addplot[style={fill=ta2orange,mark=none}]
            coordinates {(10,68.07109067) };

        \addplot[style={fill=ta2chocolate,mark=none}]
            coordinates { (11,20.14934732)};

        \addplot[style={fill=ta2chameleon,mark=none}]
            coordinates {(12,494.1411736)};

        \addplot[style={fill=ta2skyblue,mark=none}]
            coordinates {(13,45.02052756) };

        \addplot[style={fill=ta2plum,mark=none}]
            coordinates {(14, 58.63450683)};

        \addplot[style={fill=ta2scarletred,mark=none}]
            coordinates {(15,8.257200974)};

        \addplot[style={fill=ta2aluminium,mark=none}]
            coordinates {(16,58.19322964)};

        \addplot[style={fill=ta3butter,mark=none}]
            coordinates {(17,6.874436658)};

        \addplot[style={fill=ta3orange,mark=none}]
            coordinates {(18,31.67552462)};

        \addplot[style={fill=ta3chocolate,mark=none}]
            coordinates {(19,11.50705868)};

        \addplot[style={fill=ta3chameleon,mark=none}]
            coordinates {(20,31.15614007)};

        \addplot[style={fill=ta3skyblue,mark=none}]
            coordinates {(21,9.057813069)};
            
        \end{axis}
\end{tikzpicture}
\vspace*{-0.9cm}
\captionof{subfigure}{Apache -- 100,000 reqs}
\label{fig:tools-apache-2}

\end{tabularx}
\addtocounter{figure}{-1}
\captionof{figure}{Results of the Taint Rabbit and other taint systems on command-line utilities, PHPBench and Apache. Missing entries imply that the corresponding taint engine  timed-out or crashed.}
\end{figure*}

%% file: charts/spec.tex
\definecolor{tabutter}{rgb}{0.98824, 0.91373, 0.30980}		%
\definecolor{ta2butter}{rgb}{0.92941, 0.83137, 0}		%
\definecolor{ta3butter}{rgb}{0.76863, 0.62745, 0}		%

\definecolor{taorange}{rgb}{0.98824, 0.68627, 0.24314}		%
\definecolor{ta2orange}{rgb}{0.96078, 0.47451, 0}		%
\definecolor{ta3orange}{rgb}{0.80784, 0.36078, 0}		%

\definecolor{tachocolate}{rgb}{0.91373, 0.72549, 0.43137}	%
\definecolor{ta2chocolate}{rgb}{0.75686, 0.49020, 0.066667}	%
\definecolor{ta3chocolate}{rgb}{0.56078, 0.34902, 0.0078431}	%

\definecolor{tachameleon}{rgb}{0.54118, 0.88627, 0.20392}	%
\definecolor{ta2chameleon}{rgb}{0.45098, 0.82353, 0.086275}	%
\definecolor{ta3chameleon}{rgb}{0.30588, 0.60392, 0.023529}	%

\definecolor{taskyblue}{rgb}{0.44706, 0.56078, 0.81176}		%
\definecolor{ta2skyblue}{rgb}{0.20392, 0.39608, 0.64314}	%
\definecolor{ta3skyblue}{rgb}{0.12549, 0.29020, 0.52941}	%

\definecolor{taplum}{rgb}{0.67843, 0.49804, 0.65882}		%
\definecolor{ta2plum}{rgb}{0.45882, 0.31373, 0.48235}		%
\definecolor{ta3plum}{rgb}{0.36078, 0.20784, 0.4}		%

\definecolor{tascarletred}{rgb}{0.93725, 0.16078, 0.16078}	%
\definecolor{ta2scarletred}{rgb}{0.8, 0, 0}			%
\definecolor{ta3scarletred}{rgb}{0.64314, 0, 0}			%

\definecolor{taaluminium}{rgb}{0.93333, 0.93333, 0.92549}	%
\definecolor{ta2aluminium}{rgb}{0.82745, 0.84314, 0.81176}	%
\definecolor{ta3aluminium}{rgb}{0.72941, 0.74118, 0.71373}	%

\begin{figure}
\centering
\hspace*{-0.25cm}
\begin{tikzpicture}[xscale=.51, yscale=.45]

\fill[fill=gray!0!white ] (0,0) rectangle (5.4,4);
\node[above] at (2.7,4.4){\footnotesize DBI};

\fill[fill=gray!15!white ] (5.4,0) rectangle (10,4.5);
\node[above] at (7.6,4.4){\footnotesize Others};

\fill[fill=gray!30!white ] (10,0) rectangle (12.4,4.5);
\node[above] at (11.2,4.4){\footnotesize TR};

\fill[fill=gray!45!white ] (12.4,0) rectangle (15.4,4.5);
\node[above] at (13.7,4.4){\footnotesize TR - FP};

    \begin{axis}[
        height = 6 cm,
        width= 17 cm,
        clip=false,
        major x tick style = transparent,
        ybar=2*\pgflinewidth,
        bar width=3pt,
        ymajorgrids = true,
        log origin y=infty,
        ymode=log,
        label style={font=\huge},
        tick label style={font=\huge}, 
        log basis y={10},
        ylabel = {slowdown},
        axis on top,        
        scaled y ticks = true,
        xtick={1,2,3,4,5,6,7,8,9,10,11,12},
        xticklabels={,,,,,},
        x tick style={draw=none},
        enlarge x limits=0.1,
         every axis plot/.append style={
          bar width=.7,
          bar shift=0pt,
          fill
        },
        legend style={draw={none},font=\huge},
        legend style={/tikz/every even column/.append style={column sep=0.2cm}},
        legend cell align=left,        
        legend style={at={(-0.1,1.7)},anchor=north west,
                     legend columns=5,
				     cells={align=left}},]
            
        \addplot[style={fill=tabutter,mark=none}]
            coordinates {(1,1.314509933) };
           
        \addplot[style={fill=taorange,mark=none}]
           coordinates {(2,3.671687224)};

        \addplot[style={fill=tachocolate,mark=none}]
            coordinates {(3,1.124189816)};

        \addplot[style={fill=tachameleon,mark=none}]
            coordinates {(4,31.58497542)};

        \addplot[style={fill=taskyblue,mark=none}]
            coordinates {(5,10.50169179) };
 
        \addplot[style={fill=taplum,mark=none}]
           coordinates {(6,278.4378412)};
  
        \addplot[style={fill=tascarletred,mark=none}]
            coordinates {(7,14.25684241) };

        \addplot[style={fill=taaluminium,mark=none}]
            coordinates {(8,74.07885261) };
  
     \addplot[style={fill=ta2butter,mark=none}]
            coordinates {(9,39.18529889) };

     \addplot[style={fill=ta2orange,mark=none}]
            coordinates {(10,36.70010518)};

     \addplot[style={fill=ta2chocolate,mark=none}]
            coordinates {(11,22.41325807) };

     \addplot[style={fill=ta2chameleon,mark=none}] 
            coordinates {(12,17.85212521) };

        \legend{Pin-Null, Nullgrind, DR-Null, DECAF-VMI, LibDFT, Taintgrind, DrMemory, DECAF, TR-RAW-BV, TR-RAW-ID, TR-RAW-FP-BV, TR-RAW-FP-ID}
        \end{axis}
\end{tikzpicture}
\caption{Average overheads on SPECrate 2017}
\label{fig:spec_cpu}
\end{figure}
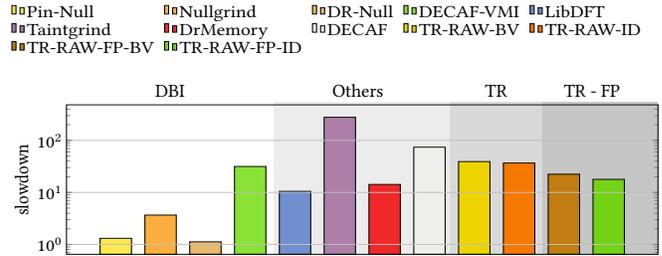

%% file: charts/staticfp.tex
\definecolor{tabutter}{rgb}{0.98824, 0.91373, 0.30980}		%
\definecolor{ta2butter}{rgb}{0.92941, 0.83137, 0}		%
\definecolor{ta3butter}{rgb}{0.76863, 0.62745, 0}		%

\definecolor{taorange}{rgb}{0.98824, 0.68627, 0.24314}		%
\definecolor{ta2orange}{rgb}{0.96078, 0.47451, 0}		%
\definecolor{ta3orange}{rgb}{0.80784, 0.36078, 0}		%

\definecolor{tachocolate}{rgb}{0.91373, 0.72549, 0.43137}	%
\definecolor{ta2chocolate}{rgb}{0.75686, 0.49020, 0.066667}	%
\definecolor{ta3chocolate}{rgb}{0.56078, 0.34902, 0.0078431}	%

\definecolor{tachameleon}{rgb}{0.54118, 0.88627, 0.20392}	%
\definecolor{ta2chameleon}{rgb}{0.45098, 0.82353, 0.086275}	%
\definecolor{ta3chameleon}{rgb}{0.30588, 0.60392, 0.023529}	%

\definecolor{taskyblue}{rgb}{0.44706, 0.56078, 0.81176}		%
\definecolor{ta2skyblue}{rgb}{0.20392, 0.39608, 0.64314}	%
\definecolor{ta3skyblue}{rgb}{0.12549, 0.29020, 0.52941}	%

\definecolor{taplum}{rgb}{0.67843, 0.49804, 0.65882}		%
\definecolor{ta2plum}{rgb}{0.45882, 0.31373, 0.48235}		%
\definecolor{ta3plum}{rgb}{0.36078, 0.20784, 0.4}		%

\definecolor{tascarletred}{rgb}{0.93725, 0.16078, 0.16078}	%
\definecolor{ta2scarletred}{rgb}{0.8, 0, 0}			%
\definecolor{ta3scarletred}{rgb}{0.64314, 0, 0}			%

\definecolor{taaluminium}{rgb}{0.93333, 0.93333, 0.92549}	%
\definecolor{ta2aluminium}{rgb}{0.82745, 0.84314, 0.81176}	%
\definecolor{ta3aluminium}{rgb}{0.72941, 0.74118, 0.71373}	%

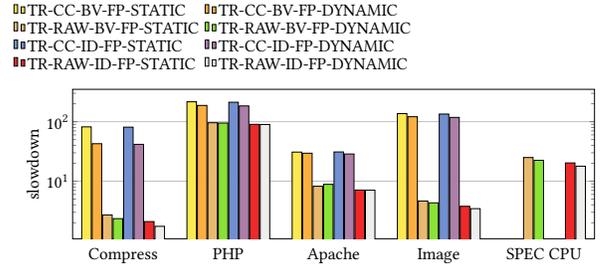
\begin{figure}
\centering
\begin{tikzpicture}[xscale=.45, yscale=.45]
    \begin{axis}[
        width  = 17cm,
        height = 6cm,
        major x tick style = transparent,
        ybar=2*\pgflinewidth,
        ymode=log,
        log basis y={10},
        bar width=8pt,
        ybar=2*\pgflinewidth,
        ymajorgrids = true,
        ylabel = {slowdown},
        symbolic x coords={Compress, PHP, Apache, Image, SPEC CPU},
        xtick={Compress, PHP, Apache, Image, SPEC CPU},
        label style={font=\huge},
        tick label style={font=\huge}, 
        scaled y ticks = false,
        enlarge x limits=0.12,    
        legend cell align=left,
        legend style={draw={none},font=\huge},
        legend style={/tikz/every even column/.append style={column sep=0.1cm}},
       legend style={at={(-0.12,1.6)},anchor=north west,
                     legend columns=2,
				     cells={align=left}},]
    ]

     \addplot[style={fill=tabutter,mark=none}]
            coordinates {(Compress,81.80207613) (PHP,215.930721) (Apache,30.81737449) (Image,136.573511)  };

     \addplot[style={fill=taorange,mark=none}]
            coordinates {(Compress,42.61439653) (PHP,187.4940496) (Apache,29.5053553) (Image,121.0457619)  };

     \addplot[style={fill=tachocolate,mark=none}]
            coordinates {(Compress,2.706781766) (PHP,95.91971663) (Apache,8.214675224) (Image,4.636972076) (SPEC CPU,24.96573636)};

     \addplot[style={fill=tachameleon,mark=none}]
            coordinates {(Compress,2.345086744) (PHP,94.88418394)  (Apache,8.849902434) (Image,4.320394627) (SPEC CPU,22.41325807)};

     \addplot[style={fill=taskyblue,mark=none}]
            coordinates {(Compress,80.86602355) (PHP,212.1336578) (Apache,30.97391234) (Image,134.2743688)  };

     \addplot[style={fill=taplum,mark=none}]
            coordinates {(Compress,41.57882084) (PHP,183.9058785) (Apache,28.62082335) (Image,117.943974)  };

     \addplot[style={fill=tascarletred,mark=none}]
            coordinates {(Compress,2.092296242) (PHP,90.16585857) (Apache,7.084932757) (Image,3.781232232) (SPEC CPU,20.24903108)};

     \addplot[style={fill=taaluminium,mark=none}]
            coordinates {(Compress,1.742765917) (PHP,89.62993093)  (Apache,7.046435393) (Image,3.449753607) (SPEC CPU,17.85212521)};

        \legend{TR-CC-BV-FP-STATIC, TR-CC-BV-FP-DYNAMIC, TR-RAW-BV-FP-STATIC, TR-RAW-BV-FP-DYNAMIC, TR-CC-ID-FP-STATIC, TR-CC-ID-FP-DYNAMIC, TR-RAW-ID-FP-STATIC, TR-RAW-ID-FP-DYNAMIC}
     \end{axis}
\end{tikzpicture}
\caption{Static vs. Dynamic Fast Path Generation}
\label{fig:staticfp_test}
\end{figure}

%% file: related_work.tex
\section{Related Work}
\label{sect:related_work}

There is a substantial body of work on improving the efficiency of bitwise
taint analysis~\cite{cheng2006tainttrace, qin2006lift, bosman2011minemu}. 
LibDFT includes carefully implemented routines so that they are
automatically inlined by DBI tools.  Minemu~\cite{bosman2011minemu} reduces
register spillage by sacrificing the SSE registers, which are assumed to be
dead, to store taint status.  Lastly, unlike our dynamic approach,
Lift~\cite{qin2006lift} uses static fast paths, and Davanian et
al.~\cite{davaniandecaf++} apply this approach system-wide.  Overall, these
works present performant bitwise solutions but lack the versatility of
generic taint analysis.

Dytan~\cite{clause2007dytan} supports generic taint analysis that enables
the user to define custom merging policies for multiple tags stored in bit
vectors.  Unfortunately, its routines are not optimized and suffer from high
overhead.  DECAF~\cite{henderson2017decaf} performs bitwise-tainting inline
to QEMU's TCG intermediate language, but maintains taint labels
asynchronously via tracing at a slower pace.  The Taint Rabbit performs
generic taint analysis that is also optimized.

Other works~\cite{saxena2008efficient,jee2012general} perform preliminary
analysis to reduce runtime overhead.  Jee et al.~\cite{jee2012general} avoid
instrumentation by means of code abstraction and
TaintEraser~\cite{zhu2011tainteraser} leverages taint summaries of standard
API functions.  These approaches are orthogonal to our work, and could
further improve performance.

While the Taint Rabbit is an \textit{online} taint tracker, other
works~\cite{ming2015taintpipe, chabbi2007efficient, cui2013flowwalker,
straighttaint} propose \textit{offline} variants where analysis is decoupled
from the application's execution.
FlowWalker \cite{cui2013flowwalker} employs DBI to log traces
and after runtime performs taint analysis. 
StraightTaint~\cite{straighttaint} takes a similar approach, but uses an
efficient multi-threaded buffer to save data required for constructing the
trace.  Chabbi et al.~\cite{chabbi2007efficient} investigate taint analysis
performed on a secondary shadow thread, which is in sync with the
application's thread.  Meanwhile, TaintPipe~\cite{ming2015taintpipe} uses
threads that perform symbolic execution on code recently executed by the
application until a concrete taint state is processed by a thread spawned
earlier. Unlike the Taint Rabbit, these approaches face issues related to
discrepancies in \textit{time of attack} versus \textit{time of detection},
or require expensive synchronization.

Iodine~\cite{banerjeeiodine} also uses dynamic information to drive static
analysis.  Instrumentation is optimistically pruned such that it avoids
roll-backs upon violations of likely runtime invariants. The
Taint Rabbit instead uses dynamic information when performing forward data-flow
analysis to generate fast paths.  Iodine does not support binaries and depends on a
prior profiling stage.

Similar to versatility, precision is also a trade-off for better
performance, and in the context of pointer tracking, has fostered several
discussions~\cite{slowinska2009pointless, dalton2010tainting,
slowinska2010pointer}.  The Taint Rabbit works at the byte-level for speed,
while Yadegari et al.~\cite{yadegari2014bit} perform bit-level taint
analysis to tackle obfuscation techniques.

Recently, Chua et al.~\cite{chuaone} investigated synthesising propagation. 
The approach aims to reduce implementation effort, but the efficiency
of the generated  analyses remains unclear. Therefore, our work provides 
reciprocal benefits.

\textbf{DBI Optimisations.} 
Kleckner~\cite{kleckner2011optimization}
reduces clean calls via partial inlining, while Wang et
al.~\cite{wang2016general} extend the applicability of persistent code
caching.  Hawkins et al.~\cite{hawkins2015optimizing} enhance the speed
of DBI for JIT applications by using parallel memory mapping.  Such
approaches could further improve the Taint Rabbit.

%% file: conclusion.tex
\section{Conclusion}

In this work,  we make several contributions towards generic taint analysis. 
First, call-avoiding instruction handlers and dynamic fast path generation
are shown to be effective optimizations. Second, we demonstrate that our approach, based on taint
primitives, is flexible enough to support a variety of taint policies.

While our results indicate that avoiding clean calls when executing
instruction handlers delivers the highest performance improvements,
fast paths also provides additional speed-ups once amortized. 
The total speed up is substantial: Dytan achieves an overhead of
237x on CPU-bound benchmarks concerning compression and image parsing when
compared to native execution times, and our optimizations enable 
the Taint Rabbit to reduce that overhead
to~1.7x. Overall, the techniques presented reduce the performance 
gap between generic and bitwise taint engines, and offer better 
scalability for difficult dynamic analyses.

%% file: acknowledgement.tex
\section*{Acknowledgement}

We would like to extend our sincere gratitude to Derek Bruening, Hendrik
Greving and the rest of the DynamoRIO team for answering any queries we had
about the DBI engine, and reviewing our pull requests.  We also thank
the creators of other taint engines for making their tools available, and
the anonymous reviewers for their invaluable feedback.  This work is
supported by the EPSRC CDT in Cyber Security, VETSS and the Endeavour
Scholarship Scheme (partly financed by the European Social Fund).

%% file: appendix.tex
\section*{Appendix}

\input{charts/instr_count_chart.tex}

\begin{algorithm}
 \SetAlgoLined
 \KwData{Taint labels $s_{1}$ and $s_{2}$}
 \KwResult{Taint label $d$}

 \uIf{$s_{1}$ = NULL and $s_{2}$ = NULL }{
    $d$ $\gets$ NULL\;
  }
  \uElseIf{$s_{1}$ $\neq$ NULL and $s_{2}$ $\neq$ NULL }{
    $d$ $\gets$ NULL\;
  }
  \uElseIf{$s_{1}$ $\neq$ NULL }{
    $d$ $\gets$ $s_{1}$\;
  }
  \Else{
    $d$ $\gets$ $s_{2}$\;
  }
  
  \Return $d$\;
\caption{Pointer tracking propagation \cite{caballero2012undangle}}
\label{algo:2}
\end{algorithm}

\noindent
\begin{minipage}{0.49\textwidth}
 \begin{lstlisting}[basicstyle=\ttfamily\scriptsize,caption={An instruction handler of LibDFT. It propagates taint using a \textit{bitwise or} for an instruction, e.g., \texttt{add},  where two registers are sources and one is also the destination.},label=code:libdfthandle,numbers=left,  xleftmargin=5ex,frame=single,breaklines=true,postbreak=\mbox{\textcolor{red}{$\hookrightarrow$}\space},breakatwhitespace=false,floatplacement=H]
static void PIN_FAST_ANALYSIS_CALL
r2r_binary_opl(thread_ctx_t *thread_ctx, uint32_t dst, uint32_t src) {
  thread_ctx->vcpu.gpr[dst] |= thread_ctx->vcpu.gpr[src];
} 
\end{lstlisting}
\end{minipage}

\noindent
\begin{minipage}{0.49\textwidth}
 \begin{lstlisting}[basicstyle=\ttfamily\scriptsize,caption={An instruction handler of DataTracker. Propagation performs union operations on the sets associated with each source byte (lines 5--8).  
 \texttt{tag\_combine}  calls  \texttt{set\_union} at line~13.},label=code:datatrackerhandle,numbers=left,  xleftmargin=5ex,frame=single,breaklines=true,postbreak=\mbox{\textcolor{red}{$\hookrightarrow$}\space},breakatwhitespace=false,floatplacement=H]
static void PIN_FAST_ANALYSIS_CALL
r2r_binary_opl(thread_ctx_t *thread_ctx, uint32_t dst, uint32_t src)
{
  ...
  RTAG[dst][0] = tag_combine(dst_tag[0], src_tag[0]);
  RTAG[dst][1] = tag_combine(dst_tag[1], src_tag[1]);
  RTAG[dst][2] = tag_combine(dst_tag[2], src_tag[2]);
  RTAG[dst][3] = tag_combine(dst_tag[3], src_tag[3]);
}

std::set<uint32_t> tag_combine(std::set<uint32_t> const & lhs, std::set<uint32_t> const & rhs) {
  std::set<uint32_t> res;
  std::set_union(lhs.begin(), lhs.end(), rhs.begin(), rhs.end(),
                 std::inserter(res, res.begin()));
...
\end{lstlisting}
\end{minipage}

\begin{algorithm}
 \SetAlgoLined
 \KwData{ID $\mathit{dst}$, ID $\mathit{src}_{1}$, Integer $\mathit{opnd\_size}$ }
 
 $\mathit{meet\_label}_{1}$ $\gets$ $\mathit{NULL}$\;
 
 \For{$i \gets 0$ to $\mathit{opnd\_size} - 1$}{
    $\mathit{label}$ $\gets$ lookup\_label($\mathit{src}_{1} + i$)\;
    $\mathit{meet\_label}_{1}$ $\gets$ meet$_\mathit{primitive}$($meet\_label_{1}$, $label$)
  }

 \For{$i \gets 0$ to $\mathit{opnd\_size} - 1$}{

    $\mathit{dst\_label}$ $\gets$ src\_dst$_\mathit{primitive}$($meet\_label_{1}$)\;

    set\_label($\mathit{dst} + i$, $\mathit{dst\_label}$)\;
  }
\caption{Taint propagation for one source operand}
\label{algo:over_approx_prim1}
\end{algorithm}

\begin{algorithm}
 \SetAlgoLined
 \KwData{ID $\mathit{dst}$, ID $\mathit{src}_{1}$, Integer $\mathit{opnd\_size}$ }
  
 \For{$i \gets 0$ to $\mathit{opnd\_size} - 1$}{
    $\mathit{src\_label}_{1}$ $\gets$ lookup\_label($src_{1} + i$)\;
    $\mathit{dst\_label}$ $\gets$ src\_dst$_\mathit{primitive}$($src\_label_{1}$)\;
    
    set\_label($\mathit{dst} + i$, $\mathit{dst\_label}$)\;
}
\caption{Optimized taint propagation for one source operand with
independent bytes}
\label{algo:under_approx_prim1}
\end{algorithm}

\input{images/perf_analysis.tex}

\input{images/supported_instr.tex}

\input{charts/pin_chart.tex}

\begin{table*} \centering
\caption{Overview of the taint engines considered in our experimental comparison. $^a$BAP Pin-Traces assigns a special constant integer value to indicate merged taint.}
\label{table:taint_engine}
\begin{small}
\begin{tabular}{@{}cccccc@{}}
\textbf{Taint Engine} & \textbf{Granularity} & \textbf{Meta-Data} &  \textbf{Union Operator}  & \textbf{Approximation} & \textbf{DBI Platform}
\vspace{-0.3cm} \\
 \\ \thickhline
\vspace{-0.2cm} \\

LibDFT \cite{kemerlis2012libdft} & Byte & Bit/Byte & Bitwise & Under & Pin
\vspace{-0.2cm} \\ \\

Triton \cite{SSTIC2015-Saudel-Salwan}  & Byte & Bool & Bitwise & Over & Pin 
\vspace{-0.2cm} \\ \\

Dytan \cite{clause2007dytan} &  Byte & Bit-Vector & Generic & Under & Pin
\vspace{-0.2cm} \\ \\

DataTracker \cite{stamatogiannakis2014looking} & Byte & Set & Set Union & Under & Pin
\vspace{-0.2cm} \\ \\

DataTracker-EWAH \cite{rawat2017vuzzer} & Byte & Compressed Set & Set Union & Under & Pin
\vspace{-0.2cm} \\ \\

BAP-Pin Traces \cite{brumley2011bap} & Byte & 32-Bit Unsigned Offset & Set to \textit{top}$^a$ & Over & Pin
\vspace{-0.2cm} \\ \\

Taintgrind \cite{weitaintgrind} & Byte & Bit & Bitwise & Under & Valgrind
\vspace{-0.2cm} \\ \\

DECAF \cite{henderson2017decaf} & Bit & Bit & Bitwise & Over & QEMU
\vspace{-0.2cm} \\ \\

Dr. Memory \cite{bruening2011practical} & Byte & 2 Bits & Bitwise & Under & DynamoRIO
\vspace{-0.2cm} \\ \\

Taint Rabbit & Byte & 32-Bit Word & Generic & Over & DynamoRIO
\vspace{-0.2cm} 
\\ \\ \thickhline

\end{tabular}
\end{small}
\end{table*}

\input{charts/fp_stats.tex}

\input{images/code_dups.tex}

\input{charts/sampling.tex}

\input{charts/param_analysis.tex}

%% file: charts/instr_count_chart.tex
\definecolor{tabutter}{rgb}{0.98824, 0.91373, 0.30980}          %
\definecolor{ta2butter}{rgb}{0.92941, 0.83137, 0}               %
\definecolor{ta3butter}{rgb}{0.76863, 0.62745, 0}               %

\definecolor{taorange}{rgb}{0.98824, 0.68627, 0.24314}          %
\definecolor{ta2orange}{rgb}{0.96078, 0.47451, 0}               %
\definecolor{ta3orange}{rgb}{0.80784, 0.36078, 0}               %

\definecolor{tachameleon}{rgb}{0.54118, 0.88627, 0.20392}       %
\definecolor{ta2chameleon}{rgb}{0.45098, 0.82353, 0.086275}     %
\definecolor{ta3chameleon}{rgb}{0.30588, 0.60392, 0.023529}     %

\definecolor{taskyblue}{rgb}{0.44706, 0.56078, 0.81176}         %
\definecolor{ta2skyblue}{rgb}{0.20392, 0.39608, 0.64314}        %
\definecolor{ta3skyblue}{rgb}{0.12549, 0.29020, 0.52941}        %

\definecolor{tascarletred}{rgb}{0.93725, 0.16078, 0.16078}      %
\definecolor{ta2scarletred}{rgb}{0.8, 0, 0}                     %
\definecolor{ta3scarletred}{rgb}{0.64314, 0, 0}

\pgfplotstableread{ %
Label     Inline Clean-Call
perlbench   5.316157205  20.18777293
gcc    3.820512821  12.02884615
omnetpp     2.787193974 8.214689266
mcf 2.890636169 10.61686919
xalancbmk 3.829101563 16.91308594
x264   2.176056338  5.769526248
deepsjeng    2.961711712  9.369369369
leela     2.81270183 8.333153929
exchange2 3.370469799 12.22281879
xz 3.118556701 12.83505155
}\testdata

\begin{figure}[!ht]
\centering
    \begin{tikzpicture}
    \begin{axis}[
    width=0.45\textwidth,
    height=5cm,
    xlabel = {slowdown factor (normalized)},
    bar width=3pt,
    xbar stacked,   %
    xmin=1,        
    ytick=data,
    legend style = {fill=black!10,draw=none},
    legend cell align=left,
    legend style={
        cells={align=left}
     },
     yticklabels from table={\testdata}{Label}  %
    ]
    \addplot [fill=tabutter] table [x=Inline, meta=Label,y expr=\coordindex] {\testdata};
    \addplot [fill=taorange] table [x=Clean-Call, meta=Label,y expr=\coordindex] {\testdata};
    \legend{Inline,Clean-Call}
    
    \end{axis}
    \end{tikzpicture}
    \caption{Peformance of clean call and inline instruction execution counters. The inline optimization is turned on/off via DynamoRIO's \texttt{-opt\_cleancall} option.}
\label{fig:ccvsin}
\end{figure}
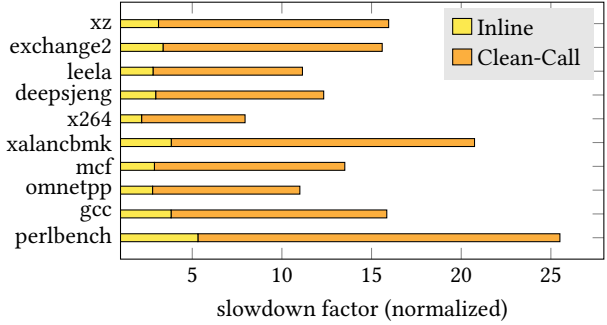

%% file: images/perf_analysis.tex
\begin{figure}[!ht] 
    \centering
  
    \includegraphics[width=0.49\textwidth]{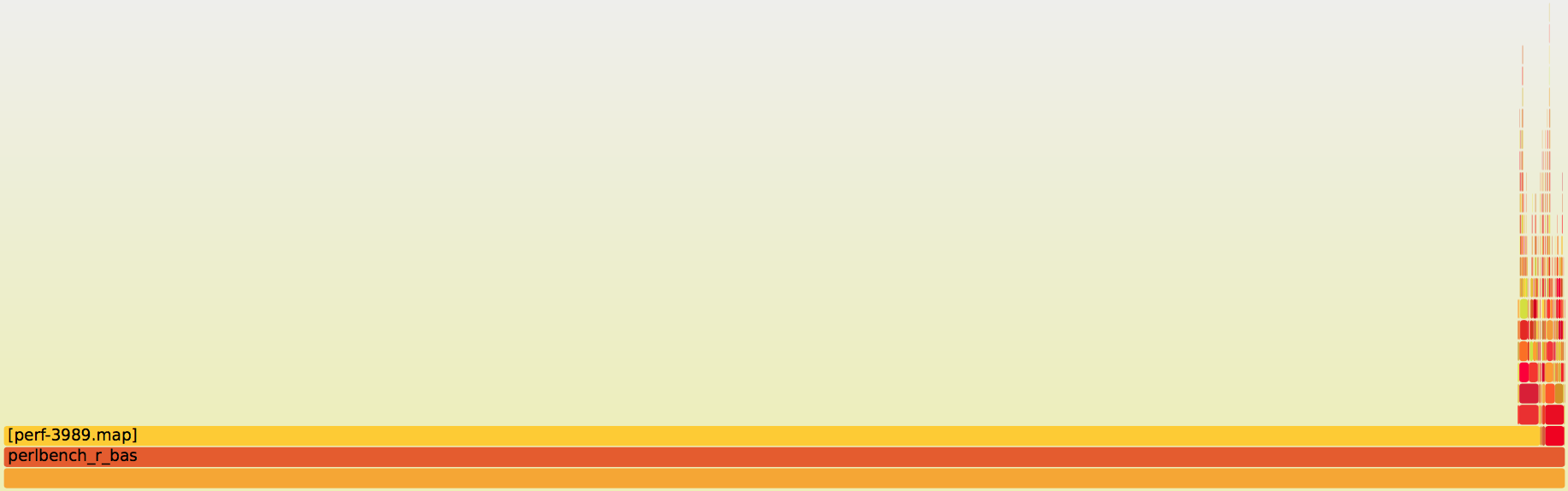}%
  
  \caption{Perf \cite{de2010new}, along with Flame Graphs \cite{Gregg}, aided the profiling of the Taint Rabbit. The figure illustrates a recording of the Taint Rabbit on \texttt{perlbench}. Most of the flames are caused by basic block instrumentation. Meanwhile, the flat area represents execution in the DBI's code cache. Its dominance is a positive result as time is not heavily spent on instrumentation (only \texttildelow 2\% in this recording). Unfortunately, symbols  required for generating the flames are not  available as this code is JIT'ed. However, profiling helped discover a bottleneck related to shadow memory during development. }
  \label{perf_example}
\end{figure}

%% file: images/supported_instr.tex
\begin{table*} \centering
\caption{The list of instructions that the Taint Rabbit supported at the
time of experimentation.  Instructions, such as \texttt{jmp} and
\texttt{prefetchnta}, that have no effect on taint propagation are not
listed.  Most of the missing instructions relate to floating-point
arithmetic and AVX.  We plan to add support for additional instructions. Therefore this list may not reflect the current state.}
\label{table:supported_instr}
\begin{small}
\begin{tabular}{@{}ccccccccccccccc@{}}\toprule

 \texttt{add} & \texttt{or} & \texttt{adc} & \texttt{sbb} & \texttt{and} & \texttt{sub} & \texttt{xor} & \texttt{inc} & \texttt{dec} & \texttt{push} & \texttt{pop} \\
 \texttt{imul} & \texttt{call} & \texttt{mov} & \texttt{lea} & \texttt{xchg} & \texttt{cwde} & \texttt{cdq} & \texttt{leave} & \texttt{rdtsc} & \texttt{cmovo} & \texttt{cmovno} \\
 \texttt{cmovb} & \texttt{cmovnb} & \texttt{cmovz} & \texttt{cmovnz} & \texttt{cmovbe} & \texttt{cmovnbe} & \texttt{cmovs} & \texttt{cmovns} & \texttt{cmovp} & \texttt{cmovnp} & \texttt{cmovl} \\
 \texttt{cmovnl} & \texttt{cmovle} & \texttt{cmovnle} & \texttt{punpcklbw} & \texttt{punpcklwd} & \texttt{punpckldq} & \texttt{packsswb} & \texttt{pcmpgtb} & \texttt{pcmpgtw} & \texttt{pcmpgtd} & \texttt{packuswb} \\
 \texttt{punpckhbw} & \texttt{punpckhwd} & \texttt{punpckhdq} & \texttt{packssdw} & \texttt{punpcklqdq} & \texttt{punpckhqdq} & \texttt{movd} & \texttt{movq} & \texttt{movdqu} & \texttt{movdqa} & \texttt{pshufw} \\
 \texttt{pshufd} & \texttt{pshufhw} & \texttt{pshuflw} & \texttt{pcmpeqb} & \texttt{pcmpeqw} & \texttt{pcmpeqd} & \texttt{seto} & \texttt{setno} & \texttt{setb} & \texttt{setnb} & \texttt{setz} \\
 \texttt{setnz} & \texttt{setbe} & \texttt{setnbe} & \texttt{sets} & \texttt{setns} & \texttt{setp} & \texttt{setnp} & \texttt{setl} & \texttt{setnl} & \texttt{setle} & \texttt{setnle} \\
 \texttt{shld} & \texttt{shrd} & \texttt{cmpxchg} & \texttt{movzx} & \texttt{bsf} & \texttt{bsr} & \texttt{movsx} & \texttt{xadd} & \texttt{pextrw} & \texttt{bswap} & \texttt{psrlw} \\
 \texttt{psrld} & \texttt{psrlq} & \texttt{paddq} & \texttt{pmullw} & \texttt{pmovmskb} & \texttt{pminub} & \texttt{pand} & \texttt{pmaxub} & \texttt{pandn} & \texttt{psraw} & \texttt{psrad} \\
 \texttt{pmulhuw} & \texttt{pmulhw} & \texttt{movntdq} & \texttt{pminsw} & \texttt{por} & \texttt{pmaxsw} & \texttt{pxor} & \texttt{psllw} & \texttt{pslld} & \texttt{psllq} & \texttt{pmaddwd} \\
 \texttt{psubb} & \texttt{psubw} & \texttt{psubd} & \texttt{psubq} & \texttt{paddb} & \texttt{paddw} & \texttt{paddd} & \texttt{psrldq} & \texttt{pslldq} & \texttt{rol} & \texttt{ror} \\
 \texttt{rcl} & \texttt{rcr} & \texttt{shl} & \texttt{shr} & \texttt{sar} & \texttt{not} & \texttt{neg} & \texttt{mul} & \texttt{div} & \texttt{idiv} & \texttt{movups} \\
 \texttt{movupd} & \texttt{movlps} & \texttt{movlpd} & \texttt{movaps} & \texttt{andps} & \texttt{andpd} & \texttt{andnps} & \texttt{andnpd} & \texttt{orps} & \texttt{orpd} & \texttt{xorps} \\
 \texttt{xorpd} & \texttt{movs} & \texttt{rep movs} & \texttt{stos} & \texttt{rep stos} & \texttt{lddqu} & \texttt{pshufb} & \texttt{palignr} & \texttt{lzcnt} & \texttt{pcmpeqq} & \texttt{movntdqa} \\
 \texttt{packusdw} & \texttt{pcmpgtq} & \texttt{pminsd} & \texttt{pminuw} & \texttt{pminud} & \texttt{pmaxsb} & \texttt{pmaxsd} & \texttt{pmaxuw} & \texttt{pmaxud} & \texttt{pmulld} & \texttt{pextrb} \\
 \texttt{pextrd} & \texttt{xgetbv} & \texttt{movq2dq} & \texttt{movdq2q} & \texttt{tzcnt} & \texttt{pext} & &  &  &  & \\
  
\midrule
\end{tabular}
\end{small}
\end{table*}

%% file: charts/pin_chart.tex
\definecolor{tabutter}{rgb}{0.98824, 0.91373, 0.30980}          %
\definecolor{ta2butter}{rgb}{0.92941, 0.83137, 0}               %
\definecolor{ta3butter}{rgb}{0.76863, 0.62745, 0}               %

\definecolor{taorange}{rgb}{0.98824, 0.68627, 0.24314}          %
\definecolor{ta2orange}{rgb}{0.96078, 0.47451, 0}               %
\definecolor{ta3orange}{rgb}{0.80784, 0.36078, 0}               %

\definecolor{tachameleon}{rgb}{0.54118, 0.88627, 0.20392}       %
\definecolor{ta2chameleon}{rgb}{0.45098, 0.82353, 0.086275}     %
\definecolor{ta3chameleon}{rgb}{0.30588, 0.60392, 0.023529}     %

\definecolor{taskyblue}{rgb}{0.44706, 0.56078, 0.81176}         %
\definecolor{ta2skyblue}{rgb}{0.20392, 0.39608, 0.64314}        %
\definecolor{ta3skyblue}{rgb}{0.12549, 0.29020, 0.52941}        %

\definecolor{tascarletred}{rgb}{0.93725, 0.16078, 0.16078}      %
\definecolor{ta2scarletred}{rgb}{0.8, 0, 0}                     %
\definecolor{ta3scarletred}{rgb}{0.64314, 0, 0}

\begin{figure}
\centering
    \begin{tikzpicture}

    \begin{axis}[
    xbar=0,
    width=0.45\textwidth,
    height=5.5cm,
    xlabel = {slowdown},
    bar width=3pt,
    xmin=0,   
    ytick=data,
    symbolic y coords={perlbench,gcc,omnetpp,x264,deepsjeng,leela,xz, exchange2,mcf,xalancbmk},
    legend cell align=left,
    legend pos = north east,
    legend style = {fill=black!10,draw=none},
    legend style={
      cells={align=left}
    },
    ]
    
    \addplot[style={fill=tabutter,mark=none}]
            coordinates {(1.706550218,perlbench) (1.896367521,gcc) (1.158684775,mcf) (1.215944758,omnetpp) (1.276367188,xalancbmk) (1.333546735,x264) (1.481981982,deepsjeng) (1.38320775,leela) (1.117225951,exchange2) (1.176116838,xz)};
        
    \addplot[style={fill=taorange,mark=none}]
            coordinates {(1.659388646,perlbench) (1.918803419,gcc) (1.223731237,mcf) (1.221594476,omnetpp) (1.25390625,xalancbmk) (1.66709347,x264) (1.418168168,deepsjeng) (1.344456405,leela) (0.9463087248,exchange2) (1.235395189,xz)};
            
        \legend{Pin 2.14, Pin 3.7}

    \end{axis}
    \end{tikzpicture}
    \caption{Performance of PinNull on Pin 2.14 and 3.7 }
    \label{fig:pin3.7}
\label{fig:pin_comparison}
\end{figure}
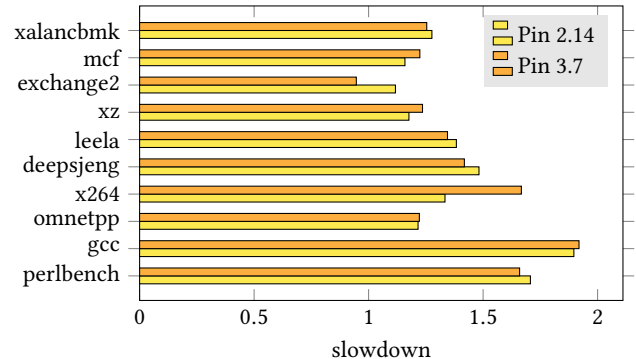

%% file: charts/fp_stats.tex
\definecolor{taskyblue}{rgb}{0.44706, 0.56078, 0.81176}		%
\definecolor{ta2skyblue}{rgb}{0.20392, 0.39608, 0.64314}	%
\definecolor{ta3skyblue}{rgb}{0.12549, 0.29020, 0.52941}	%

\input{charts/fp_stats_data.tex}

\newcommand{\genplot}[2]{%
    \begin{tikzpicture}[xscale=0.005, yscale=0.005]
        \begin{scope}[ycomb, yscale=3]
            \draw[ultra thin, taskyblue!20] (1,#2)--(300,#2);
            \draw[taskyblue!99, line width=0.001mm] plot #1;
        \end{scope}
    \end{tikzpicture}%
}

\newcolumntype{.}{D{.}{.}{2}}

\begin{table*} \centering
\caption{Statistics related to Dynamic Fast Path Generation. Generation and execution timelines are relative to each other. Generation counts are less than execution counts and therefore are hardly visible on the timeline. }
\label{table:stats_fp}
\begin{small}
\begin{tabular}{ccccccccccc}

\textbf{App.} & \thead{\textbf{\% BB} \\ \textbf{Instrum.}} & \thead{\textbf{Avg.} \\ \textbf{BB Size.}} & \thead{\textbf{Avg. Instr} \\ \textbf{Elided.}} & \thead{\textbf{\# FP} \\ \textbf{Gen.}} & \thead{\textbf{\# Revert}}  & \thead{\textbf{\# Exec.} \\ \textbf{None}} & \thead{\textbf{\# Exec.} \\ \textbf{FP}} &   \thead{\textbf{\# Exec.} \\ \textbf{Full}} & \thead{\textbf{FP Gen.} \\ \textbf{Timeline}} & \thead{\textbf{Exec. FP} \\ \textbf{Timeline}}
\vspace{-0.3cm} \\
 \\ \thickhline
\vspace{-0.2cm} \\
    
   perlbench & 81.0\% & 4 & 1 & 2009            & 1255            & 2.77E9 & 3.43E9 & 1.53E6 & \genplot{coordinates {\perlGENDATA}}{2.0} & \genplot{coordinates {\perlEXECDATA}}{2.0}\\ \hdashline[0.5pt/0.5pt]

   mcf       & 82.3\% & 5 & 4 & \hphantom{0}281 & \hphantom{00}92 & 3.42E9 & 4.03E9 & 9.32E7 & \genplot{coordinates {\mcfGENDATA}}{1.0} & \genplot{coordinates {\mcfEXECDATA}}{2.0}\\ \hdashline[0.5pt/0.5pt]

   xalancbmk & 81.2\% & 4 & 3 & \hphantom{0}213 & \hphantom{00}57 & 3.51E9 & 3.14E9 & 1.43E9 & \genplot{coordinates {\xalGENDATA}}{2.0} & \genplot{coordinates {\xalEXECDATA}}{2.0}\\ \hdashline[0.5pt/0.5pt]
   
   exchange2 & 81.2\% & 5 & 4 & \hphantom{0}791 & \hphantom{0}245 & 3.82E9 & 5.19E7 & 1.67E9 & \genplot{coordinates {\exchangeGENDATA}}{2.0} & \genplot{coordinates {\exchangeEXECDATA}}{2.0}\\ \hdashline[0.5pt/0.5pt]

  bzip2 & 87.4\% & 4 & 3 & \hphantom{0}510 & \hphantom{0}77 & 1.69E9 & 2.82E8 & 6.27E7 & \genplot{coordinates {\bzipGENDATA}}{2.0} & \genplot{coordinates {\bzipEXECDATA}}{2.0}\\ \hdashline[0.5pt/0.5pt]

   djpeg & 84\% & 4 & 1 & \hphantom{0}141 & \hphantom{0}73 & 1.49E8 & 6.42E8 & 7.33E8 & \genplot{coordinates {\djpegGENDATA}}{2.0} & \genplot{coordinates {\djpegEXECDATA}}{2.0}\\ \hdashline[0.5pt/0.5pt]

\vspace{-0.2cm} 
\\ \\ \thickhline    
      
\end{tabular}

\end{small}
\end{table*}

%% file: charts/fp_stats_data.tex
\def\exchangeEXECDATA{
(0,14.750358)
(1,18.359349)
(2,16.601901)
(3,19.945307)
(4,23.829238)
(5,22.801771)
(6,19.577009)
(7,28.012273)
(8,25.794186)
(9,26.30473)
(10,24.826982)
(11,26.290994)
(12,25.398298)
(13,24.333959)
(14,17.023107)
(15,21.232567)
(16,21.477334)
(17,21.973457)
(18,20.377158)
(19,17.273462)
(20,19.484238)
(21,19.165256)
(22,20.880794)
(23,19.776554)
(24,25.674746)
(25,23.706094)
(26,18.041662)
(27,16.463072)
(28,24.683545)
(29,19.891687)
(30,15.100889)
(31,16.064584)
(32,14.231801)
(33,10.665967)
(34,9.937351)
(35,11.537123)
(36,10.579887)
(37,12.297966)
(38,10.453755)
(39,9.425465)
(40,6.167519)
(41,10.767751)
(42,9.743051)
(43,14.186934)
(44,14.87484)
(45,10.035095)
(46,13.394644)
(47,20.381132)
(48,14.878703)
(49,11.593939)
(50,16.296055)
(51,16.084974)
(52,14.048724)
(53,11.379884)
(54,15.715256)
(55,14.582011)
(56,17.279149)
(57,12.682645)
(58,13.012491)
(59,12.960248)
(60,23.113682)
(61,41.322653)
(62,18.68309)
(63,9.339896)
(64,8.787363)
(65,11.648359)
(66,11.090807)
(67,12.49414)
(68,10.65156)
(69,10.706701)
(70,10.69541)
(71,8.833039)
(72,9.145597)
(73,9.366285)
(74,9.092812)
(75,8.644311)
(76,8.088188)
(77,7.603219)
(78,8.163746)
(79,8.926409)
(80,7.472675)
(81,12.940103)
(82,19.504356)
(83,16.475659)
(84,14.463266)
(85,16.294773)
(86,16.245463)
(87,21.205544)
(88,16.031476)
(89,22.120774)
(90,20.366912)
(91,17.069385)
(92,16.228687)
(93,15.791016)
(94,20.279434)
(95,14.575019)
(96,11.548923)
(97,17.734757)
(98,16.953312)
(99,16.322704)
(100,17.641468)
(101,15.478677)
(102,11.004414)
(103,13.949504)
(104,22.682831)
(105,10.193169)
(106,12.953289)
(107,18.77924)
(108,26.484646)
(109,36.560492)
(110,7.496589)
(111,8.123318)
(112,12.620349)
(113,10.353644)
(114,25.40868)
(115,9.830643)
(116,36.75189)
(117,21.441487)
(118,10.033609)
(119,4.547401)
(120,4.021543)
(121,4.981215)
(122,10.583738)
(123,15.272312)
(124,12.786582)
(125,10.818199)
(126,33.031586)
(127,8.06533)
(128,7.680532)
(129,8.507975)
(130,34.933239)
(131,18.088903)
(132,15.965731)
(133,35.025436)
(134,24.476592)
(135,16.681204)
(136,34.04526)
(137,36.748589)
(138,10.151372)
(139,26.481818)
(140,8.876383)
(141,23.856903)
(142,8.905964)
(143,4.104688)
(144,3.16742)
(145,1.529754)
(146,1.844761)
(147,7.896882)
(148,12.562973)
(149,12.569037)
(150,22.080683)
(151,23.776916)
(152,30.843873)
(153,20.255801)
(154,23.42009)
(155,23.858374)
(156,23.015221)
(157,17.075645)
(158,6.263276)
(159,8.284941)
(160,5.87272)
(161,7.488518)
(162,14.533096)
(163,13.180907)
(164,24.963034)
(165,13.94285)
(166,15.907647)
(167,27.977501)
(168,22.548109)
(169,48.509509)
(170,15.153914)
(171,14.047452)
(172,35.825928)
(173,41.060854)
(174,37.127967)
(175,5.96827)
(176,6.707409)
(177,6.927505)
(178,2.847551)
(179,1.752764)
(180,2.91827)
(181,2.629006)
(182,5.633931)
(183,3.926634)
(184,1.774347)
(185,3.680037)
(186,3.130894)
(187,4.637046)
(188,5.508648)
(189,4.843179)
(190,5.980754)
(191,2.997245)
(192,2.555249)
(193,1.398433)
(194,4.909726)
(195,7.197112)
(196,2.280281)
(197,33.137744)
(198,23.519645)
(199,9.398195)
(200,3.45761)
(201,20.78613)
(202,0.507062)
(203,3.035559)
(204,6.074976)
(205,7.639467)
(206,13.872684)
(207,2.106933)
(208,2.692551)
(209,10.996178)
(210,9.063041)
(211,1.732115)
(212,3.958006)
(213,3.821883)
(214,6.136429)
(215,17.401549)
(216,15.850923)
(217,15.228334)
(218,20.856746)
(219,41.221013)
(220,39.001813)
(221,19.464496)
(222,11.101061)
(223,9.396)
(224,8.253192)
(225,10.235902)
(226,9.080674)
(227,25.989203)
(228,22.437785)
(229,11.390358)
(230,22.327015)
(231,26.887744)
(232,9.544454)
(233,5.409563)
(234,4.174015)
(235,3.923734)
(236,5.161764)
(237,3.925778)
(238,5.908399)
(239,4.951108)
(240,5.2939)
(241,3.662315)
(242,2.055105)
(243,6.453082)
(244,1.867417)
(245,3.751244)
(246,0.695187)
(247,0.65858)
(248,0.943846)
(249,9.30535)
(250,5.537773)
(251,6.361741)
(252,3.493216)
(253,6.641957)
(254,4.785326)
(255,19.76999)
(256,7.452691)
(257,19.193757)
(258,14.589133)
(259,7.167836)
(260,5.085371)
(261,1.356584)
(262,0.952825)
(263,5.425182)
(264,5.42427)
(265,5.167674)
(266,4.428103)
(267,7.90383)
(268,14.621684)
(269,12.821948)
(270,32.712049)
(271,36.826553)
(272,27.129345)
(273,4.517134)
(274,4.631269)
(275,4.649858)
(276,4.679765)
(277,4.20494)
(278,5.13129)
(279,3.997791)
(280,15.357462)
(281,30.131166)
(282,25.478569)
(283,26.657251)
(284,13.829951)
(285,4.353834)
(286,2.591521)
(287,10.511388)
(288,4.836201)
(289,2.527788)
(290,11.736984)
(291,2.826329)
(292,4.763085)
(293,10.549108)
(294,4.88107)
(295,4.772527)
(296,1.76801)
(297,4.493592)
(298,5.3049)
(299,3.286958)
(300,10.14369)
(301,6.487107)
(302,3.222312)
(303,4.098037)
(304,2.420109)
(305,13.370637)
(306,19.723854)
(307,23.56036)
(308,31.322352)
(309,23.711334)
(310,30.634476)
(311,25.919307)
(312,7.695584)
(313,14.72735)
(314,9.686309)
(315,7.385877)
(316,28.699552)
(317,8.081838)
(318,5.3253)
(319,3.773082)
}

\def\exchangeGENDATA{
(0,0.000297)
(1,5.9e-05)
(2,3.2e-05)
(3,1.6e-05)
(4,0)
(5,1.1e-05)
(6,7.7e-05)
(7,8e-06)
(8,1.6e-05)
(9,1.7e-05)
(10,4e-06)
(11,4e-05)
(12,0)
(13,2e-06)
(14,0)
(15,0)
(16,2e-06)
(17,1e-06)
(18,1e-06)
(19,0)
(20,3e-06)
(21,0)
(22,1e-06)
(23,0)
(24,1e-05)
(25,2e-06)
(26,2e-06)
(27,0)
(28,6e-06)
(29,0)
(30,2e-06)
(31,4e-06)
(32,1.6e-05)
(33,1e-06)
(34,0)
(35,0)
(36,0)
(37,1e-06)
(38,1e-06)
(39,0)
(40,0)
(41,0)
(42,1e-06)
(43,1e-06)
(44,1e-06)
(45,1e-06)
(46,0)
(47,1e-06)
(48,0)
(49,1e-06)
(50,2e-06)
(51,2e-06)
(52,0)
(53,2e-06)
(54,1e-06)
(55,3.3e-05)
(56,7e-06)
(57,0)
(58,0)
(59,0)
(60,6e-06)
(61,6e-06)
(62,3e-06)
(63,0)
(64,0)
(65,0)
(66,1e-06)
(67,1e-06)
(68,0)
(69,7e-06)
(70,0)
(71,0)
(72,0)
(73,0)
(74,0)
(75,0)
(76,0)
(77,0)
(78,0)
(79,0)
(80,0)
(81,0)
(82,0)
(83,0)
(84,0)
(85,0)
(86,0)
(87,1e-06)
(88,0)
(89,1e-06)
(90,0)
(91,0)
(92,0)
(93,2e-06)
(94,1e-06)
(95,0)
(96,0)
(97,1e-06)
(98,0)
(99,0)
(100,1e-06)
(101,0)
(102,0)
(103,3e-06)
(104,1e-06)
(105,0)
(106,0)
(107,0)
(108,3e-06)
(109,0)
(110,0)
(111,0)
(112,2e-06)
(113,7e-06)
(114,1e-06)
(115,0)
(116,1e-06)
(117,0)
(118,0)
(119,0)
(120,0)
(121,0)
(122,0)
(123,0)
(124,1e-06)
(125,0)
(126,2e-06)
(127,0)
(128,0)
(129,0)
(130,0)
(131,0)
(132,2e-06)
(133,3e-06)
(134,1e-06)
(135,0)
(136,3e-06)
(137,4e-06)
(138,1e-06)
(139,0)
(140,0)
(141,0)
(142,0)
(143,0)
(144,0)
(145,0)
(146,0)
(147,0)
(148,0)
(149,1e-06)
(150,0)
(151,1e-06)
(152,0)
(153,0)
(154,0)
(155,0)
(156,0)
(157,6e-06)
(158,0)
(159,0)
(160,0)
(161,0)
(162,1e-06)
(163,0)
(164,2e-06)
(165,0)
(166,0)
(167,1e-06)
(168,0)
(169,0)
(170,0)
(171,0)
(172,1e-06)
(173,0)
(174,0)
(175,0)
(176,0)
(177,1e-06)
(178,0)
(179,0)
(180,0)
(181,0)
(182,0)
(183,0)
(184,0)
(185,0)
(186,0)
(187,0)
(188,0)
(189,0)
(190,0)
(191,1e-06)
(192,0)
(193,0)
(194,0)
(195,0)
(196,1e-06)
(197,1e-06)
(198,0)
(199,0)
(200,0)
(201,0)
(202,0)
(203,0)
(204,1e-06)
(205,0)
(206,0)
(207,0)
(208,0)
(209,0)
(210,0)
(211,0)
(212,0)
(213,0)
(214,0)
(215,0)
(216,0)
(217,0)
(218,0)
(219,0)
(220,0)
(221,2e-06)
(222,0)
(223,0)
(224,1e-06)
(225,0)
(226,0)
(227,1e-06)
(228,0)
(229,9e-06)
(230,0)
(231,0)
(232,0)
(233,0)
(234,0)
(235,0)
(236,2e-06)
(237,2e-06)
(238,0)
(239,0)
(240,0)
(241,1e-06)
(242,0)
(243,0)
(244,0)
(245,0)
(246,0)
(247,0)
(248,0)
(249,0)
(250,0)
(251,0)
(252,0)
(253,0)
(254,0)
(255,0)
(256,0)
(257,0)
(258,0)
(259,0)
(260,1e-06)
(261,0)
(262,0)
(263,0)
(264,0)
(265,0)
(266,0)
(267,0)
(268,0)
(269,0)
(270,0)
(271,1e-06)
(272,0)
(273,0)
(274,0)
(275,0)
(276,0)
(277,0)
(278,0)
(279,0)
(280,0)
(281,0)
(282,0)
(283,0)
(284,0)
(285,0)
(286,0)
(287,0)
(288,0)
(289,0)
(290,0)
(291,0)
(292,0)
(293,0)
(294,0)
(295,0)
(296,0)
(297,0)
(298,0)
(299,0)
(300,0)
(301,0)
(302,0)
(303,0)
(304,0)
(305,0)
(306,0)
(307,0)
(308,0)
(309,0)
(310,0)
(311,0)
(312,0)
(313,0)
(314,0)
(315,0)
(316,0)
(317,0)
(318,0)
(319,0)
}

\def\perlEXECDATA{
(0,1.345492)
(1,10.109388)
(2,14.100158)
(3,21.566704)
(4,21.8871)
(5,20.435874)
(6,23.090282)
(7,24.477136)
(8,24.499262)
(9,24.592828)
(10,24.728384)
(11,24.705528)
(12,24.771702)
(13,24.535138)
(14,24.514008)
(15,24.54265)
(16,24.79936)
(17,24.470604)
(18,24.624592)
(19,24.497366)
(20,24.532826)
(21,24.798662)
(22,24.666632)
(23,24.796072)
(24,24.73225)
(25,24.741942)
(26,24.790954)
(27,24.742146)
(28,24.786088)
(29,24.605102)
(30,24.6203)
(31,24.749206)
(32,24.832242)
(33,24.80128)
(34,24.807142)
(35,24.634798)
(36,24.584238)
(37,24.492496)
(38,24.837522)
(39,24.68988)
(40,24.75951)
(41,24.61616)
(42,24.70409)
(43,24.507906)
(44,24.717084)
(45,24.289702)
(46,24.753846)
(47,24.489512)
(48,24.747872)
(49,24.778946)
(50,24.567614)
(51,24.649782)
(52,24.633584)
(53,24.549312)
(54,24.73839)
(55,24.833822)
(56,24.735496)
(57,24.728324)
(58,24.600324)
(59,24.5873)
(60,24.78802)
(61,24.744922)
(62,24.800142)
(63,24.605102)
(64,24.712782)
(65,24.603804)
(66,24.694018)
(67,24.764342)
(68,24.813982)
(69,24.723038)
(70,24.752376)
(71,24.749456)
(72,24.709442)
(73,24.585622)
(74,24.654128)
(75,24.57729)
(76,24.753416)
(77,24.7288)
(78,24.708556)
(79,24.6946)
(80,24.52351)
(81,24.802168)
(82,24.843326)
(83,24.886416)
(84,24.613428)
(85,24.788316)
(86,24.708032)
(87,24.65865)
(88,24.627464)
(89,24.706704)
(90,24.575814)
(91,24.766126)
(92,24.771406)
(93,24.77406)
(94,24.602994)
(95,24.56505)
(96,24.730958)
(97,24.834588)
(98,24.687748)
(99,24.609656)
(100,24.828274)
(101,24.765084)
(102,24.656544)
(103,24.654234)
(104,24.657844)
(105,24.758084)
(106,24.653136)
(107,24.406272)
(108,23.11922)
(109,25.933764)
(110,20.346914)
(111,25.35619)
(112,25.377808)
(113,25.598248)
(114,25.569554)
(115,25.553988)
(116,25.655836)
(117,25.505868)
(118,25.250664)
(119,25.572884)
(120,25.553354)
(121,25.622972)
(122,25.482466)
(123,25.55436)
(124,25.682554)
(125,25.451734)
(126,25.567018)
(127,25.552138)
(128,25.645926)
(129,25.475632)
(130,25.565478)
(131,25.569432)
(132,25.64816)
(133,25.414846)
(134,25.525762)
(135,25.491088)
(136,25.609172)
(137,25.566306)
(138,25.421302)
(139,25.423608)
(140,25.50211)
(141,25.5651)
(142,25.596144)
(143,25.577784)
(144,25.500222)
(145,25.63079)
(146,25.400826)
(147,25.514286)
(148,25.438394)
(149,25.567306)
(150,25.263322)
(151,25.532914)
(152,25.555342)
(153,25.565638)
(154,25.536212)
(155,25.438156)
(156,25.414816)
(157,25.571794)
(158,25.51758)
(159,25.428584)
(160,25.655364)
(161,25.48104)
(162,25.525496)
(163,25.317672)
(164,25.562604)
(165,25.526282)
(166,25.64913)
(167,25.600572)
(168,25.534834)
(169,25.609214)
(170,25.376942)
(171,25.652706)
(172,25.461224)
(173,25.596804)
(174,25.532578)
(175,25.475002)
(176,24.717974)
(177,25.012862)
(178,24.798014)
(179,24.96405)
(180,24.689116)
(181,24.989392)
(182,24.87357)
(183,25.081648)
(184,25.129458)
(185,24.91762)
(186,24.877154)
(187,25.053532)
(188,24.980412)
(189,25.14468)
(190,25.052052)
(191,25.161244)
(192,24.99613)
(193,24.749924)
(194,25.099034)
(195,25.154108)
(196,25.150154)
(197,24.872076)
(198,24.953744)
(199,24.866474)
(200,24.934452)
(201,25.184444)
(202,25.003702)
(203,24.974078)
(204,25.07705)
(205,24.961024)
(206,24.859836)
(207,24.949524)
(208,24.849656)
(209,25.142976)
(210,24.937794)
(211,24.970888)
(212,25.00619)
(213,24.993996)
(214,24.916848)
(215,24.94095)
(216,24.80932)
(217,24.9135)
(218,25.087032)
(219,25.005692)
(220,24.930482)
(221,24.934264)
(222,25.084218)
(223,24.841162)
(224,25.250714)
(225,25.042244)
(226,24.999428)
(227,25.121878)
(228,24.94454)
(229,24.888164)
(230,24.872016)
(231,25.038648)
(232,25.03446)
(233,25.021606)
(234,25.007252)
(235,25.164238)
(236,25.139532)
(237,25.089224)
(238,24.876554)
(239,25.036262)
(240,25.03956)
(241,24.986642)
(242,24.992916)
(243,25.143708)
(244,24.8141)
(245,24.926152)
(246,24.863014)
(247,24.926702)
(248,24.998002)
(249,25.38502)
(250,25.19334)
(251,25.476368)
(252,25.31279)
(253,25.320162)
(254,25.046884)
(255,25.089976)
(256,25.248054)
(257,25.27089)
(258,25.271342)
(259,25.441148)
(260,25.151948)
(261,25.225652)
(262,25.350548)
(263,25.428802)
(264,25.308514)
(265,25.116264)
(266,25.248966)
(267,25.413254)
(268,25.269382)
(269,25.321886)
(270,25.52277)
(271,25.334976)
(272,25.341382)
(273,25.352588)
(274,25.072708)
(275,25.116586)
(276,23.189996)
(277,9.436784)
}

\def\perlGENDATA{
(0,0.00084)
(1,0.000544)
(2,0.000614)
(3,0.000184)
(4,0.00018)
(5,0.00022)
(6,0.000652)
(7,0.000102)
(8,0.000124)
(9,1.8e-05)
(10,1.6e-05)
(11,1.8e-05)
(12,1e-05)
(13,8e-06)
(14,2e-06)
(15,2e-06)
(16,0)
(17,6e-06)
(18,0)
(19,2e-06)
(20,2e-06)
(21,4e-06)
(22,0)
(23,2e-06)
(24,0)
(25,0)
(26,0)
(27,0)
(28,0)
(29,0)
(30,0)
(31,0)
(32,0)
(33,2e-06)
(34,0)
(35,0)
(36,0)
(37,4e-06)
(38,0)
(39,2e-06)
(40,2e-06)
(41,0)
(42,2e-06)
(43,0)
(44,0)
(45,0)
(46,2e-06)
(47,0)
(48,0)
(49,0)
(50,2e-06)
(51,2e-06)
(52,2e-06)
(53,0)
(54,0)
(55,0)
(56,2e-06)
(57,0)
(58,0)
(59,0)
(60,0)
(61,0)
(62,0)
(63,6e-06)
(64,0)
(65,0)
(66,0)
(67,0)
(68,0)
(69,0)
(70,0)
(71,0)
(72,2e-06)
(73,0)
(74,0)
(75,0)
(76,0)
(77,0)
(78,0)
(79,4e-06)
(80,2e-06)
(81,0)
(82,0)
(83,0)
(84,2e-06)
(85,2e-06)
(86,4e-06)
(87,0)
(88,0)
(89,2e-06)
(90,0)
(91,0)
(92,0)
(93,0)
(94,0)
(95,2e-06)
(96,0)
(97,0)
(98,0)
(99,2e-06)
(100,0)
(101,0)
(102,2e-06)
(103,0)
(104,2e-06)
(105,0)
(106,0)
(107,0)
(108,4e-05)
(109,2e-06)
(110,4.6e-05)
(111,2.4e-05)
(112,6.8e-05)
(113,4e-06)
(114,2e-06)
(115,0)
(116,0)
(117,4e-06)
(118,0)
(119,2e-06)
(120,0)
(121,2e-06)
(122,0)
(123,0)
(124,2e-06)
(125,2e-06)
(126,0)
(127,0)
(128,0)
(129,0)
(130,0)
(131,0)
(132,0)
(133,0)
(134,0)
(135,0)
(136,2e-06)
(137,4e-06)
(138,2e-06)
(139,6e-06)
(140,2e-05)
(141,2e-06)
(142,0)
(143,0)
(144,0)
(145,6e-06)
(146,0)
(147,0)
(148,2e-06)
(149,0)
(150,0)
(151,0)
(152,0)
(153,0)
(154,2e-06)
(155,0)
(156,2e-06)
(157,0)
(158,0)
(159,0)
(160,0)
(161,0)
(162,0)
(163,2e-06)
(164,4e-06)
(165,2e-06)
(166,0)
(167,0)
(168,0)
(169,2e-06)
(170,0)
(171,0)
(172,4e-06)
(173,1.2e-05)
(174,0)
(175,0)
(176,4.8e-05)
(177,0)
(178,0)
(179,0)
(180,0)
(181,0)
(182,2e-06)
(183,2e-06)
(184,2e-06)
(185,0)
(186,0)
(187,0)
(188,0)
(189,0)
(190,0)
(191,2e-06)
(192,2e-06)
(193,0)
(194,0)
(195,0)
(196,2e-06)
(197,0)
(198,0)
(199,0)
(200,0)
(201,0)
(202,0)
(203,0)
(204,0)
(205,0)
(206,2e-06)
(207,2e-06)
(208,8e-06)
(209,4e-06)
(210,0)
(211,0)
(212,0)
(213,0)
(214,0)
(215,0)
(216,0)
(217,0)
(218,0)
(219,0)
(220,0)
(221,0)
(222,0)
(223,0)
(224,0)
(225,0)
(226,0)
(227,0)
(228,0)
(229,0)
(230,0)
(231,2e-06)
(232,0)
(233,0)
(234,0)
(235,0)
(236,0)
(237,2e-06)
(238,0)
(239,0)
(240,0)
(241,2e-06)
(242,0)
(243,0)
(244,0)
(245,0)
(246,0)
(247,0)
(248,0)
(249,2e-06)
(250,0)
(251,0)
(252,0)
(253,0)
(254,0)
(255,0)
(256,0)
(257,0)
(258,0)
(259,0)
(260,0)
(261,0)
(262,2e-06)
(263,0)
(264,0)
(265,2e-06)
(266,0)
(267,0)
(268,0)
(269,2e-06)
(270,0)
(271,0)
(272,0)
(273,0)
(274,4e-06)
(275,0)
(276,3.8e-05)
(277,4e-06)
}

\def\xalEXECDATA{
(0,23.406395714285715)
(1,29.351147142857144)
(2,30.911164285714285)
(3,29.36528142857143)
(4,30.949512857142857)
(5,29.39583)
(6,30.052112857142856)
(7,29.764465714285713)
(8,30.250297142857143)
(9,31.80818857142857)
(10,30.21120714285714)
(11,13.042512857142857)
(12,10.471282857142857)
(13,10.780525714285714)
(14,11.771248571428572)
(15,11.717155714285715)
(16,10.483872857142858)
(17,10.033085714285715)
(18,9.250825714285714)
(19,9.120344285714285)
(20,9.406301428571428)
(21,10.487612857142857)
(22,10.449014285714286)
(23,10.458632857142858)
(24,10.414124285714285)
(25,10.411404285714285)
(26,10.436764285714286)
(27,10.444485714285713)
(28,10.453504285714287)
(29,10.486092857142857)
(30,10.517078571428572)
(31,10.502982857142857)
(32,10.556107142857142)
(33,10.464524285714285)
(34,10.48386)
(35,10.540502857142856)
(36,10.315054285714286)
(37,10.172381428571429)
(38,10.025531428571428)
(39,8.890355714285715)
(40,9.21956)
(41,9.10942)
(42,9.97717)
(43,10.556551428571428)
(44,9.940295714285714)
(45,9.421717142857142)
(46,9.476507142857143)
(47,9.975261428571429)
(48,10.286794285714286)
(49,9.58669)
(50,10.490718571428571)
(51,10.42145)
(52,10.303407142857143)
(53,9.064027142857142)
(54,9.342077142857143)
(55,9.112398571428571)
(56,10.057728571428571)
(57,10.131067142857143)
(58,10.591807142857142)
(59,10.467371428571429)
(60,10.406195714285714)
(61,10.49218857142857)
(62,10.392315714285715)
(63,10.52077)
(64,10.494518571428571)
(65,10.459705714285715)
(66,10.385334285714286)
(67,10.509315714285714)
(68,9.864577142857144)
(69,9.511024285714285)
(70,10.432685714285714)
(71,10.463488571428572)
(72,9.441951428571429)
(73,10.460221428571428)
(74,10.504665714285714)
(75,10.466207142857144)
(76,10.489042857142858)
(77,10.55473)
(78,10.41978857142857)
(79,10.45404142857143)
(80,10.492581428571429)
(81,10.410035714285714)
(82,10.50672)
(83,9.38768)
(84,8.818542857142857)
(85,10.458365714285714)
(86,10.475875714285714)
(87,10.489455714285715)
(88,10.439502857142857)
(89,10.507727142857142)
(90,10.407537142857143)
(91,10.42082)
(92,10.511164285714285)
(93,10.437837142857143)
(94,10.458562857142857)
(95,11.540115714285715)
(96,11.223234285714286)
(97,9.891012857142858)
(98,8.934422857142858)
(99,9.448002857142857)
(100,10.039955714285714)
(101,10.503767142857143)
(102,10.403115714285715)
(103,9.368524285714285)
(104,9.326401428571428)
(105,8.897237142857144)
(106,10.19586)
(107,10.393407142857143)
(108,10.401191428571428)
(109,10.483291428571428)
(110,10.47339)
(111,10.444664285714286)
(112,10.503245714285715)
(113,10.496417142857142)
(114,10.539947142857143)
(115,10.408168571428572)
(116,10.431792857142858)
(117,10.493077142857143)
(118,9.458594285714286)
(119,9.911604285714287)
(120,10.43448)
(121,10.415255714285715)
(122,10.545657142857143)
(123,10.460192857142857)
(124,10.517475714285714)
(125,10.483595714285714)
(126,9.417144285714286)
(127,8.875657142857143)
(128,9.369808571428571)
(129,9.460085714285714)
(130,9.900104285714285)
(131,9.385145714285715)
(132,9.480477142857143)
(133,10.471822857142858)
(134,10.519258571428571)
(135,10.491047142857143)
(136,10.518588571428571)
(137,10.49945857142857)
(138,10.489565714285714)
(139,10.589204285714287)
(140,10.546467142857143)
(141,10.002867142857143)
(142,9.324117142857142)
(143,8.956627142857142)
(144,9.450547142857143)
(145,9.480692857142857)
(146,10.33947)
(147,10.371488571428571)
(148,10.460164285714285)
(149,10.508165714285715)
(150,10.391604285714285)
(151,9.514964285714285)
(152,9.287632857142857)
(153,10.422018571428572)
(154,10.427415714285715)
(155,10.501735714285715)
(156,10.435672857142857)
(157,10.48378857142857)
(158,10.50447)
(159,10.520312857142857)
(160,10.428192857142857)
(161,10.454292857142857)
(162,9.95009)
(163,8.985965714285713)
(164,10.376872857142857)
(165,10.475115714285714)
(166,10.477335714285715)
(167,10.501897142857143)
(168,10.463837142857143)
(169,10.47228)
(170,10.520354285714285)
(171,10.482434285714286)
(172,9.5321)
(173,9.143401428571428)
(174,9.097154285714286)
(175,9.766347142857143)
(176,9.401952857142858)
(177,11.31022857142857)
(178,11.763358571428572)
(179,11.094744285714286)
(180,10.477585714285715)
(181,9.493607142857142)
(182,9.056072857142857)
(183,9.924007142857143)
(184,10.28493)
(185,9.437088571428571)
(186,10.441694285714286)
(187,10.450107142857142)
(188,10.440022857142857)
(189,10.413861428571428)
(190,9.853757142857143)
(191,10.451125714285714)
(192,9.954898571428572)
(193,9.141634285714286)
(194,9.769558571428572)
(195,9.483065714285715)
(196,9.885367142857143)
(197,10.204737142857143)
(198,9.727732857142858)
(199,10.505142857142857)
(200,10.516378571428572)
(201,10.481368571428572)
(202,10.445824285714286)
(203,9.98324)
(204,10.459107142857142)
(205,10.475894285714286)
(206,10.514305714285713)
(207,9.654822857142857)
(208,9.399002857142857)
(209,10.239281428571429)
(210,10.448931428571429)
(211,10.489675714285715)
(212,10.431308571428572)
(213,10.389994285714286)
(214,10.485395714285714)
(215,10.47956857142857)
(216,10.480282857142857)
(217,9.15627857142857)
(218,9.20760857142857)
(219,9.487875714285714)
(220,9.15766)
(221,10.062992857142858)
(222,9.542218571428572)
(223,10.269318571428572)
(224,10.160781428571429)
(225,9.985994285714286)
(226,10.463721428571429)
(227,10.567818571428571)
(228,10.472141428571428)
(229,10.435662857142857)
(230,9.874725714285715)
(231,9.38107142857143)
(232,10.468082857142857)
(233,10.510105714285714)
(234,10.443734285714285)
(235,10.422588571428571)
(236,10.417205714285714)
(237,10.485138571428571)
(238,10.516957142857143)
(239,10.486864285714285)
(240,10.533664285714286)
(241,10.532121428571429)
(242,10.455657142857143)
(243,10.458114285714286)
(244,10.39951)
(245,10.473837142857143)
(246,10.488841428571428)
(247,10.46475)
(248,9.457315714285714)
(249,9.397281428571429)
(250,9.835967142857143)
(251,10.466301428571429)
(252,10.503251428571428)
(253,10.493505714285714)
(254,10.37303)
(255,10.447874285714287)
(256,10.428942857142857)
(257,10.226151428571429)
(258,9.148044285714287)
(259,10.407281428571428)
(260,11.779695714285713)
(261,10.787531428571429)
(262,10.530415714285715)
(263,9.859174285714285)
(264,10.095667142857144)
(265,10.094614285714286)
(266,10.307055714285715)
(267,10.27624)
(268,10.274315714285715)
(269,10.2694)
(270,10.175294285714285)
(271,10.28207)
(272,9.82796)
(273,9.23193)
(274,9.494985714285715)
(275,10.006642857142857)
(276,10.437914285714285)
(277,10.476222857142858)
(278,10.47661)
(279,10.46356142857143)
(280,10.346714285714286)
(281,10.381435714285715)
(282,10.489687142857143)
(283,10.482305714285713)
(284,8.962411428571428)
(285,9.375785714285714)
(286,9.592107142857143)
(287,10.515862857142857)
(288,10.50286)
(289,10.436272857142857)
(290,9.356974285714285)
(291,9.412732857142856)
(292,9.507821428571429)
(293,9.469707142857143)
(294,9.914127142857144)
(295,10.410428571428572)
(296,10.454758571428572)
(297,10.48111142857143)
(298,10.509154285714287)
(299,10.505874285714286)
(300,10.47623142857143)
(301,10.494161428571429)
(302,10.577707142857143)
(303,10.039828571428572)
(304,9.458337142857143)
(305,9.397675714285715)
(306,9.972878571428572)
(307,10.415637142857143)
(308,10.552697142857143)
(309,10.4171)
(310,10.392855714285714)
(311,9.465244285714286)
(312,10.49575)
(313,10.498147142857142)
(314,10.435191428571429)
(315,9.473242857142857)
(316,9.454332857142857)
(317,10.409915714285715)
(318,9.92411)
(319,10.00556)
(320,10.51025)
(321,10.503014285714286)
(322,10.522448571428571)
(323,9.454817142857143)
(324,9.467348571428571)
(325,10.437841428571428)
(326,9.421927142857143)
(327,9.844692857142856)
(328,10.51024)
(329,10.465455714285714)
(330,10.443704285714286)
(331,10.477391428571428)
(332,9.947577142857142)
(333,9.418565714285714)
(334,10.48388142857143)
(335,10.006178571428572)
(336,10.391617142857143)
(337,10.447575714285714)
(338,10.487298571428571)
(339,10.47151)
(340,10.771477142857142)
(341,11.106892857142856)
(342,11.238268571428572)
(343,9.364972857142858)
(344,9.9887)
(345,10.501342857142857)
(346,10.520744285714287)
(347,10.456992857142858)
(348,10.454472857142857)
(349,10.4548)
(350,9.877771428571428)
(351,10.430297142857142)
(352,9.524097142857142)
(353,9.817435714285715)
(354,9.533985714285715)
(355,10.322387142857142)
(356,10.496924285714286)
(357,10.518324285714286)
(358,10.508712857142857)
(359,10.551547142857142)
(360,10.397455714285714)
(361,10.45754)
(362,10.142168571428572)
(363,9.496414285714286)
(364,10.350772857142857)
(365,10.473581428571428)
(366,10.388828571428572)
(367,10.523122857142857)
(368,10.516047142857143)
(369,10.484715714285715)
(370,10.4551)
(371,10.420455714285714)
(372,10.462751428571428)
(373,10.410954285714286)
(374,10.144985714285715)
(375,10.276902857142858)
(376,9.949411428571429)
(377,10.474601428571429)
(378,10.443228571428572)
(379,10.498354285714285)
(380,10.501687142857143)
(381,10.483964285714286)
(382,10.498002857142858)
(383,10.539971428571429)
(384,9.697577142857142)
(385,9.80037)
(386,10.417578571428571)
(387,10.483297142857143)
(388,10.440315714285715)
(389,10.546978571428571)
(390,10.486824285714286)
(391,10.48259)
(392,10.483352857142858)
(393,10.492847142857142)
(394,10.482674285714285)
(395,10.463424285714286)
(396,9.926675714285714)
(397,9.467882857142857)
(398,9.9121)
(399,10.243665714285715)
(400,10.185302857142856)
(401,10.580937142857143)
(402,10.61193)
(403,10.55103)
(404,10.499502857142858)
(405,10.441644285714286)
(406,10.512792857142857)
(407,10.446448571428572)
(408,10.476101428571429)
(409,10.450558571428571)
(410,10.4061)
(411,10.174871428571429)
(412,10.174384285714286)
(413,10.480002857142857)
(414,10.475437142857142)
(415,10.44387)
(416,10.400528571428572)
(417,10.390797142857142)
(418,7.044991428571429)
(419,0.0004014285714285714)
}

\def\xalGENDATA{
(0,0.00017142857142857143)
(1,0)
(2,1.2857142857142857e-05)
(3,1.4285714285714286e-06)
(4,5.7142857142857145e-06)
(5,7.142857142857143e-06)
(6,0)
(7,7.142857142857143e-06)
(8,0)
(9,2.8571428571428573e-06)
(10,0)
(11,3e-05)
(12,8.571428571428571e-06)
(13,1.4285714285714286e-06)
(14,2.8571428571428573e-06)
(15,8.571428571428571e-06)
(16,0)
(17,1.4285714285714286e-06)
(18,1.4285714285714286e-06)
(19,4.2857142857142855e-06)
(20,0)
(21,0)
(22,0)
(23,0)
(24,0)
(25,0)
(26,0)
(27,0)
(28,1.4285714285714286e-06)
(29,0)
(30,0)
(31,0)
(32,0)
(33,0)
(34,1.4285714285714286e-06)
(35,0)
(36,2.8571428571428573e-06)
(37,0)
(38,0)
(39,0)
(40,0)
(41,0)
(42,0)
(43,0)
(44,0)
(45,0)
(46,0)
(47,0)
(48,0)
(49,0)
(50,0)
(51,0)
(52,0)
(53,2.8571428571428573e-06)
(54,0)
(55,0)
(56,0)
(57,1.4285714285714286e-06)
(58,0)
(59,0)
(60,0)
(61,0)
(62,0)
(63,0)
(64,0)
(65,1.4285714285714286e-06)
(66,0)
(67,0)
(68,1.4285714285714286e-06)
(69,0)
(70,0)
(71,0)
(72,0)
(73,0)
(74,0)
(75,1.4285714285714286e-06)
(76,0)
(77,0)
(78,0)
(79,0)
(80,0)
(81,0)
(82,0)
(83,0)
(84,0)
(85,0)
(86,1.4285714285714286e-06)
(87,0)
(88,0)
(89,0)
(90,0)
(91,0)
(92,0)
(93,0)
(94,1.4285714285714286e-06)
(95,1.4285714285714286e-06)
(96,0)
(97,0)
(98,0)
(99,0)
(100,0)
(101,0)
(102,0)
(103,0)
(104,0)
(105,0)
(106,0)
(107,0)
(108,0)
(109,0)
(110,1.4285714285714286e-06)
(111,0)
(112,0)
(113,0)
(114,0)
(115,0)
(116,0)
(117,0)
(118,0)
(119,0)
(120,0)
(121,0)
(122,0)
(123,0)
(124,0)
(125,0)
(126,0)
(127,0)
(128,0)
(129,0)
(130,0)
(131,0)
(132,0)
(133,0)
(134,0)
(135,0)
(136,0)
(137,0)
(138,1.4285714285714286e-06)
(139,0)
(140,0)
(141,0)
(142,0)
(143,0)
(144,0)
(145,0)
(146,0)
(147,0)
(148,0)
(149,0)
(150,0)
(151,0)
(152,0)
(153,0)
(154,0)
(155,1.4285714285714286e-06)
(156,0)
(157,0)
(158,0)
(159,0)
(160,0)
(161,0)
(162,1.4285714285714286e-06)
(163,0)
(164,0)
(165,0)
(166,0)
(167,0)
(168,0)
(169,0)
(170,0)
(171,0)
(172,0)
(173,0)
(174,0)
(175,0)
(176,0)
(177,0)
(178,0)
(179,2.8571428571428573e-06)
(180,0)
(181,0)
(182,0)
(183,0)
(184,0)
(185,0)
(186,0)
(187,0)
(188,0)
(189,0)
(190,0)
(191,0)
(192,0)
(193,0)
(194,0)
(195,0)
(196,0)
(197,0)
(198,0)
(199,0)
(200,0)
(201,0)
(202,0)
(203,0)
(204,0)
(205,0)
(206,0)
(207,0)
(208,0)
(209,0)
(210,0)
(211,0)
(212,0)
(213,0)
(214,0)
(215,0)
(216,0)
(217,1.4285714285714286e-06)
(218,0)
(219,0)
(220,1.4285714285714286e-06)
(221,1.4285714285714286e-06)
(222,0)
(223,0)
(224,0)
(225,0)
(226,0)
(227,0)
(228,0)
(229,0)
(230,0)
(231,0)
(232,0)
(233,0)
(234,0)
(235,0)
(236,0)
(237,0)
(238,0)
(239,0)
(240,0)
(241,0)
(242,0)
(243,0)
(244,0)
(245,0)
(246,0)
(247,0)
(248,0)
(249,0)
(250,0)
(251,0)
(252,0)
(253,0)
(254,0)
(255,0)
(256,0)
(257,0)
(258,0)
(259,0)
(260,0)
(261,0)
(262,0)
(263,0)
(264,0)
(265,0)
(266,0)
(267,0)
(268,0)
(269,0)
(270,0)
(271,0)
(272,0)
(273,0)
(274,0)
(275,0)
(276,0)
(277,0)
(278,0)
(279,0)
(280,0)
(281,0)
(282,0)
(283,0)
(284,0)
(285,0)
(286,0)
(287,0)
(288,0)
(289,0)
(290,0)
(291,0)
(292,0)
(293,0)
(294,0)
(295,0)
(296,0)
(297,0)
(298,1.4285714285714286e-06)
(299,0)
(300,1.4285714285714286e-06)
(301,0)
(302,0)
(303,0)
(304,0)
(305,0)
(306,0)
(307,0)
(308,0)
(309,0)
(310,0)
(311,0)
(312,0)
(313,0)
(314,0)
(315,0)
(316,0)
(317,0)
(318,0)
(319,0)
(320,0)
(321,0)
(322,0)
(323,0)
(324,0)
(325,0)
(326,0)
(327,1.4285714285714286e-06)
(328,0)
(329,0)
(330,0)
(331,0)
(332,0)
(333,0)
(334,0)
(335,0)
(336,0)
(337,0)
(338,0)
(339,0)
(340,0)
(341,0)
(342,0)
(343,0)
(344,0)
(345,0)
(346,0)
(347,0)
(348,0)
(349,0)
(350,0)
(351,0)
(352,0)
(353,0)
(354,0)
(355,0)
(356,0)
(357,0)
(358,0)
(359,0)
(360,0)
(361,0)
(362,0)
(363,0)
(364,0)
(365,0)
(366,0)
(367,0)
(368,0)
(369,0)
(370,0)
(371,0)
(372,0)
(373,0)
(374,0)
(375,0)
(376,0)
(377,0)
(378,0)
(379,0)
(380,0)
(381,0)
(382,0)
(383,0)
(384,0)
(385,0)
(386,0)
(387,0)
(388,0)
(389,0)
(390,1.4285714285714286e-06)
(391,0)
(392,0)
(393,0)
(394,0)
(395,0)
(396,0)
(397,0)
(398,0)
(399,0)
(400,0)
(401,0)
(402,0)
(403,0)
(404,0)
(405,0)
(406,0)
(407,0)
(408,0)
(409,0)
(410,0)
(411,0)
(412,0)
(413,0)
(414,0)
(415,0)
(416,0)
(417,0)
(418,0)
(419,0)
}

\def\mcfEXECDATA{
(0,20.687085)
(1,19.4667125)
(2,19.1965075)
(3,18.998456)
(4,18.9161045)
(5,19.2678765)
(6,20.18984)
(7,20.5472165)
(8,20.6875275)
(9,20.7789365)
(10,20.8538695)
(11,20.850412)
(12,20.949086)
(13,20.980826)
(14,20.997794)
(15,21.05181)
(16,21.0666745)
(17,21.042206)
(18,21.4470505)
(19,22.887761)
(20,22.8088165)
(21,22.8554635)
(22,22.8298155)
(23,22.717923)
(24,22.709916)
(25,22.6639925)
(26,22.6068025)
(27,22.6305465)
(28,22.589049)
(29,22.509284)
(30,22.556769)
(31,22.5201205)
(32,22.4509385)
(33,22.4516405)
(34,22.410743)
(35,22.272972)
(36,22.268476)
(37,22.291415)
(38,22.190667)
(39,22.195717)
(40,22.1805025)
(41,22.066468)
(42,22.1073965)
(43,22.107686)
(44,22.044871)
(45,22.1041675)
(46,22.08562)
(47,21.9709075)
(48,22.0036615)
(49,21.958505)
(50,21.8644525)
(51,21.8874425)
(52,21.892509)
(53,21.839115)
(54,21.85964)
(55,21.854637)
(56,21.8433215)
(57,21.778217)
(58,21.823987)
(59,21.740456)
(60,21.7769995)
(61,21.760314)
(62,21.625652)
(63,21.7034035)
(64,20.549303)
(65,21.01168)
(66,21.521599)
(67,21.4912495)
(68,21.3614585)
(69,21.1678355)
(70,21.172442)
(71,21.0751595)
(72,21.0242065)
(73,20.9851245)
(74,20.9109955)
(75,20.873643)
(76,20.8068035)
(77,20.7409575)
(78,20.777759)
(79,20.7850565)
(80,20.7593375)
(81,20.663782)
(82,20.6544415)
(83,20.5741645)
(84,20.523291)
(85,19.699864)
(86,19.101399)
(87,19.4835385)
(88,20.0651105)
(89,20.071595)
(90,20.075914)
(91,19.9874465)
(92,19.9918535)
(93,19.93688)
(94,19.891491)
(95,19.7664695)
(96,19.7520935)
(97,19.6301195)
(98,19.4895495)
(99,19.4526285)
(100,19.342609)
(101,19.153718)
(102,19.153494)
(103,19.0510385)
(104,18.8662405)
(105,18.706828)
(106,18.661617)
(107,18.414344)
(108,18.3836415)
(109,18.292963)
(110,18.209586)
(111,18.1348245)
(112,17.9328345)
(113,17.8996035)
(114,17.829097)
(115,17.6435085)
(116,17.507773)
(117,17.3873225)
(118,17.2204495)
(119,17.0605995)
(120,16.705631)
(121,16.77481)
(122,16.7631705)
(123,16.674621)
(124,16.447196)
(125,16.1916955)
(126,15.824418)
(127,15.2684325)
(128,14.370132)
(129,12.5713985)
(130,19.665266)
(131,54.789644)
(132,55.3193015)
(133,51.9023065)
(134,47.4204185)
(135,45.134566)
(136,39.7659405)
(137,35.2464365)
(138,39.234153)
(139,38.165654)
(140,40.933862)
(141,31.2820435)
(142,31.842277)
(143,15.9940575)
(144,2.428711)
(145,23.7736535)
(146,48.367263)
(147,56.07373)
(148,56.5591395)
(149,53.6645795)
(150,53.0671305)
(151,43.7892585)
(152,26.16078)
(153,27.5785235)
(154,31.696497)
(155,31.01238)
(156,31.3096185)
(157,31.2808725)
(158,28.3359885)
(159,11.565997)
(160,13.1050515)
(161,10.1411025)
(162,11.4558705)
(163,11.4490695)
(164,12.3273225)
(165,11.3041395)
(166,9.7648465)
(167,6.046784)
(168,0.474629)
(169,12.6982785)
(170,25.722347)
(171,41.313074)
(172,38.328205)
(173,26.453885)
(174,29.4127635)
(175,32.91184)
(176,31.468568)
(177,31.570902)
(178,32.2715325)
(179,13.2964535)
(180,18.833325)
(181,18.160338)
(182,18.1128725)
(183,17.8029275)
(184,17.121104)
(185,16.8745915)
(186,16.8143475)
(187,17.229752)
(188,16.4046885)
(189,16.78742)
(190,16.8161645)
(191,13.436748)
(192,12.8315005)
(193,12.748892)
(194,15.6774375)
(195,16.091116)
(196,16.187949)
(197,16.6298525)
(198,15.8842595)
(199,15.976327)
(200,16.112698)
(201,15.860225)
(202,16.4134245)
(203,16.6356215)
(204,16.088205)
(205,16.069194)
(206,15.4103785)
(207,15.8211725)
(208,15.5709655)
(209,15.863312)
(210,15.9634675)
(211,16.0774)
(212,16.045216)
(213,15.184775)
(214,15.4752255)
(215,16.040966)
(216,15.6470055)
(217,15.2023185)
(218,15.259702)
(219,15.9805955)
(220,14.914027)
(221,15.5627025)
(222,15.661608)
(223,15.2264285)
(224,15.637417)
(225,15.648359)
(226,15.2650185)
(227,15.6347565)
(228,13.822498)
(229,15.113009)
(230,14.00845)
(231,13.826257)
(232,14.621615)
(233,14.2746195)
(234,13.3386905)
(235,13.8170665)
(236,11.9257985)
(237,12.5065385)
(238,10.1743485)
(239,11.6477895)
(240,12.442999)
(241,8.092116)
(242,6.781915)
(243,0.2511655)
(244,0.0073935)
(245,11.7345465)
(246,24.11516)
(247,26.3482745)
(248,26.359323)
(249,29.443103)
(250,31.173836)
(251,15.7497585)
(252,18.0260935)
(253,17.86088)
(254,17.4969485)
(255,16.679431)
(256,17.7222645)
(257,17.284406)
(258,17.01162)
(259,17.532316)
(260,16.412173)
(261,17.147977)
(262,17.1814695)
(263,16.927106)
(264,16.7916875)
(265,16.214813)
(266,16.952682)
(267,15.6742535)
(268,16.0847885)
(269,16.6033795)
(270,16.259988)
(271,15.9825145)
(272,15.419142)
(273,15.883983)
(274,12.190633)
(275,9.831302)
(276,4.489843)
(277,8.7950565)
(278,23.3528045)
(279,25.366952)
(280,26.4562025)
(281,21.2254615)
(282,25.0334045)
(283,26.0958015)
(284,21.956764)
(285,24.8578095)
(286,25.876499)
(287,23.145665)
(288,23.5684095)
(289,25.570463)
(290,26.4429515)
(291,21.4905645)
(292,25.078756)
(293,26.266165)
(294,22.4279275)
(295,24.95678)
(296,25.9533095)
(297,24.040325)
(298,24.8620955)
(299,25.724072)
(300,22.700743)
(301,24.3921565)
(302,25.522163)
(303,25.22702)
}

\def\mcfGENDATA{
(0,5.45e-05)
(1,1.05e-05)
(2,0)
(3,0)
(4,0)
(5,0)
(6,0)
(7,0)
(8,0)
(9,0)
(10,0)
(11,0)
(12,0)
(13,0)
(14,0)
(15,0)
(16,0)
(17,0)
(18,3e-06)
(19,1e-06)
(20,0)
(21,0)
(22,5e-07)
(23,6e-06)
(24,0)
(25,0)
(26,0)
(27,0)
(28,0)
(29,0)
(30,0)
(31,0)
(32,0)
(33,0)
(34,0)
(35,0)
(36,0)
(37,0)
(38,0)
(39,0)
(40,0)
(41,0)
(42,0)
(43,0)
(44,0)
(45,0)
(46,0)
(47,0)
(48,0)
(49,1e-06)
(50,5e-07)
(51,0)
(52,0)
(53,1.5e-06)
(54,0)
(55,5e-07)
(56,0)
(57,5e-07)
(58,0)
(59,5e-07)
(60,0)
(61,0)
(62,0)
(63,0)
(64,0)
(65,0)
(66,0)
(67,0)
(68,0)
(69,0)
(70,0)
(71,0)
(72,0)
(73,0)
(74,0)
(75,0)
(76,0)
(77,0)
(78,0)
(79,0)
(80,0)
(81,0)
(82,0)
(83,0)
(84,0)
(85,0)
(86,0)
(87,0)
(88,0)
(89,0)
(90,0)
(91,0)
(92,0)
(93,0)
(94,0)
(95,0)
(96,1e-06)
(97,0)
(98,0)
(99,0)
(100,0)
(101,0)
(102,0)
(103,0)
(104,0)
(105,0)
(106,0)
(107,0)
(108,0)
(109,0)
(110,0)
(111,0)
(112,0)
(113,0)
(114,0)
(115,0)
(116,0)
(117,0)
(118,0)
(119,0)
(120,1e-06)
(121,0)
(122,0)
(123,0)
(124,0)
(125,0)
(126,0)
(127,0)
(128,0)
(129,0)
(130,1.4e-05)
(131,1.2e-05)
(132,2e-06)
(133,5e-07)
(134,1e-06)
(135,0)
(136,0)
(137,0)
(138,0)
(139,0)
(140,5e-06)
(141,5.5e-06)
(142,0)
(143,2.5e-06)
(144,5e-07)
(145,6.5e-06)
(146,0)
(147,0)
(148,0)
(149,0)
(150,0)
(151,0)
(152,0)
(153,0)
(154,0)
(155,0)
(156,0)
(157,0)
(158,0)
(159,1.5e-06)
(160,0)
(161,0)
(162,0)
(163,0)
(164,0)
(165,0)
(166,0)
(167,0)
(168,0)
(169,1.5e-06)
(170,1.5e-06)
(171,0)
(172,0)
(173,0)
(174,0)
(175,0)
(176,0)
(177,0)
(178,0)
(179,2e-06)
(180,5e-07)
(181,0)
(182,0)
(183,0)
(184,0)
(185,0)
(186,0)
(187,0)
(188,0)
(189,0)
(190,0)
(191,0)
(192,0)
(193,0)
(194,0)
(195,0)
(196,0)
(197,0)
(198,0)
(199,0)
(200,0)
(201,0)
(202,0)
(203,0)
(204,0)
(205,0)
(206,0)
(207,0)
(208,0)
(209,0)
(210,0)
(211,0)
(212,0)
(213,0)
(214,0)
(215,0)
(216,0)
(217,0)
(218,0)
(219,0)
(220,0)
(221,0)
(222,0)
(223,0)
(224,0)
(225,0)
(226,0)
(227,0)
(228,0)
(229,0)
(230,0)
(231,0)
(232,0)
(233,0)
(234,0)
(235,0)
(236,0)
(237,0)
(238,0)
(239,0)
(240,0)
(241,0)
(242,0)
(243,0)
(244,0)
(245,0)
(246,2e-06)
(247,0)
(248,0)
(249,0)
(250,0)
(251,0)
(252,0)
(253,0)
(254,0)
(255,0)
(256,0)
(257,0)
(258,0)
(259,0)
(260,0)
(261,0)
(262,0)
(263,0)
(264,0)
(265,0)
(266,0)
(267,0)
(268,0)
(269,0)
(270,0)
(271,0)
(272,0)
(273,0)
(274,0)
(275,0)
(276,0)
(277,0)
(278,0)
(279,0)
(280,0)
(281,0)
(282,0)
(283,0)
(284,0)
(285,0)
(286,0)
(287,0)
(288,0)
(289,0)
(290,0)
(291,0)
(292,0)
(293,0)
(294,0)
(295,0)
(296,0)
(297,0)
(298,0)
(299,0)
(300,0)
(301,0)
(302,0)
(303,0)
}

\def\gzipGENDATA{
(0,0.00011)
(1,0)
(2,0)
(3,0)
(4,0)
(5,3e-05)
(6,0)
(7,0)
(8,0)
(9,0)
(10,0)
(11,0)
(12,0)
(13,0)
(14,0)
(15,0)
(16,0)
(17,0)
(18,0)
(19,0)
(20,0)
(21,0)
(22,0)
(23,0)
(24,0)
(25,0)
(26,0)
(27,0)
(28,0)
(29,0)
(30,0)
(31,0)
(32,0)
(33,0)
(34,4e-05)
(35,0)
(36,0)
(37,0)
(38,0)
(39,0)
(40,0)
(41,0)
}

\def\gzipEXECDATA{
(0,10.27446)
(1,13.86097)
(2,21.35776)
(3,7.37882)
(4,19.33209)
(5,20.72064)
(6,21.68338)
(7,24.47056)
(8,21.81354)
(9,8.18508)
(10,13.76361)
(11,23.62062)
(12,23.96804)
(13,24.44843)
(14,21.52581)
(15,22.54907)
(16,24.07105)
(17,21.84549)
(18,22.4433)
(19,22.46486)
(20,12.92595)
(21,22.5505)
(22,22.54497)
(23,22.94038)
(24,22.41702)
(25,22.29692)
(26,12.42091)
(27,22.29469)
(28,22.52743)
(29,17.38845)
(30,25.57652)
(31,20.27676)
(32,22.54571)
(33,24.22834)
(34,11.14929)
(35,22.13409)
(36,23.06609)
(37,24.31942)
(38,14.57888)
(39,22.54958)
(40,21.79746)
(41,23.41543)
}

\def\bzipGENDATA{
(0,4e-05)
(1,0.00024)
(2,0.00016)
(3,0)
(4,1e-05)
(5,0)
(6,0)
(7,0)
(8,0)
(9,0)
(10,0)
(11,0)
(12,0)
(13,0)
(14,0)
(15,0)
(16,0.00101)
(17,2e-05)
(18,0.00257)
(19,3e-05)
(20,0)
(21,0)
(22,0)
(23,3e-05)
(24,0)
(25,0)
(26,0)
(27,0)
(28,0)
(29,0)
(30,0)
(31,0)
(32,0)
(33,0)
(34,0)
(35,0)
(36,2e-05)
(37,1e-05)
(38,3e-05)
(39,0)
(40,0)
(41,2e-05)
(42,2e-05)
(43,1e-05)
(44,0)
(45,0)
(46,0)
(47,0)
(48,0)
(49,0)
(50,0)
(51,0)
(52,0)
(53,0)
(54,0)
(55,0)
(56,0)
(57,0)
(58,0)
(59,0)
(60,1e-05)
(61,0)
(62,0)
(63,0)
(64,0)
(65,0)
(66,0)
(67,0)
(68,0)
(69,0)
(70,0)
(71,0)
(72,0)
(73,0)
(74,0)
(75,0)
(76,0)
(77,1e-05)
(78,1e-05)
(79,0)
(80,1e-05)
(81,0)
(82,0)
(83,0)
(84,0)
(85,0)
(86,0)
(87,0)
(88,0)
(89,0)
(90,0)
(91,0)
(92,0)
(93,0)
(94,1e-05)
(95,0)
(96,0)
(97,0)
(98,0)
(99,0)
(100,0)
(101,0)
(102,0)
(103,0)
(104,0)
(105,0)
(106,0)
(107,0)
(108,0)
(109,0)
(110,0)
(111,0)
(112,0)
(113,0)
(114,0)
(115,0)
(116,8e-05)
(117,0)
(118,0)
(119,0)
(120,0)
(121,0)
(122,0)
(123,0)
(124,0)
(125,0)
(126,0)
(127,0)
(128,0)
(129,0)
(130,0)
(131,0)
(132,0)
(133,0)
(134,0)
(135,0)
(136,0)
(137,2e-05)
(138,0)
(139,4e-05)
(140,0)
(141,0)
(142,0)
(143,0)
(144,0)
(145,0)
(146,0)
(147,0)
(148,0)
(149,0)
(150,0)
(151,1e-05)
(152,0)
(153,2e-05)
(154,0)
(155,0)
(156,0)
(157,0)
(158,0)
(159,0)
(160,0)
(161,0)
(162,0)
(163,0)
(164,0)
(165,0)
(166,0)
(167,0)
(168,0)
(169,0)
(170,0)
(171,0)
(172,0)
(173,1e-05)
(174,0)
(175,0)
(176,0)
(177,3e-05)
(178,0)
(179,0)
(180,0)
(181,0)
(182,0)
(183,0)
(184,0)
(185,0)
(186,0)
(187,0)
(188,0)
(189,0)
(190,0)
(191,0)
(192,0)
(193,0)
(194,0)
(195,0)
(196,0)
(197,1e-05)
(198,0)
(199,0)
(200,0)
(201,0)
(202,0)
(203,0)
(204,0)
(205,0)
(206,0)
(207,0)
(208,0)
(209,0)
(210,0)
(211,0)
(212,0)
(213,0)
(214,0)
(215,0)
(216,0.00046)
(217,0)
}

\def\bzipEXECDATA{
(0,11.4399)
(1,40.17818)
(2,25.57305)
(3,17.7506)
(4,4.94231)
(5,0.74261)
(6,0.72667)
(7,0.75573)
(8,0.75023)
(9,0.75489)
(10,0.7559)
(11,0.74588)
(12,0.75444)
(13,0.75562)
(14,0.755)
(15,0.74484)
(16,15.89455)
(17,38.58246)
(18,21.99511)
(19,41.07954)
(20,31.60015)
(21,43.34445)
(22,26.41626)
(23,17.60998)
(24,3.40233)
(25,0.75368)
(26,0.75602)
(27,0.75626)
(28,0.75551)
(29,0.72986)
(30,0.75364)
(31,0.75437)
(32,0.7556)
(33,0.75673)
(34,0.75509)
(35,0.73231)
(36,29.53027)
(37,40.16317)
(38,43.55534)
(39,32.45239)
(40,36.65192)
(41,37.9249)
(42,18.5493)
(43,8.16021)
(44,0.74)
(45,0.75382)
(46,0.75468)
(47,0.75134)
(48,0.75388)
(49,0.75467)
(50,0.75619)
(51,0.75402)
(52,0.75601)
(53,0.75663)
(54,0.75458)
(55,20.74391)
(56,40.12227)
(57,42.81198)
(58,35.60631)
(59,35.30197)
(60,41.82242)
(61,19.145)
(62,11.33438)
(63,0.7564)
(64,0.75427)
(65,0.75164)
(66,0.74861)
(67,0.7524)
(68,0.75166)
(69,0.75114)
(70,0.75613)
(71,0.75513)
(72,0.75637)
(73,0.75442)
(74,9.14578)
(75,39.92966)
(76,41.24415)
(77,39.96476)
(78,33.5256)
(79,46.81142)
(80,19.59633)
(81,16.4572)
(82,1.00681)
(83,0.75614)
(84,0.75242)
(85,0.74819)
(86,0.75254)
(87,0.7487)
(88,0.75058)
(89,0.7421)
(90,0.75576)
(91,0.75795)
(92,0.75177)
(93,0.83394)
(94,40.27924)
(95,40.36349)
(96,42.77675)
(97,31.62901)
(98,42.8203)
(99,27.5046)
(100,17.53287)
(101,3.9056)
(102,0.75413)
(103,0.75528)
(104,0.75389)
(105,0.75103)
(106,0.75091)
(107,0.73955)
(108,0.75046)
(109,0.75531)
(110,0.75264)
(111,0.75557)
(112,0.75519)
(113,31.2717)
(114,40.323)
(115,43.95629)
(116,31.85331)
(117,37.86301)
(118,35.94381)
(119,18.52171)
(120,7.24552)
(121,0.75301)
(122,0.75422)
(123,0.75611)
(124,0.75511)
(125,0.75627)
(126,0.74474)
(127,0.75588)
(128,0.75236)
(129,0.75307)
(130,0.75398)
(131,0.75617)
(132,22.07668)
(133,40.30131)
(134,43.08509)
(135,34.72894)
(136,35.61043)
(137,41.21086)
(138,19.00535)
(139,10.9869)
(140,0.75416)
(141,0.75211)
(142,0.7552)
(143,0.75473)
(144,0.75353)
(145,0.75314)
(146,0.75122)
(147,0.75405)
(148,0.74938)
(149,0.72739)
(150,0.7555)
(151,7.61012)
(152,40.29354)
(153,41.10828)
(154,35.49363)
(155,30.48902)
(156,42.40455)
(157,28.35544)
(158,17.78379)
(159,4.01295)
(160,0.75509)
(161,0.75562)
(162,0.75476)
(163,0.75249)
(164,0.75637)
(165,0.75502)
(166,0.74773)
(167,0.74859)
(168,0.75381)
(169,0.75437)
(170,0.75176)
(171,29.56073)
(172,40.21579)
(173,43.61227)
(174,18.98807)
(175,6.004)
(176,6.3042)
(177,36.96091)
(178,38.26786)
(179,18.73824)
(180,8.2881)
(181,0.75469)
(182,0.75494)
(183,0.75743)
(184,0.75354)
(185,0.75507)
(186,0.7394)
(187,0.75504)
(188,0.75251)
(189,0.75414)
(190,0.7552)
(191,0.75566)
(192,18.6495)
(193,39.99323)
(194,42.71183)
(195,26.6562)
(196,18.38589)
(197,37.11284)
(198,37.19046)
(199,18.60379)
(200,7.77067)
(201,0.75393)
(202,0.74963)
(203,0.75657)
(204,0.75056)
(205,0.75556)
(206,0.75123)
(207,0.75155)
(208,0.75133)
(209,0.70174)
(210,0.75614)
(211,0.75333)
(212,14.20375)
(213,40.16445)
(214,41.61937)
(215,39.50943)
(216,18.78777)
(217,0.75373)
}

\def\djpegGENDATA{
(0,0.00022)
(1,0.00031)
(2,2e-05)
(3,0)
(4,0)
(5,0)
(6,0)
(7,0)
(8,0)
(9,0)
(10,0)
(11,0)
(12,1e-05)
(13,0)
(14,0)
(15,0)
(16,0)
(17,0)
(18,0)
(19,0)
(20,0)
(21,0)
(22,0)
(23,0)
(24,0)
(25,0)
(26,0)
(27,7e-05)
(28,1e-05)
(29,0)
(30,0)
(31,2e-05)
(32,0)
(33,0)
(34,0)
(35,0)
(36,0)
(37,0)
(38,0)
(39,0)
(40,5e-05)
(41,0)
(42,0)
(43,0)
(44,0)
(45,0)
(46,0)
(47,0)
(48,0)
(49,0)
(50,0)
(51,2e-05)
(52,4e-05)
(53,0)
(54,0)
(55,0)
(56,1e-05)
(57,0)
(58,0)
(59,0)
(60,0)
(61,0)
(62,0)
(63,0)
(64,0)
(65,0)
(66,0)
(67,0)
(68,0)
(69,0)
(70,0)
(71,0)
(72,4e-05)
(73,1e-05)
(74,0)
(75,0)
(76,0)
(77,0)
(78,0)
(79,0)
(80,0)
(81,0)
(82,3e-05)
(83,0.00047)
(84,2e-05)
(85,0)
(86,0)
(87,0)
(88,0)
(89,0)
(90,0)
(91,0)
(92,0)
(93,0)
(94,0)
(95,0)
(96,2e-05)
(97,0)
(98,0)
(99,0)
(100,0)
(101,0)
(102,0)
(103,0)
(104,0)
(105,0)
(106,0)
(107,0)
(108,0)
(109,1e-05)
(110,0)
(111,0)
(112,0)
(113,0)
(114,0)
(115,0)
(116,0)
(117,0)
(118,0)
(119,0)
(120,0)
(121,0)
(122,0)
(123,0)
(124,0)
(125,0)
(126,0)
(127,0)
(128,0)
(129,0)
(130,0)
(131,0)
(132,0)
(133,0)
(134,0)
(135,0)
(136,0)
(137,0)
(138,0)
(139,0)
(140,0)
(141,0)
(142,0)
(143,0)
(144,0)
(145,0)
(146,0)
(147,0)
(148,0)
(149,1e-05)
(150,0)
(151,0)
(152,0)
(153,0)
(154,0)
(155,0)
(156,0)
(157,1e-05)
(158,0)
(159,0)
(160,0)
(161,0)
(162,0)
(163,0)
(164,0)
(165,0)
(166,0)
(167,0)
(168,0)
(169,0)
(170,0)
(171,0)
(172,0)
(173,0)
(174,0)
(175,0)
(176,0)
(177,0)
(178,0)
(179,0)
(180,0)
(181,0)
(182,0)
(183,0)
(184,0)
(185,0)
(186,0)
(187,0)
(188,0)
(189,0)
(190,0)
(191,0)
(192,0)
(193,0)
(194,0)
(195,0)
(196,0)
(197,0)
(198,0)
(199,0)
(200,0)
(201,0)
(202,0)
(203,0)
(204,0)
(205,0)
(206,0)
(207,0)
(208,0)
(209,0)
(210,0)
(211,0)
(212,0)
(213,1e-05)
(214,0)
(215,0)
(216,0)
(217,0)
(218,0)
(219,0)
(220,0)
(221,0)
(222,0)
(223,0)
(224,0)
(225,0)
(226,0)
(227,0)
(228,0)
(229,0)
(230,0)
(231,0)
(232,0)
(233,0)
(234,0)
(235,0)
(236,0)
(237,0)
(238,0)
(239,0)
(240,0)
(241,0)
(242,0)
(243,0)
(244,0)
(245,0)
(246,0)
(247,0)
(248,0)
(249,0)
(250,0)
(251,0)
(252,0)
(253,0)
(254,0)
(255,0)
(256,0)
(257,0)
(258,0)
(259,0)
(260,0)
(261,0)
(262,0)
(263,0)
(264,0)
(265,0)
(266,0)
(267,0)
(268,0)
(269,0)
(270,0)
(271,0)
(272,0)
(273,0)
(274,0)
(275,0)
(276,0)
(277,0)
(278,0)
(279,0)
(280,0)
(281,0)
(282,0)
(283,0)
(284,0)
(285,0)
(286,0)
(287,0)
(288,0)
(289,0)
(290,0)
(291,0)
(292,0)
(293,0)
(294,0)
(295,0)
(296,0)
(297,0)
(298,0)
(299,0)
(300,0)
(301,0)
(302,0)
(303,0)
(304,0)
(305,0)
(306,0)
(307,0)
(308,0)
(309,0)
(310,0)
(311,0)
(312,0)
(313,0)
(314,0)
(315,0)
(316,0)
(317,0)
(318,0)
(319,0)
(320,0)
(321,0)
(322,0)
(323,0)
(324,0)
(325,0)
(326,0)
(327,0)
(328,0)
(329,0)
(330,0)
(331,0)
(332,0)
(333,0)
}

\def\djpegEXECDATA{
(0,1.46755)
(1,15.42884)
(2,18.32596)
(3,22.67791)
(4,21.68439)
(5,27.33197)
(6,22.96646)
(7,16.52576)
(8,14.04361)
(9,20.31706)
(10,21.76329)
(11,23.74027)
(12,16.93226)
(13,16.84412)
(14,19.01968)
(15,14.83786)
(16,12.36668)
(17,14.80234)
(18,12.78799)
(19,16.34162)
(20,23.80656)
(21,28.39476)
(22,18.97271)
(23,16.93262)
(24,18.23657)
(25,21.12744)
(26,17.43101)
(27,16.43382)
(28,17.40631)
(29,24.50638)
(30,16.89918)
(31,16.93844)
(32,20.38142)
(33,19.80666)
(34,23.76673)
(35,22.25918)
(36,19.77112)
(37,20.13393)
(38,24.34883)
(39,19.43207)
(40,21.63315)
(41,18.60169)
(42,18.90705)
(43,18.31841)
(44,16.98098)
(45,18.47904)
(46,17.64853)
(47,16.82562)
(48,16.45237)
(49,16.85857)
(50,17.0144)
(51,21.64985)
(52,29.04061)
(53,19.75339)
(54,6.1735)
(55,20.7231)
(56,20.00725)
(57,21.80253)
(58,27.03625)
(59,24.13678)
(60,16.44059)
(61,18.00952)
(62,20.00847)
(63,17.87964)
(64,17.14874)
(65,9.84128)
(66,15.83515)
(67,13.43477)
(68,30.01037)
(69,23.83588)
(70,9.92242)
(71,16.2463)
(72,20.76679)
(73,21.63694)
(74,24.05017)
(75,19.28807)
(76,16.22399)
(77,19.00928)
(78,13.83597)
(79,11.83882)
(80,11.00456)
(81,18.34096)
(82,7.6674)
(83,7.66159)
(84,18.83062)
(85,16.71645)
(86,17.02114)
(87,17.02106)
(88,16.60587)
(89,21.82744)
(90,16.37765)
(91,16.59363)
(92,17.93247)
(93,18.11665)
(94,17.01725)
(95,16.54732)
(96,16.83272)
(97,17.01416)
(98,16.68416)
(99,16.36749)
(100,16.85271)
(101,17.02304)
(102,20.94477)
(103,24.87943)
(104,21.75845)
(105,16.3574)
(106,17.4617)
(107,24.29919)
(108,22.86382)
(109,16.67976)
(110,16.53222)
(111,23.18454)
(112,20.01164)
(113,17.00344)
(114,21.61989)
(115,24.58834)
(116,23.28614)
(117,22.49957)
(118,22.11521)
(119,23.9421)
(120,16.99319)
(121,20.77297)
(122,28.56259)
(123,16.58183)
(124,16.52874)
(125,19.97852)
(126,19.03127)
(127,18.53623)
(128,20.51819)
(129,16.65978)
(130,16.76058)
(131,22.08478)
(132,21.34923)
(133,22.41172)
(134,21.20592)
(135,24.11179)
(136,20.76716)
(137,16.86707)
(138,22.04948)
(139,26.83997)
(140,23.90815)
(141,17.40229)
(142,18.9613)
(143,17.07083)
(144,28.34915)
(145,19.51739)
(146,16.30246)
(147,24.36366)
(148,25.954)
(149,20.24444)
(150,20.95348)
(151,17.56464)
(152,15.83783)
(153,21.17668)
(154,25.22529)
(155,26.02347)
(156,23.62896)
(157,16.90457)
(158,13.27547)
(159,12.3884)
(160,21.67089)
(161,16.52793)
(162,12.99434)
(163,14.93401)
(164,20.28216)
(165,19.34853)
(166,21.66232)
(167,27.01196)
(168,23.98658)
(169,23.47735)
(170,24.02214)
(171,25.27369)
(172,16.72839)
(173,16.86378)
(174,17.02967)
(175,16.46901)
(176,18.79755)
(177,16.67326)
(178,17.00675)
(179,20.0096)
(180,18.00342)
(181,17.00432)
(182,15.54637)
(183,18.21784)
(184,25.67617)
(185,21.31856)
(186,16.6779)
(187,16.63349)
(188,21.03598)
(189,23.54841)
(190,16.6868)
(191,18.82266)
(192,21.13509)
(193,13.821)
(194,21.94247)
(195,18.1657)
(196,19.26331)
(197,21.20858)
(198,20.50128)
(199,20.60617)
(200,21.89457)
(201,16.56621)
(202,17.39888)
(203,21.57586)
(204,24.64392)
(205,18.69015)
(206,19.0321)
(207,21.39145)
(208,23.44112)
(209,19.1869)
(210,23.92178)
(211,20.70391)
(212,16.89704)
(213,22.34272)
(214,16.71035)
(215,18.89805)
(216,18.56808)
(217,16.98244)
(218,16.97753)
(219,19.98331)
(220,26.64211)
(221,21.63427)
(222,16.85881)
(223,16.98955)
(224,20.01653)
(225,27.70103)
(226,16.97074)
(227,16.38904)
(228,16.90538)
(229,20.43183)
(230,25.83077)
(231,24.6306)
(232,29.59574)
(233,29.33134)
(234,26.09852)
(235,25.82699)
(236,17.91021)
(237,20.32873)
(238,20.73261)
(239,23.22896)
(240,20.05114)
(241,17.04308)
(242,20.9586)
(243,25.73503)
(244,22.98727)
(245,18.31636)
(246,10.85079)
(247,17.52362)
(248,12.34772)
(249,16.53973)
(250,14.82235)
(251,17.92791)
(252,13.93891)
(253,13.34435)
(254,16.50541)
(255,17.50614)
(256,24.18265)
(257,21.01371)
(258,27.85073)
(259,29.67841)
(260,17.97771)
(261,17.07689)
(262,17.52146)
(263,17.02949)
(264,16.21577)
(265,21.34664)
(266,21.97947)
(267,22.4318)
(268,16.88155)
(269,16.40679)
(270,16.90404)
(271,16.9517)
(272,22.13725)
(273,27.07499)
(274,21.66634)
(275,20.46298)
(276,16.23946)
(277,16.95171)
(278,20.00789)
(279,14.1766)
(280,17.91336)
(281,16.57129)
(282,17.95298)
(283,16.95334)
(284,18.61683)
(285,20.79575)
(286,25.82753)
(287,20.88125)
(288,16.73573)
(289,24.44468)
(290,18.2229)
(291,16.91834)
(292,18.45377)
(293,20.38489)
(294,21.67529)
(295,18.95447)
(296,16.9533)
(297,19.45506)
(298,16.95334)
(299,16.81341)
(300,16.44688)
(301,18.45518)
(302,18.97486)
(303,16.93311)
(304,16.45097)
(305,23.15358)
(306,21.91056)
(307,19.12021)
(308,16.78765)
(309,16.95165)
(310,22.16941)
(311,17.24309)
(312,14.60278)
(313,22.2188)
(314,16.54024)
(315,20.45652)
(316,19.45672)
(317,17.11276)
(318,16.79225)
(319,17.30781)
(320,22.10492)
(321,17.4522)
(322,14.56159)
(323,16.19697)
(324,14.87525)
(325,16.88277)
(326,16.86881)
(327,17.09209)
(328,18.90023)
(329,19.11781)
(330,12.64043)
(331,20.36506)
(332,19.77385)
(333,17.57699)
}

\def\gifGENDATA{
(0,0.00013)
(1,0.00033)
(2,0.00013)
(3,0)
(4,0.0004)
(5,0.0001)
(6,0.00017)
(7,3e-05)
(8,4e-05)
(9,0.00043)
(10,0.0001)
(11,0.00011)
(12,0.00012)
(13,0.00011)
(14,9e-05)
(15,6e-05)
(16,1e-05)
(17,1e-05)
(18,2e-05)
(19,0)
(20,1e-05)
(21,0)
(22,1e-05)
(23,0)
(24,0)
(25,0)
(26,0)
(27,1e-05)
(28,1e-05)
}

\def\gifEXECDATA{
(0,9.78758)
(1,27.15963)
(2,27.62022)
(3,20.70246)
(4,3.28745)
(5,4.70427)
(6,4.70158)
(7,5.18602)
(8,5.81398)
(9,5.59983)
(10,5.87454)
(11,4.81268)
(12,5.62444)
(13,5.72674)
(14,5.08399)
(15,5.61105)
(16,5.70906)
(17,4.61272)
(18,5.17271)
(19,6.09584)
(20,5.85981)
(21,5.64083)
(22,3.50795)
(23,4.04477)
(24,3.80544)
(25,3.87863)
(26,3.85176)
(27,3.72986)
(28,3.84842)
}

%% file: images/code_dups.tex
\captionsetup[lstlisting]{labelformat=empty, labelsep=space}

\definecolor{tabutter}{rgb}{0.98824, 0.91373, 0.30980}          %
\colorlet{Lighttabutter}{tabutter!25!white}
\definecolor{taorange}{rgb}{0.98824, 0.68627, 0.24314}          %
\colorlet{Lighttaorange}{taorange!25!white}
\definecolor{tachameleon}{rgb}{0.54118, 0.88627, 0.20392}       %
\colorlet{Lighttachameleon}{tachameleon!25!white}
\definecolor{taskyblue}{rgb}{0.44706, 0.56078, 0.81176}         %
\colorlet{Lighttaskyblue}{taskyblue!25!white}
\definecolor{tascarletred}{rgb}{0.93725, 0.16078, 0.16078}      %
\colorlet{Lighttascarletred}{tascarletred!25!white}
\definecolor{light-gray}{gray}{0.90}

\begin{figure*}[t]
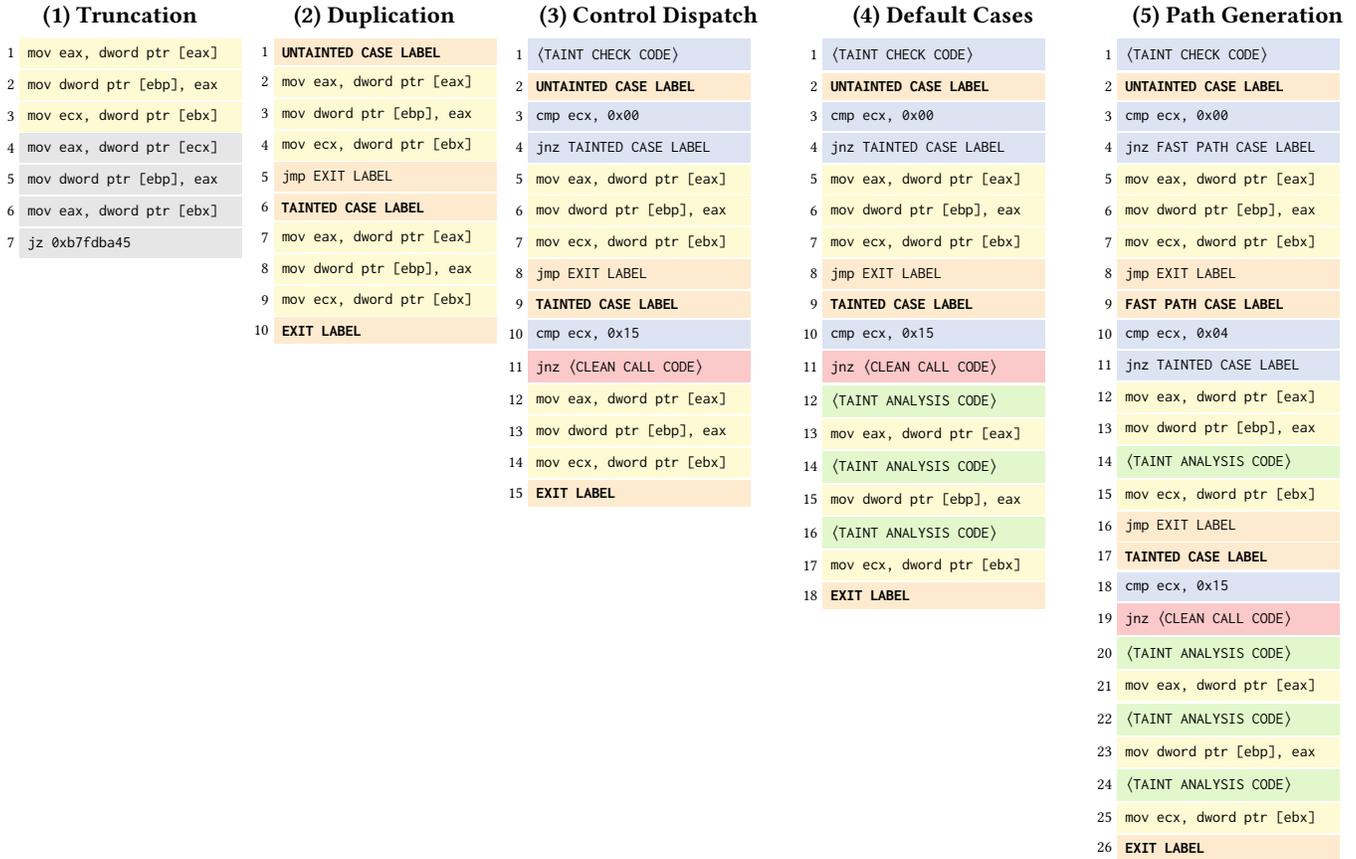

\begin{minipage}[t]{.15\textwidth}
\begin{lstlisting}[basicstyle=\fontsize{6}{6}\selectfont\ttfamily,columns=fullflexible,columns=fullflexible,breaklines=true,caption=(1) Truncation,numbers=left,breaklines=true,firstnumber=1,numbersep=2pt,postbreak=\mbox{\textcolor{red}{$\hookrightarrow$}\space},breakatwhitespace=false,,escapechar=!]{Name}
!\colorbox{Lighttabutter}{\parbox{13em}{mov eax, dword ptr [eax]}}!
!\colorbox{Lighttabutter}{\parbox{13em}{mov dword ptr [ebp], eax}}!
!\colorbox{Lighttabutter}{\parbox{13em}{mov ecx, dword ptr [ebx]}}!
!\colorbox{light-gray}{\parbox{13em}{mov eax, dword ptr [ecx]}}!
!\colorbox{light-gray}{\parbox{13em}{mov dword ptr [ebp], eax}}!
!\colorbox{light-gray}{\parbox{13em}{mov eax, dword ptr [ebx]}}!
!\colorbox{light-gray}{\parbox{13em}{jz 0xb7fdba45}}!
\end{lstlisting}
\end{minipage}\hfill
\begin{minipage}[t]{.15\textwidth}
\begin{lstlisting}[basicstyle=\fontsize{6}{6}\selectfont\ttfamily,columns=fullflexible,breaklines=true,caption=(2) Duplication,numbers=left,breaklines=true,firstnumber=1,numbersep=2pt,postbreak=\mbox{\textcolor{red}{$\hookrightarrow$}\space},breakatwhitespace=false,escapechar=!]{Name}
!\colorbox{Lighttaorange}{\parbox{13em}{\textbf{UNTAINTED CASE LABEL}}}!
!\colorbox{Lighttabutter}{\parbox{13em}{mov eax, dword ptr [eax]}}!
!\colorbox{Lighttabutter}{\parbox{13em}{mov dword ptr [ebp], eax}}!
!\colorbox{Lighttabutter}{\parbox{13em}{mov ecx, dword ptr [ebx]}}!
!\colorbox{Lighttaorange}{\parbox{13em}{jmp EXIT LABEL}}!
!\colorbox{Lighttaorange}{\parbox{13em}{\textbf{TAINTED CASE LABEL}}}!
!\colorbox{Lighttabutter}{\parbox{13em}{mov eax, dword ptr [eax]}}!
!\colorbox{Lighttabutter}{\parbox{13em}{mov dword ptr [ebp], eax}}!
!\colorbox{Lighttabutter}{\parbox{13em}{mov ecx, dword ptr [ebx]}}!
!\colorbox{Lighttaorange}{\parbox{13em}{\textbf{EXIT LABEL}}}!
\end{lstlisting}
\end{minipage}\hfill
\begin{minipage}[t]{.18\textwidth}
 \begin{lstlisting}[basicstyle=\fontsize{6}{6}\selectfont\ttfamily, columns=fullflexible,breaklines=true,caption=(3) Control Dispatch,numbers=left,breaklines=true,firstnumber=1,numbersep=2pt,postbreak=\mbox{\textcolor{red}{$\hookrightarrow$}\space},breakatwhitespace=false,,escapechar=!]{Name}
!\colorbox{Lighttaskyblue}{\parbox{13em}{$\langle$TAINT CHECK CODE$\rangle$}}!
!\colorbox{Lighttaorange}{\parbox{13em}{\textbf{UNTAINTED CASE LABEL}}}!
!\colorbox{Lighttaskyblue}{\parbox{13em}{cmp ecx, 0x00}}!
!\colorbox{Lighttaskyblue}{\parbox{13em}{jnz TAINTED CASE LABEL}}!
!\colorbox{Lighttabutter}{\parbox{13em}{mov eax, dword ptr [eax]}}!
!\colorbox{Lighttabutter}{\parbox{13em}{mov dword ptr [ebp], eax}}!
!\colorbox{Lighttabutter}{\parbox{13em}{mov ecx, dword ptr [ebx]}}!
!\colorbox{Lighttaorange}{\parbox{13em}{jmp EXIT LABEL}}!
!\colorbox{Lighttaorange}{\parbox{13em}{\textbf{TAINTED CASE LABEL}}}!
!\colorbox{Lighttaskyblue}{\parbox{13em}{cmp ecx, 0x15}}!
!\colorbox{Lighttascarletred}{\parbox{13em} {jnz $\langle$CLEAN CALL CODE$\rangle$}}!
!\colorbox{Lighttabutter}{\parbox{13em}{mov eax, dword ptr [eax]}}!
!\colorbox{Lighttabutter}{\parbox{13em}{mov dword ptr [ebp], eax}}!
!\colorbox{Lighttabutter}{\parbox{13em}{mov ecx, dword ptr [ebx]}}!
!\colorbox{Lighttaorange}{\parbox{13em}{\textbf{EXIT LABEL}}}!
\end{lstlisting}
\end{minipage}\hfill
\begin{minipage}[t]{.18\textwidth}
\begin{lstlisting}[basicstyle=\fontsize{6}{6}\selectfont\ttfamily ,columns=fullflexible,breaklines=true,caption=(4) Default Cases,numbers=left,breaklines=true,firstnumber=1,numbersep=2pt,postbreak=\mbox{\textcolor{red}{$\hookrightarrow$}\space},breakatwhitespace=false,,escapechar=!]{Name}
!\colorbox{Lighttaskyblue}{\parbox{13em}{$\langle$TAINT CHECK CODE$\rangle$}}!
!\colorbox{Lighttaorange}{\parbox{13em}{\textbf{UNTAINTED CASE LABEL}}}!
!\colorbox{Lighttaskyblue}{\parbox{13em}{cmp ecx, 0x00}}!
!\colorbox{Lighttaskyblue}{\parbox{13em}{jnz TAINTED CASE LABEL}}!
!\colorbox{Lighttabutter}{\parbox{13em}{mov eax, dword ptr [eax]}}!
!\colorbox{Lighttabutter}{\parbox{13em}{mov dword ptr [ebp], eax}}!
!\colorbox{Lighttabutter}{\parbox{13em}{mov ecx, dword ptr [ebx]}}!
!\colorbox{Lighttaorange}{\parbox{13em}{jmp EXIT LABEL}}!
!\colorbox{Lighttaorange}{\parbox{13em}{\textbf{TAINTED CASE LABEL}}}!
!\colorbox{Lighttaskyblue}{\parbox{13em}{cmp ecx, 0x15}}!
!\colorbox{Lighttascarletred}{\parbox{13em} {jnz $\langle$CLEAN CALL CODE$\rangle$}}!
!\colorbox{Lighttachameleon}{\parbox{13em}{$\langle$TAINT ANALYSIS CODE$\rangle$}}!
!\colorbox{Lighttabutter}{\parbox{13em}{mov eax, dword ptr [eax]}}!
!\colorbox{Lighttachameleon}{\parbox{13em}{$\langle$TAINT ANALYSIS CODE$\rangle$}}!
!\colorbox{Lighttabutter}{\parbox{13em}{mov dword ptr [ebp], eax}}!
!\colorbox{Lighttachameleon}{\parbox{13em}{$\langle$TAINT ANALYSIS CODE$\rangle$}}!
!\colorbox{Lighttabutter}{\parbox{13em}{mov ecx, dword ptr [ebx]}}!
!\colorbox{Lighttaorange}{\parbox{13em}{\textbf{EXIT LABEL}}}!
\end{lstlisting}
\end{minipage}\hfill
\begin{minipage}[t]{.18\textwidth}
\begin{lstlisting}[basicstyle=\fontsize{6}{6}\selectfont\ttfamily,columns=fullflexible,breaklines=true,caption=(5) Path Generation ,numbers=left,breaklines=true,firstnumber=1,numbersep=2pt,postbreak=\mbox{\textcolor{red}{$\hookrightarrow$}\space},breakatwhitespace=false,,escapechar=!]{Name}
!\colorbox{Lighttaskyblue}{\parbox{13em}{$\langle$TAINT CHECK CODE$\rangle$}}!
!\colorbox{Lighttaorange}{\parbox{13em}{\textbf{UNTAINTED CASE LABEL}}}!
!\colorbox{Lighttaskyblue}{\parbox{13em}{cmp ecx, 0x00}}!
!\colorbox{Lighttaskyblue}{\parbox{13em}{jnz FAST PATH CASE LABEL}}!
!\colorbox{Lighttabutter}{\parbox{13em}{mov eax, dword ptr [eax]}}!
!\colorbox{Lighttabutter}{\parbox{13em}{mov dword ptr [ebp], eax}}!
!\colorbox{Lighttabutter}{\parbox{13em}{mov ecx, dword ptr [ebx]}}!
!\colorbox{Lighttaorange}{\parbox{13em}{jmp EXIT LABEL}}!
!\colorbox{Lighttaorange}{\parbox{13em}{\textbf{FAST PATH CASE LABEL}}}!
!\colorbox{Lighttaskyblue}{\parbox{13em}{cmp ecx, 0x04}}!
!\colorbox{Lighttaskyblue}{\parbox{13em}{jnz TAINTED CASE LABEL}}!
!\colorbox{Lighttabutter}{\parbox{13em}{mov eax, dword ptr [eax]}}!
!\colorbox{Lighttabutter}{\parbox{13em}{mov dword ptr [ebp], eax}}!
!\colorbox{Lighttachameleon}{\parbox{13em}{$\langle$TAINT ANALYSIS CODE$\rangle$}}!
!\colorbox{Lighttabutter}{\parbox{13em}{mov ecx, dword ptr [ebx]}}!
!\colorbox{Lighttaorange}{\parbox{13em}{jmp EXIT LABEL}}!
!\colorbox{Lighttaorange}{\parbox{13em}{\textbf{TAINTED CASE LABEL}}}!
!\colorbox{Lighttaskyblue}{\parbox{13em}{cmp ecx, 0x15}}!
!\colorbox{Lighttascarletred}{\parbox{13em} {jnz $\langle$CLEAN CALL CODE$\rangle$}}!
!\colorbox{Lighttachameleon}{\parbox{13em}{$\langle$TAINT ANALYSIS CODE$\rangle$}}!
!\colorbox{Lighttabutter}{\parbox{13em}{mov eax, dword ptr [eax]}}!
!\colorbox{Lighttachameleon}{\parbox{13em}{$\langle$TAINT ANALYSIS CODE$\rangle$}}!
!\colorbox{Lighttabutter}{\parbox{13em}{mov dword ptr [ebp], eax}}!
!\colorbox{Lighttachameleon}{\parbox{13em}{$\langle$TAINT ANALYSIS CODE$\rangle$}}!
!\colorbox{Lighttabutter}{\parbox{13em}{mov ecx, dword ptr [ebx]}}!
!\colorbox{Lighttaorange}{\parbox{13em}{\textbf{EXIT LABEL}}}!
\end{lstlisting}
\end{minipage}

\caption{ A code example showing the instrumentation steps for fast path generation. The basic block is first truncated as the address in \texttt{ecx} is obtained via another memory access (at line 3). The remaining code is duplicated in the second step. Jumps and labels are also inserted. Taint checks and the control dispatcher are then inserted in step 3. At line 11, a clean call triggers path generation when control reaches the end of the compare and branch sequence. In step 4, the Taint Rabbit weaves analysis code for the two default cases (i.e.\ no taint and full taint instrumentation). Step 5 shows the inclusion of a  generated path.}
\label{fig:dfpg_stages}
\end{figure*}

%% file: charts/sampling.tex
\definecolor{tabutter}{rgb}{0.98824, 0.91373, 0.30980}		%
\definecolor{ta2butter}{rgb}{0.92941, 0.83137, 0}		%
\definecolor{ta3butter}{rgb}{0.76863, 0.62745, 0}		%

\definecolor{taorange}{rgb}{0.98824, 0.68627, 0.24314}		%
\definecolor{ta2orange}{rgb}{0.96078, 0.47451, 0}		%
\definecolor{ta3orange}{rgb}{0.80784, 0.36078, 0}		%

\definecolor{tachocolate}{rgb}{0.91373, 0.72549, 0.43137}	%
\definecolor{ta2chocolate}{rgb}{0.75686, 0.49020, 0.066667}	%
\definecolor{ta3chocolate}{rgb}{0.56078, 0.34902, 0.0078431}	%

\definecolor{tachameleon}{rgb}{0.54118, 0.88627, 0.20392}	%
\definecolor{ta2chameleon}{rgb}{0.45098, 0.82353, 0.086275}	%
\definecolor{ta3chameleon}{rgb}{0.30588, 0.60392, 0.023529}	%

\definecolor{taskyblue}{rgb}{0.44706, 0.56078, 0.81176}		%
\definecolor{ta2skyblue}{rgb}{0.20392, 0.39608, 0.64314}	%
\definecolor{ta3skyblue}{rgb}{0.12549, 0.29020, 0.52941}	%

\definecolor{taplum}{rgb}{0.67843, 0.49804, 0.65882}		%
\definecolor{ta2plum}{rgb}{0.45882, 0.31373, 0.48235}		%
\definecolor{ta3plum}{rgb}{0.36078, 0.20784, 0.4}		%

\definecolor{tascarletred}{rgb}{0.93725, 0.16078, 0.16078}	%
\definecolor{ta2scarletred}{rgb}{0.8, 0, 0}			%
\definecolor{ta3scarletred}{rgb}{0.64314, 0, 0}			%

\begin{figure}
\centering
\begin{tikzpicture}[xscale=.45, yscale=.45]
    \begin{axis}[
        width  = 18cm,
        height = 8.5cm,
        major x tick style = transparent,
        ybar=2*\pgflinewidth,
        ymode=log,
        log basis y={10},
        bar width=11.2pt,
        ybar=2*\pgflinewidth,
        ymajorgrids = true,
        ylabel = {slowdown},
        symbolic x coords={Compress, PHP, Image, Apache,  SpecCPU},
        xtick = data,
        label style={font=\huge},
        tick label style={font=\huge}, 
        scaled y ticks = false,
        enlarge x limits=0.12,    
        legend cell align=left,
        legend style={draw={none},font=\huge},
        legend style={/tikz/every even column/.append style={column sep=0.1cm}},
       legend style={at={(-0.12,1.2)},anchor=north west,
                     legend columns=7,
				     cells={align=left}},]
    ]

     \addplot[style={fill=tabutter,mark=none}]
            coordinates {(Compress,2.345086744) (PHP,94.88418394) (Image,4.320394627) (Apache,8.849902434) (SpecCPU,22.41325807)};

     \addplot[style={fill=taorange,mark=none}]
            coordinates {(Compress,2.041323051) (PHP,98.26829828) (Image,4.101958078) (Apache,9.571644847) (SpecCPU,18.34795884)};

     \addplot[style={fill=tachocolate,mark=none}]
            coordinates {(Compress,1.941393567) (PHP,98.84060133) (Image,4.145351413) (Apache,8.09643337) (SpecCPU,17.56393075)};

     \addplot[style={fill=tachameleon,mark=none}]
            coordinates {(Compress,1.500737357) (PHP,97.37756925) (Image,4.000450078) (Apache,7.490158352) (SpecCPU,12.51241288)};

     \addplot[style={fill=taskyblue,mark=none}]
            coordinates {(Compress,1.478680746) (PHP,95.51629317) (Image,1.697332514) (Apache,4.485657112) (SpecCPU,5.774999096)};

        \legend{100\% (all), 50\%, 25\%, 5\%, 0\% (none) }
    \end{axis}
\end{tikzpicture}
\caption{The Taint Rabbit achieves better performance when less taint is introduced because more fast paths are  executed. To aid validate this claim, we ran the same experiments but randomly sampled taint introduction based on various probabilities. For instance, on SPEC CPU,
overall overhead is reduced to 12.5x from 22.4x when the odds of introducing taint is set to $\frac{1}{25}$. }
\label{fig:sampling_test}
\end{figure}
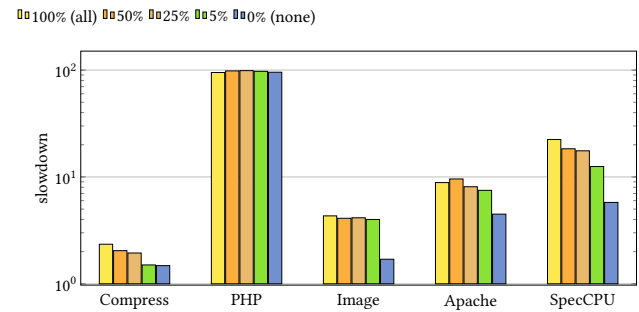

%% file: charts/param_analysis.tex
\captionsetup[subfigure]{labelformat=parens, labelsep=space}
\addtocounter{figure}{+1}

\definecolor{tabutter}{rgb}{0.98824, 0.91373, 0.30980}		%
\definecolor{ta2butter}{rgb}{0.92941, 0.83137, 0}		%
\definecolor{ta3butter}{rgb}{0.76863, 0.62745, 0}		%

\definecolor{taorange}{rgb}{0.98824, 0.68627, 0.24314}		%
\definecolor{ta2orange}{rgb}{0.96078, 0.47451, 0}		%
\definecolor{ta3orange}{rgb}{0.80784, 0.36078, 0}		%

\definecolor{tachocolate}{rgb}{0.91373, 0.72549, 0.43137}	%
\definecolor{ta2chocolate}{rgb}{0.75686, 0.49020, 0.066667}	%
\definecolor{ta3chocolate}{rgb}{0.56078, 0.34902, 0.0078431}	%

\definecolor{tachameleon}{rgb}{0.54118, 0.88627, 0.20392}	%
\definecolor{ta2chameleon}{rgb}{0.45098, 0.82353, 0.086275}	%
\definecolor{ta3chameleon}{rgb}{0.30588, 0.60392, 0.023529}	%

\definecolor{taskyblue}{rgb}{0.44706, 0.56078, 0.81176}		%
\definecolor{ta2skyblue}{rgb}{0.20392, 0.39608, 0.64314}	%
\definecolor{ta3skyblue}{rgb}{0.12549, 0.29020, 0.52941}	%

\definecolor{taplum}{rgb}{0.67843, 0.49804, 0.65882}		%
\definecolor{ta2plum}{rgb}{0.45882, 0.31373, 0.48235}		%
\definecolor{ta3plum}{rgb}{0.36078, 0.20784, 0.4}		%

\definecolor{tascarletred}{rgb}{0.93725, 0.16078, 0.16078}	%
\definecolor{ta2scarletred}{rgb}{0.8, 0, 0}			%
\definecolor{ta3scarletred}{rgb}{0.64314, 0, 0}			%

\begin{figure}[t]

\centering

\begin{tabularx}{0.49\textwidth}{X X}
\begin{tikzpicture}[xscale=.45, yscale=.45]

\begin{axis}[
        height = 5.5 cm,
        width= 8 cm,
    ylabel = {Slowdown},
    xlabel = {Possible Fast Paths},
    ymajorgrids = true,
    label style={font=\huge},
    tick label style={font=\huge},
    symbolic x coords={0, 1, 2, 4, 8},
    xtick=data]
    ]
    \addplot[line width=3pt,mark size=3pt, tabutter, mark=*, mark options={fill=tabutter}] coordinates {
        (0,2.706781766)        
        (1,2.345086744)
        (2,2.3425344)
        (4,2.347115227)
        (8,2.379194168)
    };

\end{axis}

\end{tikzpicture}
\vspace*{-0.3cm}
\captionof{subfigure}{Compression}
\label{fig:num_paths_start}
&

\begin{tikzpicture}[xscale=.45, yscale=.45]

\begin{axis}[
        height = 5.5 cm,
        width= 8 cm,
    ylabel = {Slowdown},
    xlabel = {Possible Fast Paths},
    ymajorgrids = true,
    label style={font=\huge},
    tick label style={font=\huge},
    symbolic x coords={0, 1, 2, 4, 8},
    xtick=data]
    ]
    \addplot[line width=3pt, mark size=3pt, taorange, mark=square*, mark options={fill=taorange}] coordinates {
        (0,95.91971663)        
        (1,94.88418394)
        (2,98.96889327)
        (4,98.59980565)
        (8,100.5410262)
    };
\end{axis}
\end{tikzpicture}
\vspace*{-0.3cm}
\captionof{subfigure}{PHPBench}

\\

\begin{tikzpicture}[xscale=.45, yscale=.45]
\begin{axis}[
        height = 5.5 cm,
        width= 8  cm,
    ylabel = {Slowdown},
    xlabel = {Possible Fast Paths},
    ymajorgrids = true,
    label style={font=\huge},
    tick label style={font=\huge},
    symbolic x coords={0, 1, 2, 4, 8},
    xtick=data]
    ]

    \addplot[line width=3pt, mark size=3pt, tachocolate, mark=triangle*, mark options={fill=tachocolate}] coordinates {
        (0,4.636972076)
        (1,4.320394627)
        (2,4.48437859)
        (4,4.674507013)
        (8,4.72418732)
    };
 
\end{axis}

\end{tikzpicture}
\vspace*{-0.3cm}
\captionof{subfigure}{Image Processing}

&

\begin{tikzpicture}[xscale=.45, yscale=.45]
\begin{axis}[
        height = 5.5 cm,
        width= 8 cm,
    ylabel = {Slowdown},
    xlabel = {Possible Fast Paths},
    ymajorgrids = true,
    label style={font=\huge},
    tick label style={font=\huge},
    symbolic x coords={0, 1, 2, 4, 8},
    xtick=data]
    ]
    
    \addplot[line width=3pt, mark size=3pt, taskyblue, mark=*, mark options={fill=taskyblue}] coordinates {
        (0,24.96573636) 
        (1,22.41325807)
        (2,22.28299059)
        (4,21.76606565)
        (8,21.67801951)
    };
 
\end{axis}

\end{tikzpicture}
\captionof{subfigure}{SPEC CPU}
\vspace*{-0.3cm}
\label{fig:num_paths_end}

\end{tabularx}
\addtocounter{figure}{-1}
\captionof{figure}{Overhead vs. Number of Possible Fast Paths}
\end{figure}